\documentclass[12pt]{article}
\usepackage{mathtools}
\usepackage{amsmath}
\usepackage{ntheorem}
\usepackage{amsfonts}
\usepackage{verbatim}
\usepackage{epsfig}
\usepackage[comma,sort]{natbib}

\usepackage{enumitem}
\usepackage{url}
\usepackage{floatrow}
\usepackage{multirow}
\usepackage{bm}
\usepackage{soul}
\usepackage[dvipsnames]{xcolor}
\usepackage{rotating}
\usepackage[titletoc]{appendix}
\usepackage{tikz,pgfplots}
\usepackage{threeparttable}
\usepackage{float}
\usepackage[most]{tcolorbox}
\floatstyle{plaintop}
\restylefloat{table}
\usepackage{array}
\usepackage[pdfencoding=auto]{hyperref} 
\hypersetup{hidelinks} 
\usepackage{bookmark}
\mathtoolsset{showonlyrefs}

\usetikzlibrary{calc,shapes}
\newcommand{\tikzmark}[1]{\tikz[overlay,remember picture] \node (#1) {};}

\newcommand*\widebar[1]{%
   \hbox{%
     \vbox{%
       \hrule height 0.5pt 
       \kern0.35ex
       \hbox{%
         \kern-0.0em
         \ensuremath{#1}%
         \kern-0.0em
       }%
     }%
   }%
} 

\newtheorem{theorem}{Theorem}

\newtheorem{assumption}{Assumption}
\newtheorem{lemma}[theorem]{Lemma}

\newcommand{\NewFormulation}{\textit{Weakly-targeted}}

\newcommand{\bs}{\boldsymbol}

\newcommand{\cB}{\mathcal{B}}
\newcommand{\cE}{\mathcal{E}}
\newcommand{\bI}{\mathbbm{1}}
\newcommand{\bN}{\mathbb{N}}
\newcommand{\bR}{\mathbb{R}}
\newcommand{\bE}{E}
\newcommand{\mathbbm}[1]{\text{\usefont{U}{bbm}{m}{n}#1}} 

\newcommand{\sP}{\mathcal{E}}

\newcommand{\cT}{\mathcal{T}}

\newcommand{\TcE}{T_\mathcal{E}}

\newcommand{\tbW}{\tilde{\bm{W}}}
\newcommand{\minT}{T_p}
\newcommand\Halmos{\rule{.36em}{.5em}}

\allowdisplaybreaks

\makeatletter
\def\monthname{\ifcase\month\or
January\or February\or March\or April\or May\or June\or
July\or August\or September\or October\or November\or December\fi}
\makeatother

\makeatletter
\def\@sect#1#2#3#4#5#6[#7]#8{\ifnum #2>\c@secnumdepth
     \let\@svsec\@empty\else
     \refstepcounter{#1}\edef\@svsec{\csname the#1\endcsname. \hskip 0.4em}\fi
     \@tempskipa #5\relax
      \ifdim \@tempskipa>\z@
        \begingroup #6\relax
          \@hangfrom{\hskip #3\relax\@svsec}{\interlinepenalty \@M #8\par}%
        \endgroup
       \csname #1mark\endcsname{#7}\addcontentsline
         {toc}{#1}{\ifnum #2>\c@secnumdepth \else
                      \protect\numberline{\csname the#1\endcsname}\fi
                    #7}\else
        \def\@svsechd{#6\hskip #3\relax  
                   \@svsec #8\csname #1mark\endcsname
                      {#7}\addcontentsline
                           {toc}{#1}{\ifnum #2>\c@secnumdepth \else
                             \protect\numberline{\csname the#1\endcsname}\fi
                       #7}}\fi
     \@xsect{#5}}
\makeatother

\makeatletter
\renewcommand{\section}{\@startsection{section}{1}{0mm}{-\baselineskip}{0.25\baselineskip}{\center\normalfont\normalsize\bf}}
\renewcommand{\subsection}{\@startsection{subsection}{2}{0mm}{-\baselineskip}{0.05\baselineskip}{\raggedright\normalfont\normalsize\bf}}
\renewcommand{\subsubsection}{\@startsection{subsubsection}{3}{0mm}{-\baselineskip}{0.05\baselineskip}{\raggedright\normalfont\normalsize\itshape}}
\def\@begintheorem#1#2{\trivlist \item[\hskip \labelsep{\bf #1\ #2:}]\it}
\makeatother

\makeatletter
\def\blfootnote{\xdef\@thefnmark{}\@footnotetext}
\makeatother

\makeatletter
\newtheoremstyle{MyNonumberplain}%
  {\item[\theorem@headerfont\hskip\labelsep ##1\theorem@separator]}%
  {\item[\theorem@headerfont\hskip\labelsep ##3\theorem@separator]}
\makeatother
\theoremstyle{MyNonumberplain}
\theorembodyfont{\upshape}
\newtheorem{proof}{Proof}

\renewenvironment{abstract}
 {\begin{center}\normalsize\bf\text{Abstract}
 \end{center}\begin{quote}\normalsize}
 {\end{quote}}
 
\begin{document}
\setcounter{page}{0}
\thispagestyle{empty}
\begin{center}
{\Large\bf Synthetic Controls for Experimental Design}

    {

\par}%
\blfootnote{\hspace*{-0.25in}Alberto Abadie, Department of Economics, MIT, abadie@mit.edu. Jinglong Zhao, Questrom School of Business, Boston University, jinglong@bu.edu. The authors are grateful to Victor Chernozhukov, Guido Imbens, Rahul Mazumder, Jaume Vives-i-Bastida, Yinchu Zhu and seminar participants at Amazon.com, Brown, Harvard, Princeton, Stanford, the Online Causal Inference Seminar, and USC for helpful comments and discussions. The replication codes can be found at \href{https://github.com/jinglongzhao2/SCDesign}{Github}. Abadie gratefully acknowledges NSF funding (SES-1756692).
}
    \vskip 1em
    {\large
     \lineskip .75em%
      \begin{tabular}[t]{ccc}%
       Alberto Abadie & & Jinglong Zhao\\
       MIT & & Boston University\\
      \end{tabular}

      \par}%
     \vskip 1em%
      {\large\monthname \ \number\year} \par
       \vskip 2em%
  \end{center}\par

\begin{abstract}
This article studies experimental design in settings where the experimental units are large aggregate entities (e.g., markets), and only one or a small number of units can be exposed to the treatment. In such settings, randomization of the treatment may result in treated and control groups with substantially different baseline characteristics, inducing biases. We propose a variety of experimental non-randomized synthetic
control designs \citep{abadie2003synthetic, abadie2010synthetic} 
that select the units to be treated, as well as the untreated units to be used as a control group. Average potential outcomes with treatment are estimated as weighted averages of observed outcomes for treated units, and average potential outcomes without treatment as weighted averages of observed outcomes for control units. We analyze the properties of estimators based on synthetic control designs and propose new inferential techniques. We show that in experimental settings with aggregate units, synthetic control designs can substantially reduce estimation biases in comparison to randomization of the treatment.
\end{abstract}

\setcounter{page}{1}
\addtolength{\baselineskip}{0.5\baselineskip}

\section{Introduction}
\label{section:introduction}

Consider the problem of a ride-sharing company choosing between two compensation plans for drivers (\citeauthor{doudchenkonddesigning}, n.d.; \citeauthor{jones2019uber}, \citeyear{jones2019uber}). The company can either keep the current compensation plan or adopt a new one with higher incentives. In order to estimate the effect of a change in compensation plans on profits, the company's data science unit designs an experimental evaluation where the new plan is deployed at a small scale, say, in one of the local markets (cities) in the country. In this setting, a randomized control trial\,---\,or A/B test, where drivers in a local market are randomized into the new plan (active treatment arm) or the status quo (control treatment arm)\,---\,is problematic. On the one hand, such an experiment raises equity concerns, as drivers in the same local market but in different treatment arms obtain different compensations for the same jobs. On the other hand, if drivers in the active treatment arm respond to higher incentives by working longer hours, they will effectively steal business from drivers in the control arm of the experiment, resulting in biased experimental estimates.\footnote{A randomized evaluation across many markets is a potential solution to the problem of experimental interference between drivers. In practice, however, large-scale market-level randomized evaluations are often unfeasible. In the context of the ridesharing company example, large-scale market-level randomized evaluations {\em (i)} could be prohibitively expensive, {\em (ii)} could still raise substantial equity concerns, {\em (iii)} could negatively affect morale for the large number of drivers in the treated cities if the program is rolled back after experimentation, and {\em (iv)} in some cases, the number of cities where the company operates could be too small for effective randomization.}

One possible approach to this problem is to assign an entire local market to treatment, and use the rest of the local markets, which remain under the current compensation plan during the experimental periods, as potential comparison units. In this setting, using randomization to assign the active treatment allows ex-ante (i.e., pre-randomization) unbiased estimation of the effect of the active treatment. However, ex-post (i.e., post-randomization) biases can be large if, at baseline, the treated unit differs from the untreated units in the values of the features that affect the outcomes of interest. We document the magnitude and practical relevance of these biases in Sections~\ref{sec:Walmart} and~\ref{sec:Simulation}.

As in the ride-sharing example where there is only one treated local market, large biases may arise more generally in randomized studies when either the treatment arm or the control arm contains a small number of units, so randomized treatment assignment may not produce treated and control groups that are similar in their features. In those cases, the fact that estimation biases would have averaged out over alternative treatment assignments is of little comfort to a researcher who, in practice, is limited to one assignment only. 

To address these challenges, we propose using the synthetic control method \citep{abadie2003synthetic, abadie2010synthetic} as an experimental design to select treated units in non-randomized experiments, as well as the untreated units to serve as a comparison group. We adopt the name {\em synthetic control designs} to refer to the resulting experimental designs.\footnote{While we leave the ``experimental'' qualifier implicit in ``synthetic control design'', it should be noted that the synthetic control designs proposed in this article differ from observational synthetic control designs \citep[e.g.,][]{abadie2003synthetic, abadie2010synthetic, doudchenko2016balancing}, for which the identity of the treated unit(s) is taken as given.}$^,$\footnote{See, e.g.,  \cite{abadie2021using}, \cite{amjad2018robust}, \cite{arkhangelsky2019synthetic}, \cite{doudchenko2016balancing} for background material on synthetic controls and related methods.}

In our framework, the choice of the treated unit (or treated units, if multiple treated units are desired) aims to accomplish two goals.  First, the treated units should be representative of an aggregate of interest, such as a national market, so that the estimated effect reflects the aggregate impact of the treatment. Second, the treated units should not be idiosyncratic in the sense that the untreated units cannot closely approximate their features. Otherwise, the reliability of the estimate of the effect on the treated unit may be questionable. We show how to achieve these two objectives, whenever they are possible to achieve, using synthetic control methods.

While we are aware of the extensive use of synthetic control methods for experimental design in data science units, especially in the technology industry,\footnote{See, in particular, \cite{jones2019uber}, which also provides the basis for the ride-sharing example above.} the academic literature on this subject is at a nascent stage. There are, however, a few publicly available studies that are connected to this article. Aside from the present article, to our knowledge, \citeauthor{doudchenkonddesigning}~(n.d.) and \cite{doudchenko2021synthetic} are the only other publicly available studies on the topic of experimental design with synthetic controls. The focus of \citeauthor{doudchenkonddesigning}~(n.d.) is on statistical power, which they calculate by simulating the estimated effects of placebo interventions using historical (pre-experimental) data. That is, the selection of treated units is based on a measure of statistical power implied by the distribution of the placebo estimates for each unit. As a result, estimates based on the procedure in \citeauthor{doudchenkonddesigning}~(n.d.) target the effect of the treatment for the unit or units that are most closely tracked in the placebo distribution. In the same spirit, the target parameter in \cite{doudchenko2021synthetic} is the treatment effect for a weighted average of treated units that can be closely matched in their pre-treatment outcomes by a weighted average of untreated units. In the present article, we aim to take a different perspective on the problem of unit selection in experiments with synthetic controls; one that takes into account the extent to which different sets of treated and control units approximate an aggregate causal effect of interest chosen by the analyst, such as the average treatment effect for the relevant population.\footnote{Consistent with the majority of literature on synthetic controls, our focus is primarily on average treatment effects. For an analysis of distributional effects using synthetic controls, see \cite{gunsilius2023distributional}.} The inferential methods in the present article also differ from those in the related literature. In particular, \cite{doudchenko2021synthetic} proposes a permutation procedure for inference that requires that potential outcomes without the treatment are independent and identically distributed (i.i.d.) in time. In contrast, the inferential procedure proposed in the present article allows for time series dependence and non-stationarity in outcomes, which are pervasive features of time-series data. Another important difference between the present article and \citeauthor{doudchenkonddesigning}~(n.d.) and \cite{doudchenko2021synthetic} is that \citeauthor{doudchenkonddesigning}~(n.d.) and \cite{doudchenko2021synthetic} make use of pre-treatment outcomes only to select treated and control units, while our method allows the use of other observed features of the units. 

A related literature applies synthetic control methods and nearest-neighbor matching methods to select experimental sites in multi-site designs \citep{egamidesigning,olea2024externally}. In contrast, we examine settings where treatment occurs at the aggregate (i.e., site) level: each site receives only treatment or only control, precluding the estimation of site-level treatment effects.

\cite{agarwal2021synthetic} proposes synthetic interventions, a framework related to synthetic controls, and applies it to estimate treatment effect heterogeneity in an experimental setting with multiple treatments. Their work primarily focuses on the analysis of experimental data, but not on the design of experiments. \cite{bottmer2021designbased} is also related to the present article in the sense that it studies synthetic control estimation in an experimental setting. Their article, however, considers only the case when the treatment is randomized, and is not concerned with issues of experimental design.

The research designs in this article are also related to ex-ante synthetic control designs for observational studies (see \citeauthor{abadie2021using}, \citeyear{abadie2021using}, \citeauthor{kasy2023employing}, \citeyear{kasy2023employing}, and \citeauthor{chen2023synthetic}, \citeyear{chen2023synthetic}, the latter for an online version of the problem) in that the synthetic weights are computed and can be pre-registered before post-intervention outcomes are realized. However, our methods differ significantly in one key aspect: they confront the challenge that experimenters face when selecting specific units for exposure to the intervention of interest.
In a wider context, our methods are rooted in the broader framework of experimental non-randomized designs \citep[see, e.g.,][]{kasy2016why,armstrong2018finite,thorlund2020synthetic}. Yet, they diverge by addressing a distinct challenge: estimating synthetic control-based aggregate counterfactuals in experimental settings where only a limited number of aggregate units can be treated. 

An alternative approach to control post-randomization bias involves stratifying units based on covariate values prior to randomization of treatment within each stratum. Stratification can significantly reduce post-randomization biases if units have similar covariate values within strata. However, traditional stratification methods do not adapt to the setting considered in this article, which features a limited number of large aggregate entities as units of analysis and a single unit or a handful chosen for treatment. Because every stratum in stratified designs must have at least one unit randomized into treatment, the number of strata cannot exceed the desired number of treated units in the experiment. In the case of only one treated unit, we would be limited to a single stratum. This may lead to significant variation in units' characteristics within strata, reducing the appeal of stratification procedures. Additionally, a stratified design with a small number of strata or treated units may result in the selection of a set of treated units that is not truly representative of the target population.

The rest of the article is organized as follows: Section \ref{sec:SCDesign} presents and discusses the synthetic control designs proposed in this article. Section \ref{sec:formal} details the formal properties of estimators based on synthetic control designs and proposes inferential methods. In Section \ref{sec:Walmart}, we report the findings from an empirical validation of synthetic control designs using sales data from a sample of Walmart stores. Section \ref{sec:Simulation} discusses the results of simulation studies. Finally, Section \ref{sec:Conclusion} provides concluding remarks. The Appendix contains proofs and supplemental materials.

\section{Synthetic Control Designs}
\label{sec:SCDesign}
We consider a setting with $T$ time periods and $J$ units, which may represent $J$ local markets as in the ride-sharing example in the previous section. Let $T_0$ be the number of pre-experimental periods, with $1\leq T_0<T$. 
At the end of period $T_0$, a researcher designs an experiment to conduct during periods $T_0+1, T_0+2, \ldots, T$. Using the information available at $T_0$, the experimenter aims to select the set of units that will receive the treatment (intervention) during the experimental periods.

To define causal parameters, we formally adopt a potential outcomes framework.
For any $j \in \{1,\ldots, J\}$ and any $t \in \{T_0+1, \ldots, T\}$, $Y^I_{jt}$ is the potential outcome for unit $j$ at time $t$ when unit $j$ is exposed to treatment starting at $T_0+1$. Similarly, for any $j \in \{1,\ldots, J\}$ and any $t \in \{1, \ldots, T\}$, $Y^N_{jt}$ is the potential outcome for unit $j$ at time $t$ under no treatment. 
In the ridesharing example, $Y^I_{jt}$ and $Y^N_{jt}$ could measure net revenue divided by market size under the active and the control treatment, respectively. Unit-level treatment effects are defined as
\[
Y^I_{jt} - Y^N_{jt},
\]
for $j = 1,\ldots, J$ and $t = T_0+1, \ldots, T$. They represent the effect of switching unit $j$ to the active treatment at time $T_0+1$ on the outcome of unit $j$ at time $t> T_0$.
We aim to estimate the average treatment effect
\begin{align}
\tau_t = \sum_{j=1}^J f_j \cdot (Y_{jt}^I-Y_{jt}^N), \label{eqn:Estimand}
\end{align}
for $t=T_0+1, \ldots, T$. In this expression, $f_1, \ldots, f_J$ are known positive weights that define the average of interest. In the ride-sharing example from the previous section, $f_j$ may represent the size of local market $j$ as a share of the national market. Without loss of generality, and because it is often the case in applications, we can assume that the weights $f_j$ sum to one,
\[
\sum_{j=1}^J f_j = 1. 
\]
When units are equally weighted, we set $f_j=1/J$ for $j=1,\ldots, J$. We use the notation $\bm f$ for a vector that collects the values of $f_j$ for all the units, i.e., $\bm f=(f_1,\ldots, f_J)$. 

At time $T_0$, in order to estimate the treatment effect $\tau_t$ for $t=T_0+1, \ldots, T$, the experimenter chooses $\bm w=(w_1, \ldots, w_J)$ and $\bm v = (v_1, \ldots, v_J)$, such that 
\begin{gather}
\sum_{j=1}^J w_j = 1, \nonumber\\
\sum_{j=1}^J v_j = 1, \label{equation:wrestrictions}\\
w_j \geq 0, \ v_j\geq 0, \mbox { and } w_j v_j=0,\ \forall j=1,\ldots, J.\nonumber 
\end{gather}
Units with $w_j>0$ are units that will be assigned to the intervention of interest from $T_0+1$ to $T$, and will be used to estimate average outcomes under the intervention. Units with $w_j=0$ constitute an untreated reservoir of potential control units (a ``donor pool''). Among units with $w_j=0$, those with $v_j>0$ will be used to estimate average outcomes under no intervention. 

The first goal of the experimenter is to choose $w_1,\ldots, w_J$ such that
\begin{align}
\label{equation:match1}
    \sum_{j=1}^J w_jY_{jt}^I=
    \sum_{j=1}^J f_jY_{jt}^I, 
\end{align}
for $t=T_0+1, \ldots, T$. If equation \eqref{equation:match1} holds, a weighted average of outcomes for the units selected for treatment reproduces the average outcome with treatment for the entire population of $J$ units. In practice, however, the choice of $w_1,\ldots, w_J$ cannot directly rely on matching the population average of $Y_{jt}^I$, as in equation \eqref{equation:match1}. The quantities $Y_{jt}^I$ are unobserved before time $T_0+1$, and will remain unobserved in the experimental periods for the units that are not exposed to the treatment. Instead, we aim to approximate equation (\ref{equation:match1}) using predictors observed at $T_0$ of the values of $Y_{jT_0+1}^I, \ldots, Y_{jT}^I$. Note also that it is not possible to use the weights $w_1=f_1, \ldots, w_J=f_J$, because it would leave no units in the donor pool, making the set of units with $v_j>0$ empty and violating equation \eqref{equation:wrestrictions}.

The second goal of the experimenter is to choose $v_1,\ldots, v_J$ such that
\begin{align}
\label{equation:match0p}
    \sum_{j=1}^J v_jY_{jt}^N=
    \sum_{j=1}^J f_jY_{jt}^N,  
\end{align}
or, alternatively, 
\begin{align}
\label{equation:match0t}
    \sum_{j=1}^J v_jY_{jt}^N=
    \sum_{j=1}^J w_j Y_{jt}^N. 
\end{align}

If equations \eqref{equation:match0p} or \eqref{equation:match0t} hold, a weighted average of outcomes for the units in the donor pool reproduces the average outcome without treatment for the entire population of $J$ units (equation \eqref{equation:match0p}), or for the units selected for treatment (equation \eqref{equation:match0t}). Like in the previous case with treated outcomes, it is not feasible to directly choose $v_1, \ldots, v_J$ so that equation \eqref{equation:match0p} or \eqref{equation:match0t} is satisfied. Instead, we propose a variety of methods to approximate either (\ref{equation:match0p}) or (\ref{equation:match0t}) based on predictors of $Y_{jT_0+1}^N, \ldots, Y_{jT}^N$.

For the treated units, we define
$Y_{jt}=Y^N_{jt}$ if $t=1,\ldots, T_0$, and $Y_{jt}=Y^I_{jt}$ if $t=T_0+1,\ldots, T$. 
For the untreated units, we define $Y_{jt}=Y^N_{jt}$, for all $t=1, \ldots, T$.
That is, $Y_{jt}$ is the outcome observed for unit $j=1,\ldots, J$ at time $t=1,\ldots, T$. We say that
\[
\sum_{j=1}^J w_jY_{jt}\quad\mbox{ and }\quad \sum_{j=1}^J v_jY_{jt}
\]
are the synthetic treated and synthetic control outcomes, respectively. The difference between these two quantities is 
\[
\tau_t(\bm w,\bm v) = \sum_{j=1}^J w_jY_{jt}- \sum_{j=1}^J v_jY_{jt},
\]
for $t=T_0+1, \ldots, T$. Suppose that equations (\ref{equation:match1}) and (\ref{equation:match0p}) hold. Then,
$\tau_t(\bm w,\bm v)$ is equal to the average treatment effect, $\tau_t$. If equation (\ref{equation:match0t}) holds instead, then $\tau_t(\bm w,\bm v)$ is equal to the average effect of the treatment on the treated ($\bm w$-weighted),
\[
\tau^T_t = \sum_{j=1}^J w_j \cdot (Y^I_{jt}-Y^N_{jt})
\]
\citep{doudchenko2021synthetic}. 

We choose $\bm w=(w_1, \ldots, w_J)$ and $\bm v=(v_1, \ldots, v_J)$ to match the pre-intervention values of predictors of the potential outcomes $Y_{jt}^N$ and $Y_{jt}^I$ for $t > T_0$.

Let $\bm X_j$ be a column vector of pre-intervention features of unit $j$. We view the features in $\bm X_j$ as predictors of the values of $Y^N_{jt}$ and $Y^I_{jt}$ in the experimental periods, in a sense that will be made precise in Section \ref{sec:formal}. We use the notation
\[
\widebar{\bm X} =\sum_{j=1}^J f_j \bm X_j.
\]
That is, $\widebar{\bm X}$ is the vector of population values for the predictors in $\bm X_j$.
For any real vector $\bm x$, $\|\bm x\|$ is the Euclidean norm of $\bm x$, and $\|\bm x\|_0$ is the number of non-zero coordinates of $\bm x$. Let $\underline m$ and $\widebar m$ be positive integers such that $1\leq \underline m\leq \widebar m\leq J-1$. A simple selector of $\bm w=(w_1, \ldots, w_J)$ and $\bm v=(v_1, \ldots, v_J)$ is
\begin{align}
\min_{\substack{w_1, \ldots, w_J,\\ v_1, \ldots, v_J}}\quad & \left\|\,\widebar{\bm X}-\sum_{j=1}^J w_j\bm{X}_j   \right\|^2 + \left\|\, \widebar{\bm X} - \sum_{j=1}^J v_j \bm{X}_j \right\|^2 \nonumber\\[1ex]
\mbox{\ \ \ s.t.\ \ } 
& \sum_{j=1}^J w_j = 1,\nonumber \\
& \sum_{j=1}^J v_j = 1, \nonumber \\
& w_j, v_j \geq 0,\quad \forall j=1,\ldots, J, \nonumber \\
& w_j v_j = 0,\quad \forall j=1,\ldots, J, \nonumber \\
& \underline{m}\leq \|\bm w\|_0\leq \widebar{m}. \label{eqn:OptOriginal}
\end{align}
The first term of the objective function in (\ref{eqn:OptOriginal}) measures the discrepancies between the population average of the features in $\bm X_j$ ($\bm f$-weighted) and the averages of the features for units assigned to the treatment group ($\bm w$-weighted). The second term is analogous but with the second average taken over the units assigned to no intervention ($\bm v$-weighted). The first four constraints require that the weights in $\bm w$, as well as the weights in $\bm v$, are non-negative and sum to one. They also require that any unit selected for treatment cannot be utilized as a control unit --- so, if $w_j>0$, then $v_j=0$. The last constraint allows a minimum and maximum number of units assigned to treatment. This restriction is of practical importance in a variety of contexts, especially when experimentation is costly and the experimenter is restricted in the number of units that may receive the treatment. We say that the design is \textit{Unconstrained} if $\underline m=1$ and $\widebar m=J-1$; otherwise, we say the design is \textit{Constrained}. 
The last constraint in (\ref{eqn:OptOriginal}) is not the only conceivable restriction to the size or cost of the experiment. An explicit upper bound on the cost of an experiment would be given by $\bm c'\bm d\leq B$, where the $j$-th coordinate of $\bm c$ is equal to the cost of assigning unit $j$ to treatment, $\bm d$ is a $J$-dimensional vector with ones at coordinates where $w_j>0$, and zeros otherwise, and $B$ is the experimenter's budget.

Let $\bm w^*=(w^*_1, \ldots, w^*_J)$ and $\bm v^*=(v^*_1, \ldots, v^*_J)$ be a solution to the optimization problem in (\ref{eqn:OptOriginal}). In practice, we do not require optimality of $(\bs w^*, \bs v^*)$, as long as $(\bs w^*, \bs v^*)$ is feasible and satisfies $\widebar{\bm X}-\sum_{j=1}^J w^*_j\bm{X}_j\approx \bs 0$ and $\widebar{\bm X}-\sum_{j=1}^J v^*_j\bm{X}_j\approx \bs 0$, where $\bs 0$ is a vector of zeros of the same dimension as $\bs X_j$. Suppose that units with $w^*_j>0$ are assigned to treatment in the experiment, and units with $w^*_j=0$ are kept untreated. A synthetic control estimator of $\tau_t$ is $\widehat\tau_t=\tau_t(\bm w^*,\bm v^*)$, i.e., 
\begin{equation}
\label{equation:synth}
\widehat\tau_t = \sum_{j=1}^J w^*_j Y_{jt} - \sum_{j=1}^J v^*_j Y_{jt}.
\end{equation}
This estimator is based on approximations to equations (\ref{equation:match1}) and (\ref{equation:match0p}) that rely on $\bm X_j$, the observed predictors of the potential outcomes, $Y_{jt}^N$, and $Y_{jt}^I$. Note that for every solution to (\ref{eqn:OptOriginal}) with $\underline m \leq \|\bm v\|_0 \leq \widebar m$, there exists another solution that swaps the roles of the treated and the untreated in the experiment without altering the value of the objective function. 

In what follows, we take the weight selector in (\ref{eqn:OptOriginal}) as a starting point for synthetic control designs and modify it in several ways. 
A second formulation of the synthetic control design is based on equations \eqref{equation:match1} and \eqref{equation:match0t},
\begin{align}
\min_{\substack{w_1, \ldots, w_J,\\ v_1, \ldots, v_J}}\quad & \left\|\,\widebar{\bm X}-\sum_{j=1}^J w_j\bm{X}_j   \right\|^2 + \beta \left\|\, \sum_{j=1}^J w_j\bm{X}_j - \sum_{j=1}^J v_j \bm{X}_j \right\|^2 \nonumber\\
\mbox{\ \ \ s.t.\ \ } 
& \sum_{j=1}^J w_j = 1,\nonumber \\
& \sum_{j=1}^J v_j = 1, \nonumber \\
& w_j, v_j \geq 0,\quad \forall j=1,\ldots, J, \nonumber \\
& w_j v_j = 0,\quad \forall j=1,\ldots, J, \nonumber \\
& \underline{m}\leq \|\bm w\|_0\leq \widebar{m}. \label{eqn:OptTargeted}
\end{align}
The parameter $\beta > 0$ reflects the trade-off between selecting treated units to fit the aggregate value of the predictors $\widebar{\bm{X}}$, and selecting control units to fit the aggregate value of the treated units.
A small value of $\beta$ favors designs with treated units that closely match $\widebar{\bm{X}}$. 
A large value of $\beta$, on the other hand, favors designs with aggregate treated and aggregate control units that closely match each other.
While it is possible to use data-driven selectors of $\beta$, the rule of thumb $\beta = 1$ provides a natural choice that equally weights the two terms in the objective function in \eqref{eqn:OptTargeted}. For this formulation of the synthetic control design, the treatment assignment and estimation procedures follow the same steps as those used in the previous formulation. Large values for $\beta$ produce estimators that target the $\bm w$-weighted average effect of the treatment on the treated, $\tau_t^T$ of \cite{doudchenko2021synthetic}. Small values of $\beta$ prioritize estimation of the average treatment effect, $\tau_t$.

In our third formulation of the synthetic control design, the experimenter selects a synthetic treated unit to match the average values of the characteristics in the population. However, unlike the design in \eqref{eqn:OptTargeted}, the experimenter chooses multiple synthetic controls, one for each unit that contributes to the synthetic treated unit.   
For any $J$-dimensional vector of non-negative coordinates, $\bm w=(w_1, \ldots, w_J)$, let $\mathcal J_{\bm w}$ be the set of the indices with non-zero coordinates, $\mathcal J_{\bm w} = \{j: w_j>0\}$. Our next version of the synthetic control design is:
\begin{align}
\label{eqn:OptTreated}
\min_{\substack{w_j, \forall j=1,2,..,J,\\ v_{ij}, \forall i,j = 1,2,...,J}}\quad & \left\|\,\widebar{\bm X}-\sum_{j=1}^J w_j\bm{X}_j \right\|^2 + \xi \sum_{j=1}^J w_j \left\|\bm{X}_j- \sum_{i=1}^J v_{ij} \bm{X}_i \right\|^2 \nonumber\\[1ex]
\mbox{\ \ \ s.t.\ \ } & \sum_{j=1}^J w_j = 1,\nonumber \\
& w_j\geq 0,\quad \forall j=1,\ldots, J, \nonumber \\
& \sum_{i=1}^J v_{ij} = 1,\quad \forall j\in \mathcal J_{\bm w}, \nonumber \\
& v_{ij} = 0,\quad \forall i \in \mathcal J_{\bm w}, \ j=1,\ldots,J, \nonumber \\
& v_{ij} \geq 0, \quad \forall j\in \mathcal J_{\bm w},\ i=1,\ldots, J,\nonumber\\
& v_{ij} = 0, \quad \forall j \notin \mathcal J_{\bm w},\  i=1,\ldots,J, \nonumber \\
& \underline{m}\leq \|\bm w\|_0\leq \widebar{m}. 
\end{align}
The parameter $\xi>0$ arbitrates potential trade-offs between selecting treated units to fit the aggregate value of the predictors $\widebar{\bm X}$ and selecting control units to fit the values of the predictors for the treated units. A small value of $\xi$ favors experimental designs with treated units that closely match $\widebar{\bm X}$. A large value of $\xi$, on the other hand, favors designs where the values of the predictors for the treated units are closely matched by those of their synthetic controls. 

Let $\{w^*_j, v^*_{ij}\}_{i,j=1,\ldots,J}$ be a solution of the optimization problem in (\ref{eqn:OptTreated}). As before, we do not strictly require optimality of $\{w^*_j, v^*_{ij}\}_{i,j=1,\ldots,J}$, provided $\{w^*_j, v^*_{ij}\}_{i,j=1,\ldots,J}$ is feasible and $\widebar{\bm X}-\sum_{j=1}^J w^*_j\bm{X}_j\approx \bs 0$ and $\bm X_j-\sum_{j=1}^J v^*_{ij}\bm{X}_j\approx \bs 0$ for all $j$ such that $w^*_j>0$. Assign units with $w^*_j>0$ to treatment in the experiment, and keep units with $w^*_j=0$ untreated. Let
\begin{align}
v^*_j = \sum_{i=1}^J w^*_i v^*_{ij}. \label{eqn:OptTreated:AggregateWeights}
\end{align}
Then,
\begin{align}
\label{equation:synth:ATET}
\widehat\tau_t 
&=\sum_{j=1}^J w^*_j Y_{jt} - \sum_{j=1}^J v^*_j Y_{jt}\nonumber\\
&= \sum_{j=1}^J w^*_j \left(Y_{jt} - \sum_{i=1}^J v^*_{ij} Y_{it}\right).
\end{align}

\begin{figure}[t]
\caption{Clustering in a synthetic control design}
\label{figure:clusters}
\begin{center}
\begin{tikzpicture}[scale=.7]
\draw (0,0) -- (10,0) -- (10,10) -- (0,10) -- (0,0);
\draw (11,0) -- (11,10) -- (21,10) -- (21,0) -- (11,0);
\filldraw [black] (2.2,2.2) circle (2pt);
\filldraw [black] (2.5,2.5) circle (2pt);
\filldraw [black] (3,2.8) circle (2pt);
\filldraw [black] (2.8,2) circle (2pt);
\filldraw [black] (3,2.4) circle (2pt);
\filldraw [black] (2.6,2.8) circle (2pt);
\filldraw [black] (2.3,2.9) circle (2pt);
\filldraw [black] (2.1,2.7) circle (2pt);

\filldraw [red] (4.54,5.29) circle (2pt);
\filldraw [black] (5.35,5.30) circle (2pt);
\filldraw [black] (5.67,5.09) circle (2pt);
\filldraw [black] (4.50,5.60) circle (2pt);
\filldraw [red] (5.25,4.85) circle (2pt);
\filldraw [black] (4.89,4.51) circle (2pt);
\filldraw [black] (5.50,4.55) circle (2pt);
\filldraw [black] (4.56,4.81) circle (2pt);
\filldraw [red] (5,5) circle (2pt);

\filldraw [black] (7.8,7.8) circle (2pt);
\filldraw [black] (7.5,7.5) circle (2pt);
\filldraw [black] (7,7.2) circle (2pt);
\filldraw [black] (7.2,8) circle (2pt);
\filldraw [black] (7,7.6) circle (2pt);
\filldraw [black] (7.4,7.2) circle (2pt);
\filldraw [black] (7,7.6) circle (2pt);
\filldraw [black] (7.7,7.1) circle (2pt);
\filldraw [black] (7.9,7.3) circle (2pt);

\filldraw [black] (13.2,2.2) circle (2pt);
\filldraw [red] (13.5,2.5) circle (2pt);
\filldraw [black] (14,2.8) circle (2pt);
\filldraw [black] (13.8,2) circle (2pt);
\filldraw [black] (14,2.4) circle (2pt);
\filldraw [black] (13.6,2.8) circle (2pt);
\filldraw [black] (13.3,2.9) circle (2pt);
\filldraw [black] (13.1,2.7) circle (2pt);

\filldraw [black] (18.8,7.8) circle (2pt);
\filldraw [red] (18.5,7.5) circle (2pt);
\filldraw [black] (18,7.2) circle (2pt);
\filldraw [black] (18.2,8) circle (2pt);
\filldraw [black] (18,7.6) circle (2pt);
\filldraw [black] (18.4,7.2) circle (2pt);
\filldraw [black] (18,7.6) circle (2pt);
\filldraw [black] (18.7,7.1) circle (2pt);
\filldraw [black] (18.9,7.3) circle (2pt);

\filldraw [black] (15.54,5.29) circle (2pt);
\filldraw [black] (16.35,5.30) circle (2pt);
\filldraw [black] (16.67,5.09) circle (2pt);
\filldraw [black] (15.50,5.60) circle (2pt);
\filldraw [black] (16.25,4.85) circle (2pt);
\filldraw [black] (15.89,4.51) circle (2pt);
\filldraw [black] (16.50,4.55) circle (2pt);
\filldraw [black] (15.56,4.81) circle (2pt);
\filldraw [red] (16,5) circle (2pt);

\draw[rotate around={-45:(13.5,2.5)},dashed] (13.5,2.5) ellipse (1.5 and .9);
\draw[rotate around={-45:(18.5,7.5)},dashed] (18.5,7.5) ellipse (1.5 and .9);
\draw[rotate around={-45:(16,5)},dashed] (16,5) ellipse (1.5 and .9);
\draw[rotate around={45:(5,5)},dashed] (5,5) ellipse (5.2 and 1.2);

\node at (5,-1) {(a)};
\node at (16,-1) {(b)};
\end{tikzpicture}
\end{center}
\floatfoot{{\it Note:} Panels (a) and (b) plot the values of the predictors in $\bm X_j$, which is bivariate in this simple example. Units assigned to treatment are drawn in red. In panel (a), we treat the entire sample as a single cluster. In panel (b), we divide the sample into three clusters and assign one unit in each cluster to the treatment.}
\end{figure}

Our next adjustment to the synthetic control design is motivated by settings where experimental units may be naturally divided into clusters with similar values in the predictors, $\bm X_1, \ldots, \bm X_J$. For example, weather patterns, which may be highly dependent across cities in the same region (e.g., Northeast, Midwest, etc., in the US), may influence the seasonality of the demand for ride-sharing services. In those cases, it is natural to treat each cluster (each region, in our example) as a distinct experimental design to ameliorate interpolation biases. Figure~\ref{figure:clusters} illustrates this point. Panels (a) and (b) depict identical samples in the space of the predictors. In this simple example, we have two predictors only, and their values for each unit are represented by the coordinates of the dots in the figure. Red dots represent units assigned to treatment. All other units are plotted as black dots. Panel (a) visualizes the result of treating the entire sample as one cluster. Three units are assigned to treatment. They closely reproduce the value of $\widebar{\bm X}$, but they all fall in the same central cluster, far away from observations in other clusters. In panel (b), assignment to treatment takes into account the clustered nature of the data, and one unit is treated per cluster. This provides a better approximation of the distribution of the predictor values for the entire sample, ameliorating concerns of interpolation biases. 

Suppose we divide the set of $J$ available units into $K$ clusters. Let $\mathcal I_k$ be the set of indices for the units in cluster $k$. The cluster mean is
\[
\widebar{\bm X}_k=\sum_{j\in \mathcal I_k} f_j \bm X_j \Big/\sum_{j\in \mathcal I_k} f_j,
\]
for each cluster $k=1,\ldots, K$. 
For each index $i = 1,\ldots,J$, let $k(i)$ be the cluster to which unit $i$ belongs, i.e., $i \in \mathcal{I}_{k(i)}$.
A clustered version of the synthetic control design in \eqref{eqn:OptTreated} is given by:
\begin{align}
\label{eqn:OptCluster}
\min_{\substack{w_j, \forall j=1,2,..,J,\\ v_{ij}, \forall i,j = 1,2,...,J}}\quad & \sum_{k=1}^K\Bigg(\sum_{j\in \mathcal I_k} f_j\Bigg)\Bigg\{\Bigg\|\,\widebar{\bm X}_k-\sum_{j\in\mathcal I_k} w_j\bm{X}_j \Bigg\|^2 + \xi \sum_{j\in\mathcal I_k} w_j \Bigg\|\bm{X}_j- \sum_{i,j\in\mathcal I_k} v_{ij} \bm{X}_i \Bigg\|^2\Bigg\} \nonumber\\[1ex]
\mbox{\ \ \ s.t.\ \ } & \sum_{j\in\mathcal I_k} w_j = 1, \quad \forall k=1, \ldots, K,\nonumber \\
& w_j\geq 0,\quad \forall j=1,\ldots, J, \nonumber \\
& \sum_{i=1}^J v_{ij} = 1,\quad \forall j\in \mathcal J_{\bm w}\nonumber\\
& v_{ij}\geq 0, \quad \forall j \in \mathcal J_{\bm w},\ i=1,\ldots, J,\nonumber\\
& v_{ij} = 0, \quad \forall j \notin \mathcal J_{\bm w},\ i=1,\ldots, J,\nonumber\\
& v_{ij} = 0,\quad \forall i \in \mathcal{J}_{\bm{w}}, \ j = 1, \ldots, J, \nonumber\\
& v_{ij} = 0,\quad \forall i, j, \text{ such that } k(i)\neq k(j), \nonumber\\
& \underline{m}\leq \|\bm w\|_0\leq \widebar{m}.
\end{align}

We conclude this section by discussing other possible extensions to the synthetic control design.
First, it is well known that synthetic control estimators may not be unique. Lack of uniqueness is typical in settings where the values of the predictors that a synthetic control is targeting  (i.e., $\widebar{\bm X}$ in equation \eqref{eqn:OptOriginal}, or $\bm X_j$ for a treated unit in equation \eqref{eqn:OptTreated}) fall inside the convex hull of the values of $\bm X_j$ for the units in the donor pool. To address the potential lack of uniqueness, we adapt the penalized estimator of \cite{abadie2021penalized} to the synthetic control designs proposed in this article. The penalized synthetic control estimator of \cite{abadie2021penalized} is unique provided that predictor values for the units in the donor pool are in general quadratic position \citep[see][for details]{abadie2021penalized}. Moreover, penalized synthetic controls favor solutions where the synthetic units are composed of units that have predictor values, $\bm X_j$, similar to the target values.
Applying the penalized synthetic control of \cite{abadie2021penalized} to the objective function of \eqref{eqn:OptOriginal}, we obtain
\begin{align}
\label{eqn:Penalized}
\min_{\substack{w_1, \ldots, w_J,\\ v_1, \ldots, v_J}}\quad & \bigg\|\,\widebar{\bm X}-\sum_{j=1}^J w_j\bm{X}_j \bigg\|^2 + \bigg\|\, \widebar{\bm X} - \sum_{j=1}^J v_j \bm{X}_j \bigg\|^2 \nonumber \\
& \hspace{1cm}+ \lambda_1 \sum_{j=1}^J w_j \Big\|\widebar{\bm X}-\bm X_j\Big\|^2 + \lambda_2 \sum_{j=1}^J v_j \Big\|\widebar{\bm X}-\bm X_j\Big\|^2 \nonumber\\
\mbox{\ \ \ s.t.\ \ } 
& \sum_{j=1}^J w_j = 1,\nonumber \\
& \sum_{j=1}^J v_j = 1, \nonumber \\
& w_j, v_j \geq 0,\quad \forall j=1,\ldots, J, \nonumber \\
& w_j v_j = 0,\quad \forall j=1,\ldots, J, \nonumber \\
& \underline{m}\leq \|\bm w\|_0\leq \widebar{m}. 
\end{align}
Here, $\lambda_1$ and $\lambda_2$ are positive constants that penalize discrepancies between the target values of the predictor $\widebar{\bm{X}}$ and the values of the predictors for the units that contribute to their synthetic counterparts.\footnote{See \cite{abadie2021penalized} for details on penalized synthetic control estimators. The synthetic control design in \eqref{eqn:Penalized} is a penalized version of \eqref{eqn:OptOriginal}. Section~\ref{sec:adaptdesigns} in the Online Appendix discusses how to apply the \citeauthor{abadie2021penalized} penalty to the other synthetic designs proposed in this article.}

Other types of penalization are possible. In particular, \cite{doudchenko2016balancing}, \cite{doudchenko2021synthetic}, and others have proposed synthetic control estimators that use ridge or elastic net regularization on the synthetic control weights (e.g., on $w_j$ and $v_j$ in design \eqref{eqn:OptOriginal}). The synthetic control designs proposed in this article can be modified to incorporate regularization on the weights.

Finally, \cite{abadie2021penalized}, \cite{arkhangelsky2019synthetic}, and \cite{augmented2021feller} have proposed bias-correction techniques for synthetic control methods. Section~\ref{sec:adaptdesigns} in the Online Appendix provides details on how to apply bias correction techniques in a synthetic control design.

\section{Formal Results}
\label{sec:formal}

We introduce an extension of the linear factor model commonly employed in the synthetic control literature and use it to analyze the properties of estimators based on synthetic control designs. 

\begin{assumption}
\label{asp:FactorModel}
Potential outcomes follow a linear factor model,
\begin{subequations}
\begin{align}
Y^N_{jt} & = \delta_t + \bm{\theta}_t' \bm{Z}_j + \bm{\lambda}_t' \bm{\mu}_j + \epsilon_{jt}, \label{eqn:FactorModelN} \\
Y^I_{jt} & = \upsilon_t + \bm{\gamma}_t' \bm{Z}_j + \bm{\eta}_t' \bm{\mu}_j + \xi_{jt}, \label{eqn:FactorModelI}
\end{align}
\end{subequations}
where $\bm{Z}_j$ is a $(R \times 1)$ vector of observed covariates,
$\bm{\theta}_t$ and $\bm{\gamma}_t$ are $(R \times 1)$ vectors of unknown parameters,
$\bm{\mu}_j$ is a $(F \times 1)$ vector of unobserved covariates,
$\bm{\lambda}_t$ and $\bm{\eta}_t$ are $(F \times 1)$ vectors of unknown parameters, and
$\epsilon_{jt}$ and $\xi_{jt}$ are unobserved random shocks. 
\end{assumption}
Equation (\ref{eqn:FactorModelN}) is the linear factor model for potential outcomes under no treatment, a benchmark commonly used in the literature to analyze the properties of synthetic control estimators \citep[see, e.g.,][]{abadie2010synthetic, ferman2021properties}. Equation (\ref{eqn:FactorModelI}) extends the linear factor structure to potential outcomes under treatment. The reason for this extension is that, in contrast to synthetic control estimation with observational data, synthetic control designs require the choice of a treatment group in addition to the choice of a comparison group. 

We employ the covariates in $\bm Z_j$ as well as pre-experimental values of the outcome variable $Y_{jt}$ to construct the vectors of predictors, $\bm X_j$. In particular, let $\mathcal E\subseteq \{1,\ldots, T_0\}$, let $\TcE=|\mathcal E|$, and let $\bm Y^{\mathcal E}_j$ be the $(\TcE\times 1)$ vector of $\TcE$ pre-experimental outcomes for unit $j$ and time indices in $\mathcal E$. We define
\[
\bm X_j = \left(\begin{array}{c}\bm Y^{\mathcal E}_j\\\bm Z_j\end{array}\right),
\]
for $j=1, \ldots, J$. That is, the vector of predictors $\bm X_j$ collects the covariates in $\bm Z_j$ and the pre-experimental outcome values $Y_{jt}$ for the {\em fitting periods} in $\mathcal E$. In practice, the values in $\bm X_j$ are often scaled to make them independent of units of measurement or to reflect the relative importance of each of the predictors \citep[see, e.g.,][]{abadie2021using}. 

The next assumption gathers regularity conditions on model primitives.
\begin{assumption}
\leavevmode
\label{asp:ModelPrimitives}
\begin{enumerate}[label=(\roman*)]
\item $F \leq \TcE$. Moreover, let $\bm{\lambda}_{\mathcal E}$ be the $(\TcE\times F)$ matrix with rows equal to the $\bm\lambda_t$'s indexed by $\mathcal E$. Let $\zeta_{\mathcal E}$ be the smallest eigenvalue of $\bm{\lambda}_{\mathcal E}' \bm{\lambda}_{\mathcal E}$. Then, $\underline{\zeta}=\zeta_{\mathcal E}/\TcE>0$.  

\item For each $j=1, \ldots, J$, $\epsilon_{j1}, \ldots, \epsilon_{jT}$ is a sequence of i.i.d.~sub-Gaussian random variables with mean zero and variance proxy $\widebar{\sigma}^2$. For any $j=1, \ldots, J$,  $\xi_{jT_0+1}, \ldots, \xi_{jT}$ is a sequence of i.i.d.~sub-Gaussian random variables with mean zero,  variance proxy $\widebar{\sigma}^2$, and independent of $\epsilon_{j1}, \ldots, \epsilon_{jT}$. 
\end{enumerate}
\end{assumption}

Assumption~\ref{asp:ModelPrimitives}{\it (i)} is similar to conditions in \citet{abadie2010synthetic}.
Assumption~\ref{asp:ModelPrimitives}{\it (ii)} is similar to conditions in  \citet{abadie2010synthetic}, \cite{doudchenko2016balancing}, \cite{chernozhukov2021exact}, and \cite{arkhangelsky2019synthetic}. Sub-Gaussianity is not strictly necessary, but it simplifies the form of our results. It can be relaxed by assuming bounded finite-order moments (instead of bounding the entire moment generating function). 
At the same time, sub-Gaussianity is a relatively mild assumption. It holds for any Gaussian distribution, as well as any distribution with a bounded support. Distributions with heavy tails, such as the Cauchy distribution, are not sub-Gaussian. Notably, Assumption~\ref{asp:ModelPrimitives}{\it (ii)} allows for dependence of $\epsilon_{jt}$ and $\xi_{jt}$ across units. 

Unless otherwise noted, all probability statements are over the joint distribution of $\epsilon_{jt}$ and $\xi_{jt}$ and conditional on the values of the other components on the right-hand sides of equations \eqref{eqn:FactorModelN} and \eqref{eqn:FactorModelI}. The next assumption pertains to the quality of the synthetic control fit. For concreteness, we focus on the base design in \eqref{eqn:OptOriginal}, and choose where $\bm w^*=(w^*_1, \ldots, w^*_J)$ and $\bm v^*=(v^*_1, \ldots, v^*_J)$ so that the synthetic treated and synthetic control units reproduce the average values of $\bm X_j$. 

\begin{assumption}
\label{asp:PerfectFit}
With probability one, {\it (i)}
\begin{subequations}
\begin{align}
\sum_{j = 1}^J w^*_j \bm{Z}_{j} = \sum_{j = 1}^J v^*_j \bm{Z}_{j}=\sum_{j = 1}^J f_j \bm{Z}_{j}, \label{eqn:Optimum:AssumptionZ}
\end{align}
and {\it (ii)} 
\begin{align}
\sum_{j = 1}^{J} w^*_j \bm{Y}^{\mathcal E}_j=\sum_{j = 1}^{J} v^*_j \bm{Y}^{\mathcal E}_j = \sum_{j = 1}^{J} f_j \bm{Y}^{\mathcal E}_j.\label{eqn:Optimum:AssumptionY}
\end{align}
\end{subequations}
\end{assumption}

Assumption~\ref{asp:PerfectFit} implies that the synthetic treated and control units defined by $\bm w^*$ and $\bm v^*$ provide a perfect fit for $\widebar{\bm X}$. Assumption \ref{asp:PerfectFit} is a strong restriction, which may only hold approximately in practice. 
The next assumption relaxes the perfect fit condition in Assumption \ref{asp:PerfectFit}. 

\begin{assumption}
\label{asp:ApproximateFit}
There exists a positive constant $d > 0$, such that with probability one, {\it (i)} 
\begin{subequations}
\begin{align}
\Big\| \sum_{j = 1}^J w^*_j \bm{Z}_{j} - \sum_{j = 1}^J f_j \bm{Z}_{j} \Big\|_2^2 \leq R d^2, \qquad \Big\| \sum_{j = 1}^J v^*_j \bm{Z}_{j} - \sum_{j = 1}^J f_j \bm{Z}_{j} \Big\|_2^2 \leq R d^2,\label{eqn:Approximate:AssumptionP}
\end{align}
and {\it (ii)}
\begin{align}
\Big\|\sum_{j = 1}^{J} w^*_j \bm{Y}^{\mathcal E}_j - \sum_{j = 1}^{J} f_j \bm{Y}^{\mathcal E}_j \Big\|_2^2 \leq T_\mathcal{E} d^2, \qquad \Big\|\sum_{j = 1}^{J} v^*_j \bm{Y}^{\mathcal E}_j - \sum_{j = 1}^{J} f_j \bm{Y}^{\mathcal E}_j \Big\|_2^2 \leq T_\mathcal{E} d^2. \label{eqn:Approximate:AssumptionN}
\end{align}
\end{subequations}
\end{assumption}

Let $\lambda_{t,f}$ be the $f$-th coordinate of $\bm\lambda_t$, and
\[
\widebar\lambda = \max_{\substack{t=1, \ldots , T\\f=1, \ldots, F}} |\lambda_{tf}|.
\]
We define $\eta_{t f}$, $\theta_{t r}$, $\gamma_{t r}$, $\widebar{\eta}$, $\widebar{\theta}$ and $\widebar{\gamma}$ analogously, so 
$|\eta_{t f} | \leq \widebar{\eta}$ for $t=T_0+1,\ldots, T$, $f=1, \ldots ,F$, 
$| \theta_{t r} | \leq \widebar{\theta}$  for $t=1,\ldots, T$, $r=1, \ldots ,R$, and $| \gamma_{t r} | \leq \widebar{\gamma}$, for $t=T_0+1,\ldots, T$, $r=1, \ldots ,R$. 
Next theorem extends results on the bias of synthetic control estimators \citep[see, e.g.,][]{abadie2010synthetic, vives2022predictor} to the experimental set-up of Section \ref{sec:SCDesign}.

\begin{theorem}
\label{thm:ATEUnbiasedEstimator}
If Assumptions~\ref{asp:FactorModel} --~\ref{asp:PerfectFit} hold, then for any $t \geq T_0+1$,
\begin{align}
\label{equation:biasbound}
|\bE \left[ \widehat\tau_t - \tau_t \right]| \leq 
\frac{\widebar{\lambda}(\widebar{\eta} + \widebar{\lambda}) F}{\underline{\zeta}} \sqrt{2\log{(2J)}}\frac{\widebar{\sigma}}{\sqrt{T_\mathcal{E}}}. 
\end{align}
If Assumptions~\ref{asp:FactorModel},~\ref{asp:ModelPrimitives}, and~\ref{asp:ApproximateFit} hold, then for any $t \geq T_0+1$, 
\begin{align}
\label{equation:biasbound_with_discrepancy}
|\bE \left[ \widehat\tau_t - \tau_t \right]| \leq 
\Big((\widebar{\gamma} + \widebar{\theta})R + \frac{\widebar{\lambda} (\widebar{\eta} + \widebar{\lambda}) F}{\underline{\zeta}} (1+\widebar{\theta} R)\Big)   d + 
\frac{\widebar{\lambda}(\widebar{\eta} + \widebar{\lambda}) F}{\underline{\zeta}} \sqrt{2\log{(2J)}}\frac{\widebar{\sigma}}{\sqrt{T_\mathcal{E}}}.
\end{align}
\end{theorem}

Note that, while the factor model in equations \eqref{eqn:FactorModelN} and \eqref{eqn:FactorModelI} leave the sign and scale of $\bm{\lambda}_t$ and $\bm{\eta}_t$ free (e.g., multiplying $\bm{\lambda}_t$ and dividing $\bm{\mu}_t$ by the same non-zero constant does not change the value of $\bm{\lambda}_t'\bm{\mu}_j$), the value of the bound in Theorem \ref{thm:ATEUnbiasedEstimator} is invariant to changes in the sign or the scale of $\bm{\lambda}_t$ and $\bm{\eta}_t$. Moreover, the bound in \eqref{equation:biasbound_with_discrepancy} does not depend on the scale of ${\bm Z}_j$, because changing the scale of $\bm{Z}_j$ leaves the product $\widebar\theta d$ unchanged. The scale of $Y_{jt}$ does affect the bound in \eqref{equation:biasbound_with_discrepancy} because the treatment effect $\tau_t$ is measured in the same units as $Y_{jt}$.
The results in Theorem~\ref{thm:ATEUnbiasedEstimator} do not depend on the specific formulation of the synthetic control design (e.g., {\it Constrained} vs.~{\it Unconstrained}). 

The bias bounds \eqref{equation:biasbound} and \eqref{equation:biasbound_with_discrepancy} depend on the ratio between the scale of $\epsilon_{jt}$, represented by $\widebar\sigma$, and the number of fitting periods $\TcE$. Intuitively, the bias of the synthetic control estimator is small when a good fit in pre-experimental outcomes (Assumption \ref{asp:PerfectFit}) is obtained by implicitly fitting the values of the latent variables, $\mu_j$. Overfitting happens when pre-experimental outcomes are instead fitted out of the variability in the individual transitory shocks, $\epsilon_{jt}$. A small number of fitting periods $\TcE$ combined with large variability in $\epsilon_{jt}$ increases the risk of overfitting and, as a result, increases the bias bound. Similarly, for any fixed value of $\TcE$, the bias bound increases with $J$, reflecting the increased risk of over-fitting caused by increased variability in $\epsilon_{jt}$ over larger donor pools. Finally, the number of unobserved factors $F$ enters the bound \eqref{equation:biasbound} linearly, which highlights the importance of including the observed predictors $\bm Z_j$ ---\,other than pre-experimental outcomes\,--- in the vector of fitting variables $\bm X_j$. Under the factor model in equations \eqref{eqn:FactorModelN} and \eqref{eqn:FactorModelI}, observed predictors not included in ${\bm Z}_j$ are shifted to ${\bm\mu}_j$, increasing $F$ and the magnitude of the bound.\footnote{Shifting predictors from ${\bm Z}_j$ to ${\bm\mu}_j$ changes the bias bound \eqref{equation:biasbound} in a more complex manner than what might be inferred from a cursory look at the bias formula. First, moving predictors from ${\bm Z}_j$ to ${\bm\mu}_j$ also means shifting components of ${\bm\theta}_t$ to ${\bm\lambda}_t$, which can change the value of $\underline\zeta$. Poincar\'{e}'s separation theorem implies that $\underline\zeta$ cannot increase as a result of this shift. Moreover, moving predictors from ${\bm Z}_j$ to ${\bm\mu}_j$ cannot decrease the values of $\widebar\lambda$ and $\widebar\eta$. Overall, the value of the bias bound in \eqref{equation:biasbound} cannot decrease and will typically increase by moving predictors from ${\bm Z}_j$ to ${\bm\mu}_j$. This is not necessarily true for the bound in \eqref{equation:biasbound_with_discrepancy}, because a shift of components from ${\bm Z}_j$ to ${\bm\mu}_j$ decreases the value of $R$.} 

We next turn our attention to inference. We utilize a set of {\em blank periods}, $\mathcal B\subseteq \{1,\ldots, T_0\}\setminus {\mathcal E}$, which comprise pre-experimental periods whose outcomes $Y_{jt}$ have not been used to calculate $\bm w^*$ or $\bm v^*$. Because pre-experimental periods that are not in $\mathcal E$ or $\mathcal B$ are not used in our procedure, without loss of generality, we consider $\mathcal B= \{1,\ldots, T_0\}\setminus {\mathcal E}$. We, therefore, assume that the number of elements of $\mathcal B$ is $T_{\mathcal B}=|\mathcal B|=T_0-\TcE$. We aim to test the null hypothesis:\vspace*{.2cm}

\begin{tcolorbox}[breakable, boxrule=.5pt, colback=white, right=0pt, left=-.4cm, enlarge left by=.8cm, enlarge right by=-.1cm, width=\linewidth-.7cm, arc=0pt]
\begin{quote}
    For $t=T_0+1, \ldots, T$, and $j=1, \ldots, J$,\end{quote} 
    \begin{equation}
    \label{equation:null}
    Y^I_{jt} = \delta_t+\bm\theta_t'\bm Z_j + \bm\lambda_t'\bm\mu_j + \xi_{jt},
    \end{equation}
    \begin{quote}where $\xi_{jt}$ has the same distribution as $\epsilon_{jt}$.\hspace*{8.8cm}\tikzmark{br}
\end{quote}
\end{tcolorbox}

Under the null hypothesis in \eqref{equation:null}, the distribution of $Y^{I}_{jt}$ is the same as the distribution of $Y^{N}_{jt}$, for $t=T_0+1, \ldots, T$, and $j=1, \ldots, J$. But the realized values of $Y^{I}_{jt}$ and $Y^{N}_{jt}$ may differ. 

Recall from \eqref{equation:synth} that, for $t \in \{T_0+1, \ldots, T\}$, a synthetic control estimator is defined as
\begin{align*}
\widehat{\tau}_t = \sum_{j=1}^J w_j^* Y_{jt}-\sum_{j=1}^J v_j^* Y_{jt}.
\end{align*}
Let $\widehat{u}_t = \widehat{\tau}_t, \forall t \in \{T_0+1, \ldots, T\}$ be the synthetic control estimator in the experimental periods. 
Similarly, for each $t\in \mathcal B$ in the blank periods, let
\begin{align*}
\widehat u_t = \sum_{j=1}^J w_j^* Y_{jt}-\sum_{j=1}^J v_j^* Y_{jt}.
\end{align*}
Such $\widehat{u}_t$ for $t \in \mathcal{B}$ are placebo treatment effects estimated for the blank periods. 
We study the properties of a test based on combinations from the set $\{\widehat{u}_t : t\in\mathcal B\cup \{T_0+1, \ldots, T\}\}$. 

We define $\Pi$ as the set of all $(T-T_0)$-combinations of $\mathcal{B} \cup \{T_0+1, \ldots, T\}$.
That is, for each $\pi\in\Pi$, $\pi$ is a subset of indices from the blank periods and the experimental periods $\mathcal{B} \cup \{T_0+1, \ldots, T\}$, such that $|\pi| = T-T_0$. 
The cardinality of $\Pi$ is $|\Pi| = (T-\TcE)!/((T-T_0)!(T_0-\TcE)!)$.
For each $\pi \in \Pi$, let $\pi(i)$ be the $i^{\text{th}}$ smallest value in $\pi$, and
\begin{align*}
\widehat{\bm{e}}_{\pi} = (\widehat{u}_{\pi(1)}, \widehat{u}_{\pi(2)}, ..., \widehat{u}_{\pi(T-T_0)}).
\end{align*}
In addition, let $\widehat{\bm e} = (\widehat{u}_{T_0+1},\ldots, \widehat{u}_{T}) = (\widehat \tau_{T_0+1}, \ldots, \widehat \tau_{T})$. 
This is a vector of treatment effect estimates from the experimental periods.
For any $(T-T_0)$-dimensional vector $\bm e =(e_1,\ldots, e_{T-T_0})$, we adopt the test statistic,
\begin{align}
S(\bm{e}) = \frac{1}{T-T_0} \sum_{t=1}^{T-T_0} \left| e_t \right|.
\label{eqn:TestStatistic}
\end{align}
Other choices of test statistics are possible, such as those based on an $L_p$-norm of $\bm e$ \citep{chernozhukov2021exact} and one-sided versions of the resulting test statistics (i.e., with the positive or the negative parts of $e_t$ replacing $|e_t|$ in equation (\ref{eqn:TestStatistic})).

The $p$-value of a permutation test on \eqref{eqn:TestStatistic} is
\begin{align}
\widehat{p} & = \frac{1}{\left| \Pi \right|} \sum_{\pi \in \Pi} \bI\{S(\widehat{\bm{e}}_\pi) \geq S(\widehat{\bm{e}})\} \label{eqn:pValue}
\end{align}
Theorem~\ref{thm:ExactPValue} below shows that if $\bm{\lambda}_t$ are exchangeable random variables for $t\in \mathcal B\,\cup\,\{T_0+1, \ldots, T\}$, then a test of the null hypothesis in \eqref{equation:null} based on the $p$-value in \eqref{eqn:pValue} is exact.

\begin{theorem}
\label{thm:ExactPValue}
Suppose that Assumptions~\ref{asp:FactorModel},~\ref{asp:ModelPrimitives}{\it (ii)}, and \ref{asp:PerfectFit}{\it (i)} hold. 
Assume that $\{\bm{\lambda}_t\}_{t \in \mathcal{B} \cup \{T_0+1,...,T\}}$ is a sequence of exchangeable random variables independent of $\{\epsilon_{jt}\}_{t \in \mathcal{B} \cup \{T_0+1,...,T\}}$ and $\{\xi_{jt}\}_{t \in \{T_0+1,...,T\}}$.
Then under the null hypothesis \eqref{equation:null}, we have
\begin{align}
\alpha - \frac{1}{|\Pi|} \leq \Pr(\widehat{p} \leq \alpha) \leq \alpha,
\label{equation:valid_p_value}
\end{align}
for any $\alpha\in [0,1]$, where $\Pr(\widehat{p} \leq \alpha)$ is taken over the distribution of $\{\xi_{jt},\epsilon_{jt},\bm{\lambda}_t\}$.
\end{theorem}

Note that, under the assumptions of Theorem \ref{thm:ExactPValue}, the potential outcome series $Y^N_{jt}$ is allowed to be non-stationary through the term $\delta_t+\bm{\theta}_t' \bm{Z}_j$ in equation \eqref{eqn:FactorModelN}. 
This is in contrast to a related result in \cite{doudchenko2021synthetic}, which requires that the potential outcomes $Y^N_{jt}$ are i.i.d.~over time. 

The assumptions in Theorem~\ref{thm:ExactPValue} build upon those in Theorem~\ref{thm:ATEUnbiasedEstimator}. Although these assumptions are simple and sufficient for the result of the theorem, they can be substantially relaxed.
Under exchangeability of $\bm{\lambda}_t$, if Assumption~\ref{asp:ModelPrimitives}{\it (ii)} is violated, the result for Theorem~\ref{thm:ExactPValue} holds if for each $j=1, \ldots, J$, $\{\epsilon_{jt}\}_{t \in \mathcal{B} \cup \{T_0+1,...,T\}}$ and $\{\xi_{jt}\}_{t \in \{T_0+1,...,T\}}$ are sequences of exchangeable random variables.
Second, if Assumption~\ref{asp:PerfectFit}{\it (i)} is violated, the result for Theorem~\ref{thm:ExactPValue} holds if $\{({\bm \theta}_t, {\bm\lambda}_t)\}_{t \in \mathcal{B} \cup \{T_0+1,...,T\}}$ is a sequence of exchangeable random variables independent of $\{\epsilon_{jt}\}_{t \in \mathcal{B} \cup \{T_0+1,...,T\}}$ and $\{\xi_{jt}\}_{t \in \{T_0+1,...,T\}}$.
In the above two cases under exchangeability of $\bm{\lambda}_t$, we still have exact $p$-value.
Finally, exchangeability of $\bm{\lambda}_t$ is a strong restriction. Theorem~\ref{thm:AsymptoticValidity} in the Online Appendix relaxes this restriction by showing that for fixed $\bm{\lambda}_t$ (i.e., without resorting to exchangeability of $\bm{\lambda}_t$), the $p$-value in \eqref{eqn:pValue} is still approximately valid for large $\TcE$.

In some settings, the number of possible combinations, $|\Pi|$, could be very large, making exact calculation of $\widehat p$ computationally expensive. In those instances, random samples from $\Pi$ can be used to approximate the $p$-value in equation \eqref{eqn:pValue}.

The inferential technique proposed in this article is related to, but distinct from, the permutation methods in \cite{abadie2010synthetic, chernozhukov2021exact, chernozhukov2019distributional,
lei2021conformal, firpo2018synthetic}, and others. Inferential methods that reassign treatment across units \citep[e.g.,][]{abadie2010synthetic} are not appropriate for the designs of Section \ref{sec:SCDesign}, which explicitly select treated and control units to satisfy an optimality criterion.

Similar to \cite{chernozhukov2021exact}, our method is based on rearrangements of estimated treatment effects across time periods. But unlike \cite{chernozhukov2021exact}, which proposes permutations over all periods, including the pre-intervention periods, our inferential method permutes only over the blank periods and post-intervention periods, which are not used to estimate the weights in the synthetic control design. Relative to \cite{chernozhukov2021exact}, the generative models of equations \eqref{eqn:FactorModelN} and \eqref{eqn:FactorModelI}, which allow for unobserved factors, and the finite sample nature of the results require a novel testing procedure that, similar to split conformal prediction methods \citep{vovk2005algorithmic, lei2018distribution}, takes advantage of the availability of blank periods. 

Confidence intervals for $\tau_t$ can be constructed using split conformal inference methods.
For any $\alpha \in (0,1)$, let 
\begin{align}
\widehat{q}_{1-\alpha} = \inf_{z \in \bR} \Bigg\{ \frac{1}{T_0-\TcE} \sum_{t \in \cB} \bI\bigg\{ \Big\vert \sum_{j=1}^J w^*_j Y_{jt} - \sum_{j=1}^J v^*_j Y_{jt} \Big\vert \leq z \bigg\} \geq 1-\alpha \Bigg\} \label{eqn:EmpiricalQuantile}
\end{align}
be the empirical $(1-\alpha)$-quantile on the absolute values of placebo treatment effects in the blank periods, and 
\begin{align}
\widehat{C}_{1-\alpha}(Y_{1t}, Y_{2t}, ..., Y_{Jt}) = \bigg[ \sum_{j=1}^J w^*_j Y_{jt} - \sum_{j=1}^J v^*_j Y_{jt} - \widehat{q}_{1-\alpha}, \ \sum_{j=1}^J w^*_j Y_{jt} - \sum_{j=1}^J v^*_j Y_{jt} + \widehat{q}_{1-\alpha} \bigg]. \label{eqn:CI}
\end{align}
We next show that the confidence interval defined in \eqref{eqn:CI} approximately achieves correct point-wise coverage in large samples if treatment does not change the distribution of the idiosyncratic noises.

\begin{theorem}
\label{thm:CI:Coverage}
Assume that Assumptions~\ref{asp:FactorModel}--~\ref{asp:PerfectFit} hold. 
Assume there exists a constant $\kappa<\infty$, such that for all $j=1, \ldots, J$, $t=1, \ldots, T$, $\epsilon_{jt}$ are continuously distributed with the probability density function upper bounded by $\kappa$.
Assume that for $t=T_0+1, \ldots, T$, and $j=1, \ldots, J$, $\xi_{jt}$ has the same distribution as $\epsilon_{jt}$.
Then the confidence interval defined in \eqref{eqn:CI} approximately achieves point-wise coverage, i.e., for any $\alpha \in (0,1)$ and any $t \in \{T_0+1,...,T\}$, as $(T_0-\TcE), \TcE \to +\infty$,
\begin{multline*}
\bigg\vert \Pr\Big( \tau_t \in \widehat{C}_{1-\alpha}(Y_{1t}, Y_{2t},...,Y_{Jt}) \Big) - (1-\alpha)\bigg\vert \\
= O\Big(\big(\log{(T_0-\TcE)}/(T_0-\TcE)\big)^{1/2} + \big(\log{\TcE}/\TcE\big)^{1/2}\Big) \longrightarrow 0. 
\end{multline*}
\end{theorem}

\section{Empirical Illustration Using Walmart Data} \label{sec:Walmart}

In this section, we illustrate the applicability of the methods in this article using store-level data from Walmart \citep{prakash2023}.
The dataset is a balanced panel of weekly sales for $J=45$ Walmart stores and $T=143$ weeks, spanning the period from the week of February 5, 2010, to the week of October 26, 2012. We estimate the effect of a placebo intervention and show that, in the presence of a good pre-intervention fit, the methods of Section~\ref{sec:formal} produce point estimates that are close to zero and a test result that does not reject the null hypothesis in \eqref{equation:null} for the placebo intervention.

We consider the design of a fictitious experiment across stores taking place on July 20, 2012 (week $129$ in the data).
Out of the $T_0 = 128$ pre-experimental weeks, we take the first $T_\mathcal{E} = 100$ weeks as the fitting period, and the last $(T_0-T_\mathcal{E}) = 28$ weeks as the blank period. The number of weeks in the experimental period is $T-T_0=15$. The outcomes $\{Y_{jt}\}_{j=1,...,J, t=1,...,T}$ are weekly sales (units of revenue are undisclosed in the data).
We use uniform weights $f_j = 1/J$ for $j=1,...,J$, to average sales across all stores. For the purpose of estimating the synthetic treated and synthetic control weights, we normalize each of the 100 pre-experimental outcomes to have a unit variance. 

\begin{figure}[t]\centering
\caption{Synthetic Treatment Unit and Synthetic Control Unit, $\widebar{m} = 2$}
\label{fig:Walmart2Treat}
\includegraphics[width=0.9\textwidth]{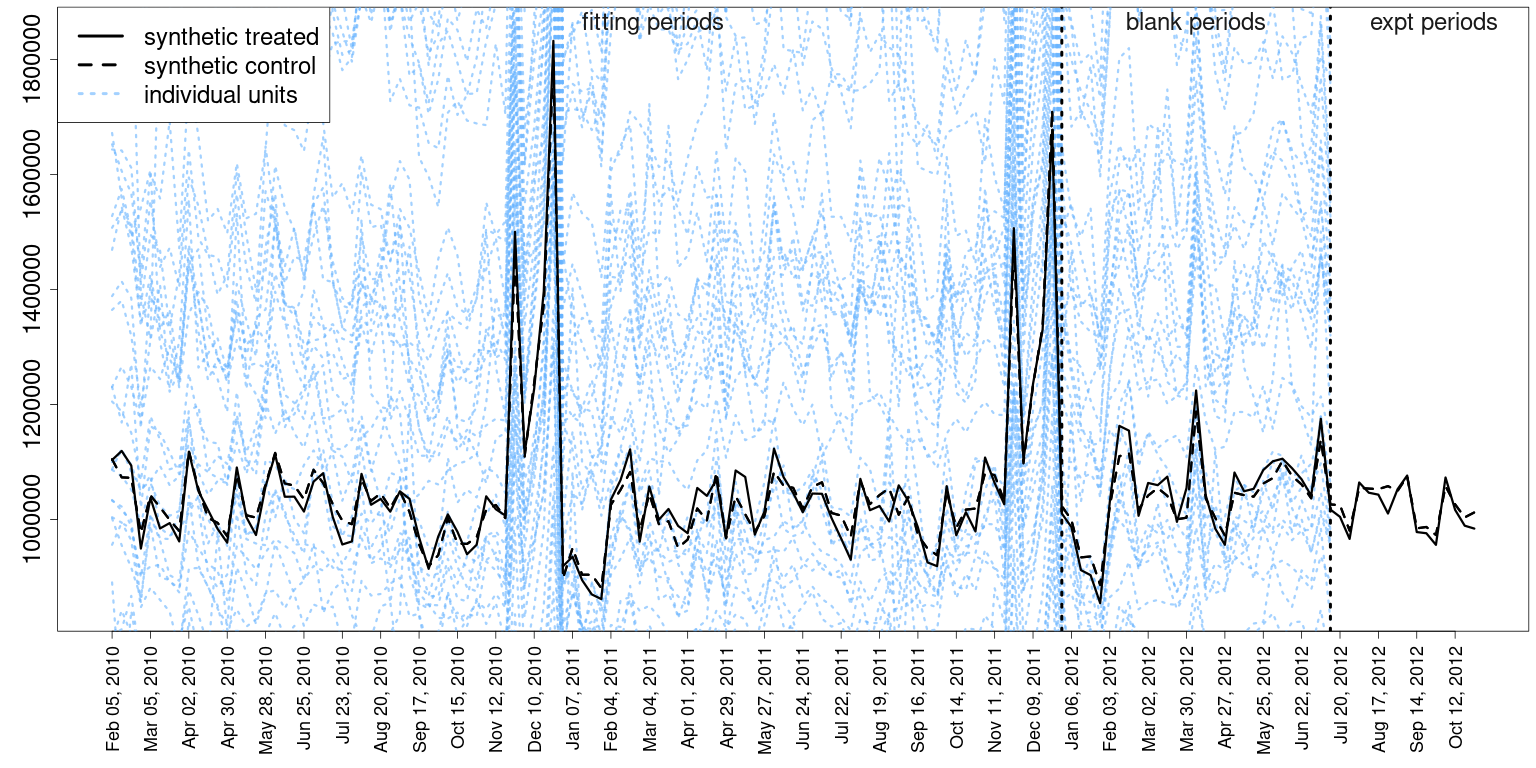}
\floatfoot{{\it Note:} The black solid line represents the synthetic treated outcome. The black dashed line represents the synthetic control outcome. The blue dashed lines are individual stores' sales.}
\end{figure}

\begin{figure}[h]\centering
\caption{Treatment Effect Estimate, when $\widebar{m} = 2$.}
\label{fig:WalmartResiduals2Treat}
\includegraphics[width=0.9\textwidth]{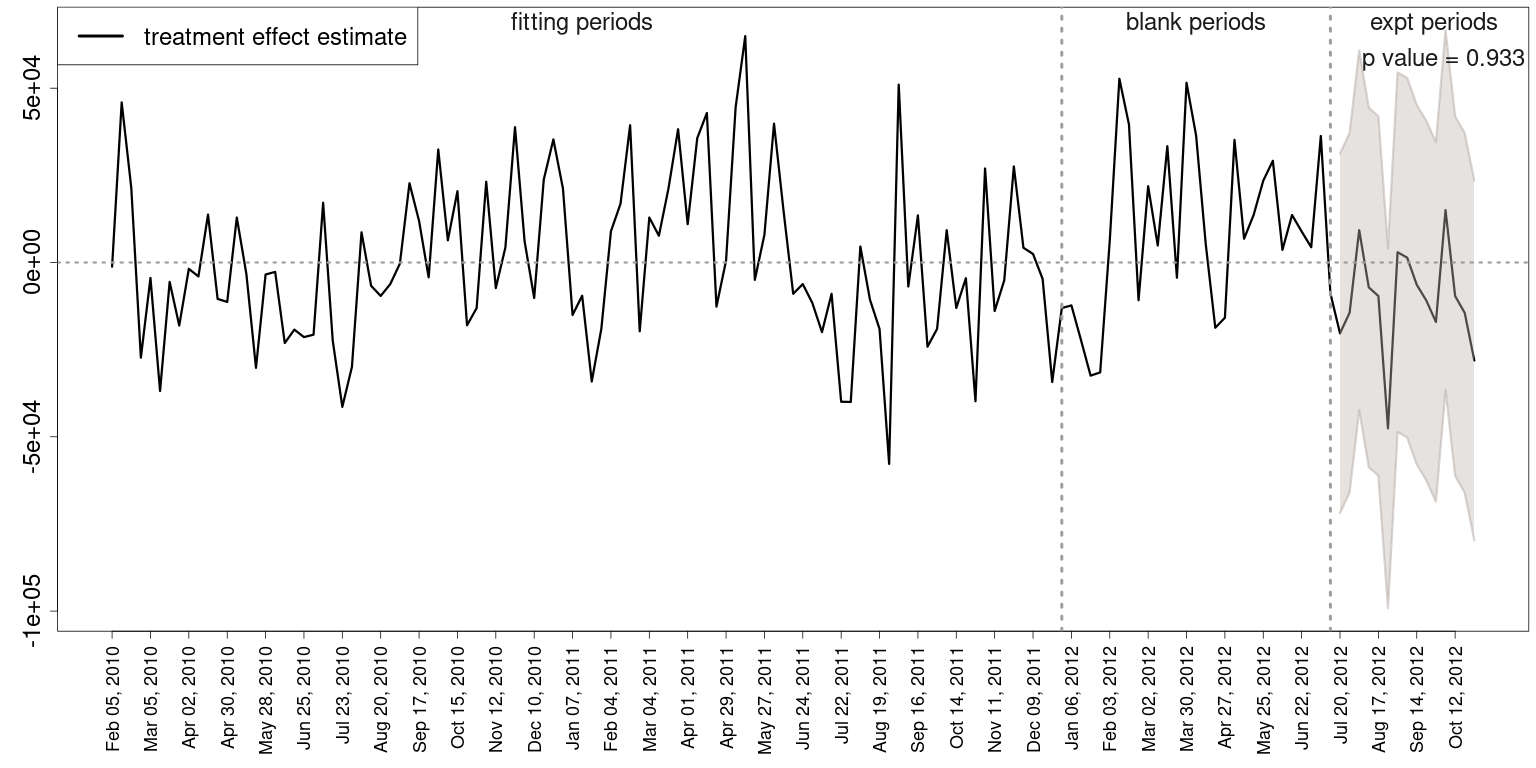}
\floatfoot{{\it Note:} This figure reports the difference between the synthetic treated and synthetic control outcomes of Figure~\ref{fig:Walmart2Treat}. For the experimental periods, this is the treatment effect estimate. The shaded region indicates the 95\% confidence interval for each of the experimental periods.}
\end{figure}

We compute synthetic treated and control units that apply the {\it Constrained} formulation in \eqref{eqn:OptOriginal} with $\widebar{m} = 2$. We adopt $\widebar m =2$ because using only one store for the synthetic treated fails to produce a good fit between the resulting synthetic treated and synthetic control units during the fitting period. Increasing to $\widebar m =3$ brings only marginal improvements in fit. 
Figures~\ref{fig:Walmart2Treat} and~\ref{fig:WalmartResiduals2Treat} report results for $\widebar m=2$. Results for $\widebar m=1$ and $\widebar m=3$ appear in the Online Appendix.

Figure~\ref{fig:Walmart2Treat} reports the time series of weekly sales for the synthetic treated unit (black solid line), the synthetic control unit (black dashed line), and for each individual store in the dataset (blue dashed lines). Weekly sales for the synthetic treated and the synthetic control units closely follow each other during the fitting period. The gap between the two synthetic units remains small after the fitting period, indicating good out-of-sample predictive power in the absence of intervention.  

Figure~\ref{fig:WalmartResiduals2Treat} reports the difference in weekly sales between the synthetic treated and the synthetic control units. The $p$-value of equation \eqref{eqn:pValue}, calculated over the residuals of Figure~\ref{fig:WalmartResiduals2Treat}, is equal to $0.933$, which results in a failure to reject the null hypothesis \eqref{equation:null}.
Confidence intervals based on equation~\eqref{eqn:CI} cover zero for all $t$ in the experimental period.

Table~\ref{tbl:Walmart:Randomization} compares the performance of the synthetic control design to those of straight randomization followed by difference-in-means, randomization after stratification on pre-intervention outcomes followed by difference-in-means, and 1- and 5-nearest neighbor adjustment after randomization. In particular, Table~\ref{tbl:Walmart:Randomization} reports out-of-sample root mean square error (RMSE) over the post-intervention period, normalized by the post-intervention outcome mean (see Section \ref{section:randomization} for a precise definition of the estimators and RMSE performance metric). For each of the three randomization-based estimators, the reported RMSE is the average over 1000 randomized treatment assignments. The synthetic control design dominates all other alternatives, even when it uses only the outcomes in the fitting periods to construct the synthetic treated and synthetic control units, whereas stratification and nearest-neighbor adjustment utilize all pre-intervention outcomes. 

\begin{table}[t]
\caption{Out-of-Sample Normalized Root Mean Square Error} 
\label{tbl:Walmart:Randomization}
\begin{threeparttable}
\begin{tabular}{lccccc}
\hline \hline
                & SC            & RND           & STR           & 1-NN           & 5-NN           \\ [1ex]
$\widebar{m}=1$ & 0.052         & 0.452         & 0.452         & 0.096          & 0.082          \\
$\widebar{m}=2$ & 0.018         & 0.312         & 0.299         & 0.070          & 0.063          \\
$\widebar{m}=3$ & 0.019         & 0.254         & 0.173         & 0.059          & 0.053          \\
$\widebar{m}=4$ & 0.027         & 0.223         & 0.181         & 0.052          & 0.048          \\
$\widebar{m}=5$ & 0.012         & 0.202         & 0.164         & 0.047          & 0.043          \\ \hline
\end{tabular}
\end{threeparttable}
\begin{minipage}{.75\textwidth}
\medskip
\footnotesize {\it Note:}~Root mean square error divided by the average outcome in the experimental periods.
$\widebar{m}$ stands for the maximum number of treated units. 
SC: \textit{Constrained} formulation of the synthetic control design.
RND: Randomized treatment assignment followed by the difference-in-means estimator.
STR: Stratified randomization, followed by difference in means in each stratum.
1-NN: Randomized treatment assignment followed by $1$-nearest neighbor matching, using all pre-experimental outcomes.
5-NN: Randomized treatment assignment followed by $5$-nearest neighbor matching, using all pre-experimental outcomes.
\end{minipage}
\end{table}

\section{Simulation Study} \label{sec:Simulation}

This section presents simulation results that showcase the behavior of estimators based on synthetic control designs. We consider a setting with $J = 15$ units, $R=7$ observable covariates, and $F=11$ unobservable covariates. We simulate data for a total of $T=30$ periods, comprising $T_0=25$ pre-experimental periods and $T-T_0=5$ experimental or post-intervention periods. We compute weights during the first $\TcE=20$ periods and leave periods $t=21, \ldots, 25$ as blank periods.  
We set the weights $f_j$ in expression \eqref{eqn:Estimand} to be $f_j = 1/J$, for all $j=1,...,J$.

For our baseline simulation design, we use the factor model in Assumption \ref{asp:FactorModel} to generate potential outcomes. For $t = 1, \ldots, T$, we generate the series $\delta_t$ and $\upsilon_t$ as small-to-large re-arrangements of $T$ i.i.d.~Uniform $(0,20)$ random variables.
For $j = 1, \ldots, J$, we set both $\bm{Z}_j$ and $\bm{\mu}_j$ to be random vectors of i.i.d.~Uniform $(0,1)$ random variables. For $t = 1, \ldots, T$, we set $\bm{\theta}_t$, $\bm{\gamma}_t$, $\bm{\lambda}_t$, and $\bm{\eta}_t$ to be random vectors of i.i.d.~Uniform $(0,10)$ random variables.
Finally, for $j = 1, \ldots, J$, and any $t = 1, \ldots, T$, we set $\epsilon_{jt}$ and $\xi_{jt}$ to be i.i.d. Normal $(0,\sigma^2)$ random variables, with $\sigma^2=1$. We present additional simulation results of alternative values of the noise parameter $\sigma^2$ in Section~\ref{sec:additional:simulation} in the Online Appendix.

\subsection{Results for a Single Simulation}
\label{sec:simu:main}

Using the data generating process described above, we draw a single sample and conduct the synthetic control design in \eqref{eqn:OptOriginal}, with parameters $\underline{m}=1$ and $\widebar{m}=14$, i.e., no constraint on the number of treated units.
We report the results in Figures~\ref{fig:ComputationalN(0,1)Noise} and~\ref{fig:ResidualsN(0,1)Noise}.
In Figure~\ref{fig:ComputationalN(0,1)Noise}, each blue dashed line represents an outcome trajectory $Y_{jt}$, for $t=1,\ldots,T$ and $j=1, \ldots J$. The solid black line represents the trajectory of the synthetic treated unit $\sum_{j=1}^J w^*_j Y_{jt}$, for $t=1,\ldots,T$. The black dashed line represents the trajectory of the synthetic control unit $\sum_{j=1}^J v^*_j Y_{jt}$, for $t=1,\ldots,T$.
The synthetic treated and synthetic control units closely track each other in the pre-experimental periods. They diverge during the experimental periods, when a treatment effect emerges as a result of the differences in the parameters of the data-generating processes for $Y^N_{jt}$ and $Y^I_{jt}$. Figure~\ref{fig:ResidualsN(0,1)Noise} reports the difference between the synthetic treated and the synthetic control outcomes. The inferential procedure of Section~\ref{sec:formal} produces $p$-value equal to $0.004$ for the null hypothesis of no treatment effect in \eqref{equation:null}. 

\begin{figure}[t]\centering
\caption{Synthetic Treatment Unit and Synthetic Control Unit, $\sigma^2 = 1$}
\label{fig:ComputationalN(0,1)Noise}
\includegraphics[width=0.9\textwidth]{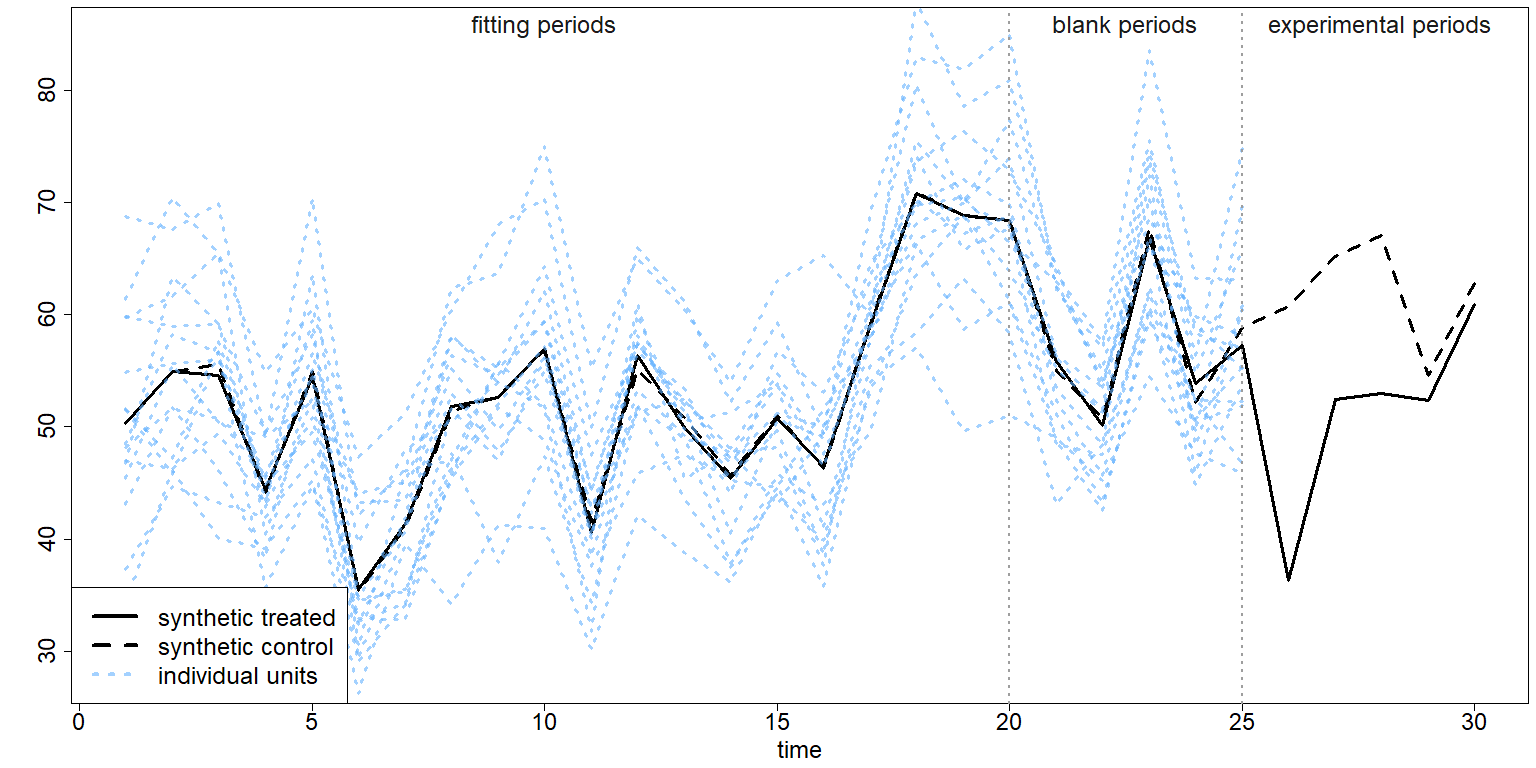}
\floatfoot{{\it Note:} The black solid line represents the synthetic treated outcome ($\bm{w}^*$-weighted). The black dashed line represents the synthetic control outcome ($\bm{v}^*$-weighted).}
\end{figure}

\begin{figure}[t]\centering
\caption{Treatment Effect Estimate, when $\sigma^2 = 1$.}
\label{fig:ResidualsN(0,1)Noise}
\includegraphics[width=0.9\textwidth]{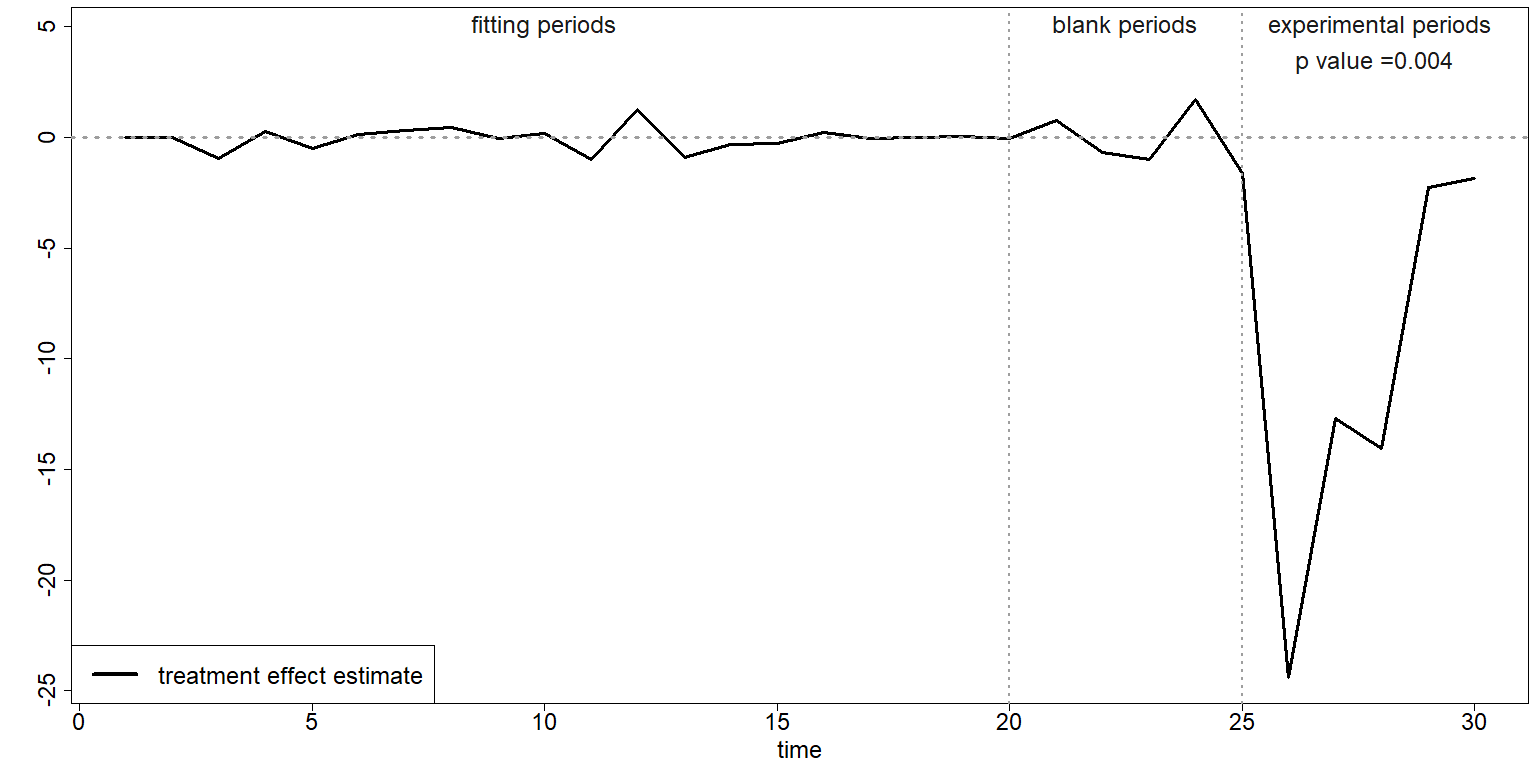}
\floatfoot{{\it Note:} This figure reports the difference between the synthetic treated and synthetic control outcomes of Figure~\ref{fig:ComputationalN(0,1)Noise}. For the experimental periods, this is the treatment effect estimate.}
\end{figure}

\subsection{Performance Across Many Simulations}
\label{sec:designs}

This section compares the performance of the different varieties of the synthetic control designs over 1000 simulations that independently generate the model primitives (i.e., the factor loadings, covariates, and error terms) of Assumption~\ref{asp:FactorModel}. The data generating process is the same as in Section~\ref{sec:simu:main}.

We consider five varieties of the synthetic control design:
\begin{enumerate}[noitemsep,topsep=0pt]
\item {\it Unconstrained design}: This is the design in \eqref{eqn:OptOriginal} without a cardinality constraint, so $\underline{m}=1$ and $\widebar{m}=J-1=14$.
\item {\it Constrained design}: Same as the design in \eqref{eqn:OptOriginal}, but with $\underline{m}=1$ and $\widebar{m}=1,\ldots, 7$.
\item {\it $\NewFormulation$ design}: This is the design in \eqref{eqn:OptTargeted}. We vary $\beta$ from $0.01$ to $100$.
\item {\it Unit-level design}: This is the design in \eqref{eqn:OptTreated}, which fits a different synthetic control to each unit assigned to treatment.
We vary $\xi$ from $0.01$ to $100$.
\item {\it Penalized design}: This is the design in \eqref{eqn:Penalized}, with $\lambda=\lambda_1=\lambda_2$. We vary $\lambda$ from $0.01$ to $100$.
\end{enumerate}

The \textit{Constrained} design imposes sparsity in the synthetic treatment weights through a hard cardinality constraint specified by the integer $\widebar{m}$. 
The \textit{Weakly-targeted} design targets the average treatment effect for small values of $\beta$ and a weighted average effect for the treated for large values of $\beta$. For the \textit{Unit-level} design, large values of $\xi$ generate sparsity in the synthetic treated weights. A sufficiently large value of $\xi$ produces a \textit{Unit-level} design where the only single treated unit can be closely fitted by a convex combination of the other units. 
For large values of $\lambda$, the \textit{Penalized} design behaves like a one-to-one matching design, assigning all the weight to one treated and one control unit.

For the {\it Unit-level} design, synthetic control weights are aggregated as in \eqref{eqn:OptTreated:AggregateWeights}. For the {\it Unconstrained} and {\it Penalized} designs, the synthetic treated and synthetic control weights can always be swapped without changing the objective values for their respective designs. For the {\it Constrained} design, the weights can be swapped when $\|\bm{v}^*\|_0\leq \widebar m$. When it is possible to swap synthetic treated and synthetic control weights, we choose the treated units so that the number of units with positive weights in $\bm w^*$ is smaller than the number of units with positive weights in $\bm v^*$.
When $\|\bm w^*\|_0=\|\bm v^*\|_0$, we determine whether to swap using a specific rule described in Section~\ref{sec:SwappingRule} of the Online Appendix.


\subsubsection{Average Treatment Effects}

\begin{table}[t!]
\caption{Average Treatment Effects (Averages over 1000 Simulations)}
\label{tbl:MultiRuns:DifferentFormulations:ATE}
\begin{threeparttable}
\resizebox{\textwidth}{!}{
\begin{tabular}{llrrrrrcccc}
\hline\hline
                         &                  & \multicolumn{5}{c}{$\tau_t$}               &                &                 &               &                      \\ \cline{3-7}
                         &                  & $t=26$ & $t=27$ & $t=28$ & $t=29$ & $t=30$ &                &                 &               &                      \\ [1ex]   
                         &                  & -13.58 & -10.99 & -8.35  & -5.00  & -2.50  &                &                 &               &                      \\ [1ex] \hline \noalign{\vskip 0.1cm}
                         &                  & \multicolumn{5}{c}{$\widehat\tau_t$}       & $\mathit{MAE}$ & $\mathit{RMSE}$ & $\widehat{p}$ & $\widehat{p} < 0.05$ \\ \cline{3-7}
                         &                  & $t=26$ & $t=27$ & $t=28$ & $t=29$ & $t=30$ &                &                 &               &                      \\ [1ex]
$\textit{Unconstrained}$ &                  & -13.57 & -10.97 & -8.37  & -5.06  & -2.52  & 0.83           & 0.97            & 0.014         & 0.946                \\ [1ex]
$\textit{Constrained}$   & $\widebar{m}=1$  & -13.61 & -10.97 & -8.39  & -4.86  & -2.41  & 2.93           & 3.45            & 0.057         & 0.668                \\
                         & $\widebar{m}=2$  & -13.58 & -10.90 & -8.43  & -5.01  & -2.40  & 1.69           & 2.00            & 0.028         & 0.854                \\
                         & $\widebar{m}=3$  & -13.56 & -11.00 & -8.38  & -5.05  & -2.52  & 1.26           & 1.49            & 0.019         & 0.916                \\
                         & $\widebar{m}=4$  & -13.59 & -11.06 & -8.40  & -4.99  & -2.50  & 1.06           & 1.25            & 0.016         & 0.935                \\
                         & $\widebar{m}=5$  & -13.57 & -11.01 & -8.37  & -5.02  & -2.48  & 0.93           & 1.09            & 0.015         & 0.933                \\
                         & $\widebar{m}=6$  & -13.51 & -10.95 & -8.29  & -5.01  & -2.47  & 0.87           & 1.02            & 0.015         & 0.942                \\
                         & $\widebar{m}=7$  & -13.57 & -10.96 & -8.37  & -5.06  & -2.52  & 0.83           & 0.97            & 0.014         & 0.946                \\ [1ex]
$\NewFormulation$        & $\beta = 0.01$   & -13.58 & -10.95 & -8.38  & -4.99  & -2.53  & 1.18           & 1.38            & 0.018         & 0.920                \\
                         & $\beta = 0.1$    & -13.57 & -11.00 & -8.34  & -4.98  & -2.52  & 0.93           & 1.08            & 0.014         & 0.949                \\
                         & $\beta = 1$      & -13.56 & -10.98 & -8.32  & -4.93  & -2.44  & 0.86           & 1.01            & 0.013         & 0.951                \\
                         & $\beta = 10$     & -13.57 & -10.98 & -8.38  & -5.01  & -2.51  & 0.94           & 1.10            & 0.013         & 0.955                \\
                         & $\beta = 100$    & -13.60 & -10.98 & -8.39  & -5.07  & -2.52  & 1.01           & 1.18            & 0.013         & 0.951                \\ [1ex]
$\textit{Unit-level}$    & $\xi = 0.01$     & -13.60 & -10.95 & -8.39  & -5.04  & -2.53  & 0.95           & 1.13            & 0.014         & 0.938                \\
                         & $\xi = 0.1$      & -13.58 & -10.97 & -8.35  & -4.97  & -2.47  & 0.91           & 1.07            & 0.015         & 0.942                \\
                         & $\xi = 1$        & -13.57 & -10.99 & -8.39  & -4.99  & -2.49  & 1.34           & 1.58            & 0.020         & 0.899                \\
                         & $\xi = 10$       & -13.60 & -10.93 & -8.45  & -5.05  & -2.52  & 2.16           & 2.57            & 0.030         & 0.829                \\
                         & $\xi = 100$      & -13.61 & -10.86 & -8.48  & -5.02  & -2.54  & 2.76           & 3.27            & 0.040         & 0.770                \\ [1ex]
$\textit{Penalized}$     & $\lambda = 0.01$ & -13.59 & -10.98 & -8.35  & -5.05  & -2.48  & 0.88           & 1.02            & 0.014         & 0.950                \\
                         & $\lambda = 0.1$  & -13.64 & -11.03 & -8.43  & -5.03  & -2.50  & 1.21           & 1.43            & 0.019         & 0.904                \\
                         & $\lambda = 1$    & -13.67 & -10.96 & -8.41  & -4.87  & -2.45  & 2.08           & 2.46            & 0.037         & 0.791                \\
                         & $\lambda = 10$   & -13.68 & -11.04 & -8.37  & -4.79  & -2.45  & 3.72           & 4.40            & 0.091         & 0.542                \\
                         & $\lambda = 100$  & -13.64 & -10.94 & -8.42  & -4.86  & -2.50  & 4.17           & 4.93            & 0.111         & 0.490                \\ [1ex] \hline
\end{tabular}
}
\begin{tablenotes}[para,flushleft]
{\it Note:}~Unless otherwise noted, all designs use $\underline{m}=1$ and $\widebar{m}=14$.
\end{tablenotes}
\end{threeparttable}
\end{table}

The first panel of Table~\ref{tbl:MultiRuns:DifferentFormulations:ATE} reports average treatment effects, $\tau_t$, over $1000$ simulations.
The second panel reports estimates of the average treatment effects, mean absolute error, root mean square error, and $p$-value, all averaged over 1000 simulations, as well as rejection rates.
Mean absolute error (MAE) and root mean square error are defined as
\begin{align}
\mbox{MAE}=\frac{1}{T-T_0}\sum_{t=T_0+1}^T |\widehat\tau_t-\tau_t|, \qquad \mbox{RMSE}=\sqrt{\frac{1}{T-T_0}\sum_{t=T_0+1}^T (\widehat\tau_t-\tau_t)^2},
\label{equation:MAERMSE}
\end{align}
and the $p$-value is defined as in \eqref{eqn:pValue}.
Because the treatment effect is not equal to zero in the simulation of Table~\ref{tbl:MultiRuns:DifferentFormulations:ATE}, smaller $p$-values and larger rejection rates reflect better performance of the testing procedure for a particular design. 

In Table~\ref{tbl:MultiRuns:DifferentFormulations:ATE}, the \textit{Unconstrained} design has a strong relative performance.
The performance of the \textit{Constrained} design improves for larger $\widebar{m}$, and is virtually identical to the performance of the \textit{Unconstrained} design when $\widebar{m} = 7$.  
The performance of $\NewFormulation$ and \textit{Unit-level} designs is best when $\beta$ and $\xi$ take intermediate values.
The \textit{Penalized} design yields results similar to those of the \textit{Unconstrained} design for small values of the penalization parameter $\lambda$. 

\subsubsection{Performance with Nonlinearities}
\label{sec:designs:nonlinear}

We now examine the behavior of estimators based on synthetic control designs under deviations from the linear model in \eqref{eqn:FactorModelN} and \eqref{eqn:FactorModelI}.  
We consider a nonlinear data generating process,
\begin{subequations}
\begin{align}
Y^N_{jt} & = \delta_t + \exp{(\bm{\theta}_t' \bm{Z}_j)} + \exp{(\bm{\lambda}_t' \bm{\mu}_j)} + \epsilon_{jt}, \label{eqn:NonlinearN} \\
Y^I_{jt} & = \upsilon_t + \exp{(\bm{\gamma}_t' \bm{Z}_j)} + \exp{(\bm{\eta}_t' \bm{\mu}_j)} + \xi_{jt}. \label{eqn:NonlinearI}
\end{align}
\end{subequations}
The motivation to study a nonlinear model is that nonlinearities may induce interpolation biases, affecting the relative performance of the different designs. All parameter values are the same as in the simulation setup of section \ref{sec:simu:main}, except for the values of $\bm{\theta}_t$, $\bm{\gamma}_t$, $\bm{\lambda}_t$, and $\bm{\eta}_t$, which are chosen to be random vectors of i.i.d.~Uniform $(0,3)$ random variables, instead of Uniform $(0,10)$, to control the magnitude of the exponential components in the nonlinear design. 

\begin{table}[t!]
\caption{Average Treatment Effects, Nonlinear Model (Averages over 1000 Simulations)}
\label{tbl:MultiRuns:Nonlinear:ATE}
\begin{threeparttable}
\resizebox{\textwidth}{!}{
\begin{tabular}{llrrrrrcccc}
\hline\hline
                         &                  & \multicolumn{5}{c}{$\tau_t$}               &                &                 &               &                      \\ \cline{3-7} 
                         &                  & $t=26$ & $t=27$ & $t=28$ & $t=29$ & $t=30$ &                &                 &               &                      \\ [1ex]  
                         &                  & -13.18 & -10.72 & -7.96  & -5.47  & -2.43  &                &                 &               &                      \\ [1ex] \hline \noalign{\vskip 0.1cm}
                         &                  & \multicolumn{5}{c}{$\widehat\tau_t$}       & $\mathit{MAE}$ & $\mathit{RMSE}$ & $\widehat{p}$ & $\widehat{p} < 0.05$ \\ \cline{3-7}
                         &                  & $t=26$ & $t=27$ & $t=28$ & $t=29$ & $t=30$ &                &                 &               &                      \\ [1ex]
$\textit{Unconstrained}$ &                  & -13.44 & -10.92 & -8.18  & -5.85  & -2.78  & 1.99           & 2.54            & 0.059         & 0.741                \\ [1ex]
$\textit{Constrained}$   & $\widebar{m}=1$  & -15.70 & -13.18 & -10.50 & -7.76  & -4.78  & 3.51           & 4.27            & 0.061         & 0.717                \\
                         & $\widebar{m}=2$  & -14.27 & -11.86 & -8.90  & -6.44  & -3.34  & 2.64           & 3.29            & 0.061         & 0.725                \\
                         & $\widebar{m}=3$  & -13.69 & -11.38 & -8.38  & -5.95  & -2.97  & 2.23           & 2.83            & 0.058         & 0.745                \\
                         & $\widebar{m}=4$  & -13.58 & -11.09 & -8.23  & -5.89  & -2.75  & 2.10           & 2.67            & 0.058         & 0.754                \\
                         & $\widebar{m}=5$  & -13.37 & -10.97 & -8.14  & -5.79  & -2.88  & 2.05           & 2.61            & 0.060         & 0.747                \\
                         & $\widebar{m}=6$  & -13.54 & -11.03 & -8.31  & -5.86  & -2.86  & 2.00           & 2.56            & 0.060         & 0.738                \\
                         & $\widebar{m}=7$  & -13.49 & -10.94 & -8.17  & -5.86  & -2.78  & 1.98           & 2.53            & 0.058         & 0.743                \\ [1ex]
$\NewFormulation$        & $\beta = 0.01$   & -11.66 & -9.02  & -6.37  & -3.87  & -1.00  & 2.59           & 3.24            & 0.116         & 0.604                \\
                         & $\beta = 0.1$    & -12.08 & -9.60  & -6.87  & -4.31  & -1.47  & 2.15           & 2.74            & 0.083         & 0.680                \\
                         & $\beta = 1$      & -12.51 & -10.13 & -7.35  & -4.81  & -1.91  & 1.97           & 2.51            & 0.057         & 0.761                \\
                         & $\beta = 10$     & -13.03 & -10.51 & -7.81  & -5.25  & -2.32  & 2.19           & 2.76            & 0.031         & 0.854                \\
                         & $\beta = 100$    & -13.28 & -10.72 & -8.00  & -5.43  & -2.59  & 2.45           & 3.11            & 0.024         & 0.886                \\ [1ex]
$\textit{Unit-level}$    & $\xi = 0.01$     & -11.76 & -9.15  & -6.51  & -3.91  & -1.15  & 2.57           & 3.22            & 0.118         & 0.593                \\
                         & $\xi = 0.1$      & -13.11 & -10.59 & -7.82  & -5.15  & -2.29  & 2.06           & 2.64            & 0.060         & 0.754                \\
                         & $\xi = 1$        & -13.74 & -11.12 & -8.42  & -5.75  & -2.84  & 2.37           & 3.02            & 0.029         & 0.850                \\
                         & $\xi = 10$       & -13.74 & -11.20 & -8.55  & -5.89  & -3.09  & 3.02           & 3.77            & 0.028         & 0.866                \\
                         & $\xi = 100$      & -13.79 & -11.16 & -8.54  & -5.90  & -3.08  & 3.20           & 4.00            & 0.029         & 0.863                \\ [1ex]
$\textit{Penalized}$     & $\lambda = 0.01$ & -13.40 & -10.93 & -8.32  & -5.82  & -2.82  & 1.97           & 2.53            & 0.055         & 0.759                \\
                         & $\lambda = 0.1$  & -13.33 & -10.79 & -8.13  & -5.56  & -2.65  & 2.07           & 2.65            & 0.045         & 0.779                \\
                         & $\lambda = 1$    & -13.32 & -10.84 & -8.15  & -5.39  & -2.60  & 3.08           & 3.84            & 0.056         & 0.738                \\
                         & $\lambda = 10$   & -13.39 & -10.82 & -7.95  & -5.34  & -2.58  & 3.85           & 4.80            & 0.103         & 0.595                \\
                         & $\lambda = 100$  & -13.35 & -10.82 & -8.00  & -5.29  & -2.57  & 4.10           & 5.11            & 0.117         & 0.562                \\ [1ex] \hline
\end{tabular}
}
\begin{tablenotes}[para,flushleft]
{\it Note:}~Unless otherwise noted, all designs use $\underline{m}=1$ and $\widebar{m}=14$.
\end{tablenotes}
\end{threeparttable}
\end{table}

Table~\ref{tbl:MultiRuns:Nonlinear:ATE} reports the results for $\tau_t$. In comparison to the results in Table~\ref{tbl:MultiRuns:DifferentFormulations:ATE}, we now see that the \textit{Unit-level} and \textit{Penalized} designs can easily match and in some cases improve the performance of the \textit{Unconstrained} design. 
By fitting each treated unit with a unit-specific synthetic control, the \textit{Unit-level} design can ameliorate interpolation biases induced by the aggregation of ${\bm X}_j$. The \textit{Penalized} design selects synthetic treated and control units close to $\widebar{\bm X}$ in the space of the predictors, which can reduce interpolation biases at the potential cost of lower precision for large values of $\lambda$ (in which case, the \textit{Penalized} design employs a small number of units in the synthetic treated and synthetic control).

\subsubsection{Test size}

\begin{table}[t!]
\caption{Average Treatment Effects Under the Null Hypothesis \eqref{equation:null} (Averages over 1000 Simulations)}
\label{tbl:MultiRuns:DifferentFormulations:NullHypothesis}
\begin{threeparttable}
\resizebox{\textwidth}{!}{
\begin{tabular}{llrrrrrcccc}
\hline\hline
                         &                  & \multicolumn{5}{c}{$\tau_t$}               &                &                 &               &                      \\ \cline{3-7}         
                         &                  & $t=26$ & $t=27$ & $t=28$ & $t=29$ & $t=30$ &                &                 &               &                      \\ [1ex]                         
                         &                  & -0.01  & 0.00   & 0.00   & 0.01   & -0.01  &                &                 &               &                      \\ [1ex] \hline \noalign{\vskip 0.1cm}
                         &                  & \multicolumn{5}{c}{$\widehat\tau_t$}       & $\mathit{MAE}$ & $\mathit{RMSE}$ & $\widehat{p}$ & $\widehat{p} < 0.05$ \\ \cline{3-7}
                         &                  & $t=26$ & $t=27$ & $t=28$ & $t=29$ & $t=30$ &                &                 &               &                      \\ [1ex]
$\textit{Unconstrained}$ &                  & -0.01  & 0.00   & -0.03  & -0.04  & -0.07  & 0.97           & 1.13            & 0.495         & 0.060                \\ [1ex]
$\textit{Constrained}$   & $\widebar{m}=1$  & 0.20   & 0.14   & 0.02   & -0.08  & 0.05   & 3.00           & 3.55            & 0.495         & 0.056                \\
                         & $\widebar{m}=2$  & -0.02  & -0.01  & -0.02  & -0.09  & -0.03  & 1.80           & 2.13            & 0.497         & 0.038                \\
                         & $\widebar{m}=3$  & -0.09  & -0.07  & -0.02  & -0.05  & -0.02  & 1.37           & 1.62            & 0.505         & 0.048                \\
                         & $\widebar{m}=4$  & -0.02  & -0.02  & 0.00   & -0.01  & -0.01  & 1.19           & 1.41            & 0.494         & 0.054                \\
                         & $\widebar{m}=5$  & 0.01   & -0.02  & 0.03   & 0.00   & -0.05  & 1.07           & 1.25            & 0.496         & 0.057                \\
                         & $\widebar{m}=6$  & 0.07   & 0.06   & 0.10   & -0.01  & -0.03  & 0.99           & 1.17            & 0.484         & 0.054                \\
                         & $\widebar{m}=7$  & -0.01  & 0.00   & -0.02  & -0.04  & -0.07  & 0.96           & 1.13            & 0.495         & 0.059                \\ [1ex]
$\NewFormulation$        & $\beta = 0.01$   & 0.01   & 0.03   & -0.06  & 0.00   & -0.03  & 1.27           & 1.50            & 0.503         & 0.042                \\
                         & $\beta = 0.1$    & 0.02   & -0.01  & 0.01   & 0.04   & -0.03  & 1.03           & 1.21            & 0.498         & 0.055                \\
                         & $\beta = 1$      & 0.00   & 0.01   & 0.03   & 0.10   & 0.04   & 0.95           & 1.11            & 0.501         & 0.044                \\
                         & $\beta = 10$     & -0.07  & -0.01  & -0.03  & 0.04   & 0.00   & 0.94           & 1.09            & 0.485         & 0.061                \\
                         & $\beta = 100$    & -0.09  & -0.08  & -0.06  & -0.05  & -0.04  & 0.95           & 1.11            & 0.493         & 0.051                \\ [1ex]
$\textit{Unit-level}$    & $\xi = 0.01$     & 0.00   & 0.03   & -0.04  & -0.04  & -0.02  & 1.05           & 1.25            & 0.511         & 0.053                \\
                         & $\xi = 0.1$      & 0.00   & 0.00   & 0.03   & 0.02   & 0.02   & 1.05           & 1.24            & 0.500         & 0.049                \\
                         & $\xi = 1$        & 0.01   & 0.02   & -0.06  & -0.05  & -0.03  & 1.38           & 1.63            & 0.499         & 0.046                \\
                         & $\xi = 10$       & 0.18   & 0.00   & -0.02  & -0.15  & -0.02  & 1.97           & 2.33            & 0.496         & 0.038                \\
                         & $\xi = 100$      & 0.19   & -0.03  & -0.02  & -0.18  & -0.03  & 2.34           & 2.77            & 0.502         & 0.053                \\ [1ex]
$\textit{Penalized}$     & $\lambda = 0.01$ & 0.00   & 0.00   & 0.02   & -0.04  & -0.01  & 1.01           & 1.18            & 0.494         & 0.051                \\
                         & $\lambda = 0.1$  & -0.07  & -0.05  & -0.07  & -0.11  & -0.10  & 1.32           & 1.56            & 0.505         & 0.041                \\
                         & $\lambda = 1$    & 0.02   & 0.07   & -0.07  & -0.07  & 0.01   & 2.17           & 2.57            & 0.495         & 0.045                \\
                         & $\lambda = 10$   & 0.16   & 0.03   & -0.11  & -0.08  & -0.08  & 3.79           & 4.48            & 0.514         & 0.045                \\
                         & $\lambda = 100$  & 0.22   & 0.15   & -0.14  & -0.14  & -0.08  & 4.22           & 5.00            & 0.515         & 0.041                \\ [1ex] \hline
\end{tabular}
}
\begin{tablenotes}[para,flushleft]
{\it Note:}~Unless otherwise noted, all designs use $\underline{m}=1$ and $\widebar{m}=14$.
\end{tablenotes}
\end{threeparttable}
\end{table}

In this section, we generate the model primitives under the null hypothesis \eqref{equation:null}. That is, we employ a data generating process such that the values of the common factors and the distributions of the idiosyncratic error variables are unaffected by the intervention.

We report the simulation results in Table~\ref{tbl:MultiRuns:DifferentFormulations:NullHypothesis}, which organizes information in the same way as in Table~\ref{tbl:MultiRuns:DifferentFormulations:ATE}. 
Because the data are generated from the same distribution under treatment and under no treatment, the average treatment effects in Table~\ref{tbl:MultiRuns:DifferentFormulations:NullHypothesis} are close to zero. The same is true for the averages of $\widehat\tau_t$ for all designs.
Under the null hypothesis \eqref{equation:null}, the $p$-value should approximately follow a uniform distribution between zero and one.
The results in Table~\ref{tbl:MultiRuns:DifferentFormulations:NullHypothesis} show good behavior of our testing procedure under the null hypothesis: average $p$-values and rejection rates are close to $0.5$ and $0.05$, respectively.

\subsubsection{Comparison to Randomized Treatment Assignment}
\label{section:randomization}

Randomized treatment assignment produces ex-ante (pre-randomization) unbiased estimation of the average treatment effect. As we show below, however, ex-post (post-randomization) biases can be large, especially when only a small number of units are treated.

\begin{table}[t!]
\caption{RMSE for  Different Experimental Designs and Estimators (Averages over 1000 Simulations)} 
\label{tbl:MultiRuns:Randomization}
\begin{threeparttable}
\begin{tabular}{lccccccccc}
\hline \hline
                & SC            & RND            & STR              & REG            & 1-NN             & 5-NN             \\ [1ex]
$\widebar{m}=1$ & 3.45          & 6.35           & 6.35             & 8.14           & 5.28             & 4.40             \\ 
$\widebar{m}=2$ & 2.00          & 4.70           & 3.53             & 6.00           & 3.69             & 3.20             \\
$\widebar{m}=3$ & 1.49          & 3.91           & 2.75             & 5.11           & 3.02             & 2.66             \\
$\widebar{m}=4$ & 1.25          & 3.49           & 2.44             & 4.49           & 2.67             & 2.40             \\
$\widebar{m}=5$ & 1.09          & 3.22           & 2.07             & 4.12           & 2.38             & 2.28             \\
$\widebar{m}=6$ & 1.02          & 3.04           & 1.95             & 3.87           & 2.24             & 2.23             \\
$\widebar{m}=7$ & 0.97          & 3.01           & 1.85             & 3.90           & 2.18             & 2.32             \\ \hline
\end{tabular}
\end{threeparttable}
\begin{minipage}{.75\textwidth}
\medskip
\footnotesize {\it Note:}~
SC:~\textit{Constrained} formulation of the synthetic control design.
RND:~Randomized treatment assignment followed by the difference-in-means estimator.
STR:~Stratified randomization, followed by difference in means in each stratum.
REG:~Randomized treatment assignment followed by regression adjustment. 
1-NN:~Randomized treatment assignment followed by $1$-nearest neighbor matching.
5-NN:~Randomized treatment assignment followed by $5$-nearest neighbor matching.
SC uses outcomes in the fitting periods and covariates as predictors. STR, 1-NN, and 5-NN use all pre-intervention outcomes and covariates. REG adjusts for the covariates only.
\end{minipage}
\end{table}

In this section, we adopt the same set-up as for
Table~\ref{tbl:MultiRuns:DifferentFormulations:ATE}. 
We consider randomized treatment assignment with $\widebar m$ treated units. 
$D_j$ is a treatment indicator that equals one if unit $j$ is randomized into the treated group and zero otherwise. 
We study the performance of the following estimation strategies:\vspace{-0.3cm}
\begin{enumerate}
\item \textit{SC}:~\textit{Constrained} formulation of the synthetic control design. The results reproduce those of Table~\ref{tbl:MultiRuns:DifferentFormulations:ATE}. 
\item \textit{RND}: Randomized assignment of $\widebar m$ units to treatment followed by the difference in means estimator,
\begin{align*}
 \frac{1}{\widebar m} \sum_{j=1}^J D_j Y_{jt} - \frac{1}{J-\widebar m} \sum_{j=1}^J (1-D_j)Y_{jt}.
\end{align*}
\item \textit{STR}: Divide the sample in $\widebar m$ strata, such that each stratum has at least two units. In each stratum, one unit is assigned to treatment at random. The composition of the strata is chosen to minimize the maximal within-strata discrepancy in the covariates, $\bm{Z}_j$, and pre-experimental outcomes (all normalized to have unit variance). Let $B_{jk}$ be a binary variable that equals one if and only if unit $j$ belongs to cluster $k$. Let $J_k$ be the number of units in stratum $k$. 
\begin{align*}
\sum_{k=1}^{\widebar m} \frac{J_k}{J}\bigg(\sum_{j=1}^J B_{jk}D_jY_{jt} - \frac{1}{J_k-1}\sum_{j=1}^k B_{jk}(1-D_j)Y_{jt}\bigg),
\end{align*}
where $J_k$ represents the number of units within the $k$-th block.
\item \textit{REG}:~Randomized assignment of $\widebar m$ units to treatment followed by regression adjustment on the covariates, $\bm{Z}_j$.  Ordinary least-squares adjustment on all pre-treatment outcomes is unfeasible as the number of pre-treatment outcomes exceeds the number of units in the sample.
\item \textit{1-NN} and \textit{5-NN}:~Randomized assignment of $\widebar m$ units to treatment followed by $1$-nearest neighbor and $5$-nearest neighbor matching, respectively, on all pre-experimental outcomes and covariates. In both cases, predictors are rescaled to have unit variance.
\end{enumerate}
Results are reported in Table 5. Across all values of $\widebar m$, the synthetic control design outperforms randomized assignment, including variants that incorporate pre-stratification, post-stratification, or regression adjustment. Taken together with the findings in Table \ref{tbl:Walmart:Randomization}, these results underscore the potential of synthetic controls as a more effective design strategy in experiments involving aggregate units and a limited number of treated units.

\section{Conclusions}
\label{sec:Conclusion}

Experimental design methods have largely been concerned with settings where a large number of experimental units are randomly assigned to a treatment arm, and a similarly large number of experimental units are assigned to a control arm. This focus on large samples and randomization has proven to be enormously useful in various classes of problems but becomes inadequate when treating more than a few units is unfeasible, as is often the case in experimental studies with large aggregate units (e.g., markets). In that case, randomized designs may produce estimators that are substantially biased (post-randomization) relative to the average treatment effect or to the average treatment effect on the treated. Large biases can be expected when the unit or units assigned to treatment fail to approximate average outcomes under treatment for the entire population or when the units in the control arm fail to approximate the outcomes that treated units would experience without treatment.  

In this article, we have proposed synthetic control techniques, widely used in observational studies, to design experiments when the treatment can only be applied to a small number of experimental units. The synthetic control design optimizes jointly over the identities of the units assigned to the treatment and the control arms and over the weights that determine the relative contribution of those units to reproduce the counterfactuals of interest. 
We propose various designs to estimate average treatment effects, analyze the properties of such designs and the resulting estimators, and devise inferential methods to test a null hypothesis of no treatment effects and construct confidence intervals. In addition, we report results from an application to retail sales data and simulation results that demonstrate the applicability and computational feasibility of the methods proposed in this article. We show that synthetic control design can substantially outperform randomized designs in experimental settings with a small number of treated units. 

Corporate researchers, policymakers, and academic investigators are often confronted with settings where interventions at the micro-unit level (e.g., customers, workers, or families) are unfeasible, impractical, or ineffective \citep[see, e.g.,][]{duflo2007using, jones2019uber}. Consequently, there is broad scope for experimental design methods targeting large aggregate entities (such as regional markets, school districts, or states), a setting where synthetic control designs offer a powerful tool for data-driven evaluation of treatment effects.

{\small
\bibliographystyle{aea}
\bibliography{bibliography}
}

\clearpage

\begin{appendices}
\begin{center}
{\bf Appendix}\vspace*{-0.4cm}
\end{center}

\parindent 0in
\setcounter{equation}{0}
\renewcommand{\theequation}{\thesection.\arabic{equation}}
\setcounter{theorem}{0}
\renewcommand{\thetheorem}{\Alph{section}.\arabic{theorem}}

\section{Proofs}
\label{sec:proof:thm:ATEUnbiasedEstimator}
\setcounter{equation}{0}

\subsection{Proof of Theorem~\ref{thm:ATEUnbiasedEstimator}}
\begin{proof}[Proof of Theorem~\ref{thm:ATEUnbiasedEstimator}.]
For any period $t = T_0+1, \ldots, T$ we decompose $(\widehat\tau_t - \tau_t)$ as follows,
\begin{align}
\widehat\tau_t - \tau_t = & \left( \sum_{j=1}^J w^*_j Y^I_{jt} - \sum_{j=1}^J v^*_j Y^N_{jt} \right) - \left( \sum_{j=1}^J f_j Y^I_{jt} - \sum_{j=1}^J f_j Y^N_{jt} \right) \nonumber \\
= & \left( \sum_{j=1}^J w^*_j Y^I_{jt} - \sum_{j=1}^J f_j Y^I_{jt} \right) - \left( \sum_{j=1}^J v^*_j Y^N_{jt} - \sum_{j=1}^J f_j Y^N_{jt} \right). \label{eqn:ATE_final}
\end{align}
The first term in (\ref{eqn:ATE_final}) measures the difference between the synthetic treatment outcome and the aggregated treatment outcomes. The second term measures the difference between the synthetic control outcome and the aggregate control outcomes.
We bound these two terms separately.
From \eqref{eqn:FactorModelI}, we obtain
\begin{multline}
\sum_{j = 1}^J  w^*_j Y^I_{jt} - \sum_{j = 1}^J f_j Y^I_{jt} = \bm{\gamma}_t' \Bigg(\sum_{j = 1}^J w^*_j\bm{Z}_j - \sum_{j = 1}^J f_j\bm{Z}_j\Bigg) \\
+ \bm{\eta}_t' \Bigg(\sum_{j = 1}^J w^*_j \bm{\mu}_j - \sum_{j = 1}^J f_j \bm{\mu}_j\Bigg) + \Bigg(\sum_{j = 1}^J w^*_j \xi_{j t} - \sum_{j = 1}^J f_j \xi_{j t}\Bigg) \label{eqn:interStep:E1}
\end{multline}
Similarly, using expression  \eqref{eqn:FactorModelN}, we obtain
\begin{multline*}
\sum_{j = 1}^J w^*_j \bm{Y}_{j}^\sP - \sum_{j = 1}^J f_j \bm{Y}_{j}^\sP = \bm{\theta}_\sP \Bigg(\sum_{j = 1}^J w^*_j \bm{Z}_j - \sum_{j = 1}^J f_j \bm{Z}_j\Bigg) \\
+ \bm{\lambda}_\sP \Bigg(\sum_{j = 1}^J w^*_j \bm{\mu}_j - \sum_{j = 1}^J f_j \bm{\mu}_j\Bigg) + \Bigg(\sum_{j = 1}^J w^*_j \bm{\epsilon}_{j}^\sP - \sum_{j = 1}^J f_j \bm{\epsilon}_j^\sP\Bigg),
\end{multline*}
where $\bm{\theta}_\sP$ is the $(\TcE\times R)$ matrix with rows equal to the $\bm\theta_t$'s indexed by $\mathcal E$, and $\bm{\epsilon}_{j}^\sP$ is defined analogously. 
Pre-multiplying by $\bm{\eta}_t' (\bm{\lambda}_{\sP}' \bm{\lambda}_\sP)^{-1} \bm{\lambda}_{\sP}'$ yields
\begin{align}
\bm{\eta}_t' (\bm{\lambda}_{\sP }' \bm{\lambda}_\sP)^{-1} \bm{\lambda}_{\sP }' & \Bigg(\sum_{j = 1}^J w^*_j \bm{Y}_{j}^\sP - \sum_{j = 1}^J f_j \bm{Y}_{j}^\sP\Bigg) \\
& =  \bm{\eta}_t' (\bm{\lambda}_{\sP }' \bm{\lambda}_\sP)^{-1} \bm{\lambda}_{\sP }' \bm{\theta}_\sP \Bigg(\sum_{j = 1}^J w^*_j \bm{Z}_j - \sum_{j = 1}^J f_j \bm{Z}_j\Bigg) \nonumber \\
& + \bm{\eta}_t' \Bigg(\sum_{j = 1}^J w^*_j \bm{\mu}_j - \sum_{j = 1}^J f_j \bm{\mu}_j\Bigg) \nonumber \\
& + \bm{\eta}_t' (\bm{\lambda}_{\sP }' \bm{\lambda}_\sP)^{-1} \bm{\lambda}_{\sP }' \Bigg(\sum_{j = 1}^J w^*_j \bm{\epsilon}_{j}^\sP - \sum_{j = 1}^J f_j \bm{\epsilon}_j^\sP\Bigg). \label{eqn:interStep:E2}
\end{align}
Equations \eqref{eqn:interStep:E1} and \eqref{eqn:interStep:E2} imply
\begin{align}
\label{equation:bias5terms}
\sum_{j = 1}^J w^*_j Y^I_{jt} - \sum_{j = 1}^J f_j Y^I_{jt} = \ & (\bm{\gamma}'_t - \bm{\eta}_t' (\bm{\lambda}_{\sP }' \bm{\lambda}_\sP)^{-1} \bm{\lambda}_{\sP }' \bm{\theta}_\sP) \Bigg(\sum_{j = 1}^J w^*_j \bm{Z}_j - \sum_{j = 1}^J f_j \bm{Z}_j\Bigg) \nonumber \\
& + \bm{\eta}_t' (\bm{\lambda}_{\sP }' \bm{\lambda}_\sP)^{-1} \bm{\lambda}_{\sP }' \Bigg(\sum_{j = 1}^J w^*_j \bm{Y}^\sP_j - \sum_{j = 1}^J f_j \bm{Y}^\sP_j\Bigg) \nonumber \\
& - \bm{\eta}_t' (\bm{\lambda}_{\sP }' \bm{\lambda}_\sP)^{-1} \bm{\lambda}_{\sP }' \sum_{j = 1}^J w^*_j \bm{\epsilon}_{j}^\sP\nonumber\\ 
& + \bm{\eta}_t' (\bm{\lambda}_{\sP }' \bm{\lambda}_\sP)^{-1} \bm{\lambda}_{\sP }' \sum_{j = 1}^J f_j \bm{\epsilon}_j^\sP \nonumber\\
& + \Bigg(\sum_{j = 1}^J w^*_j \xi_{j t} - \sum_{j = 1}^J f_j \xi_{j t}\Bigg).
\end{align}

If Assumption~\ref{asp:PerfectFit} holds, \eqref{equation:bias5terms} becomes
\begin{align}
\label{equation:bias3terms}
\sum_{j = 1}^J w^*_j Y^I_{jt} - \sum_{j = 1}^J f_j Y^I_{jt}& =
 - \bm{\eta}_t' (\bm{\lambda}_{\sP }' \bm{\lambda}_\sP)^{-1} \bm{\lambda}_{\sP }' \sum_{j = 1}^J w^*_j \bm{\epsilon}_{j}^\sP\nonumber\\ & + \bm{\eta}_t' (\bm{\lambda}_{\sP }' \bm{\lambda}_\sP)^{-1} \bm{\lambda}_{\sP }' \sum_{j = 1}^J f_j \bm{\epsilon}_j^\sP \nonumber\\
& + \Bigg(\sum_{j = 1}^J w^*_j \xi_{j t} - \sum_{j = 1}^J f_j \xi_{j t}\Bigg).
\end{align}
Only the first term on the right-hand side of \eqref{equation:bias3terms} has a non-zero mean (because the weights, $w^*_j$, depend on the error terms $\bm{\epsilon}_{j}^\sP$). Therefore,
\begin{align}
\left| E\Bigg[\sum_{j = 1}^J w^*_j Y^I_{jt} - \sum_{j = 1}^J f_j Y^I_{jt} \Bigg]\right| = \left| \bE\left[\bm{\eta}_t' (\bm{\lambda}_{\sP }' \bm{\lambda}_\sP)^{-1} \bm{\lambda}_{\sP }' \sum_{j = 1}^J w^*_j \bm{\epsilon}_{j}^\sP \right] \right|.\label{eqn:TreatmentBracket}
\end{align}
Using the same line of reasoning for the second term on the right-hand side of \eqref{eqn:ATE_final}, we obtain
\begin{align}
\left| E\Bigg[\sum_{j = 1}^J v^*_j Y^N_{jt} - \sum_{j = 1}^J f_j Y^N_{jt} \Bigg]\right| = \left| \bE\left[ \bm{\lambda}_t' (\bm{\lambda}_{\sP }' \bm{\lambda}_\sP)^{-1} \bm{\lambda}_{\sP }' \sum_{j = 1}^J v^*_j \bm{\epsilon}_{j}^\sP \right] \right|. \label{eqn:ControlBracket}
\end{align}
For any $t\geq T_0+1$ and $s \in \mathcal{E}$, under Assumption \ref{asp:ModelPrimitives} (i), we apply Cauchy-Schwarz inequality and the eigenvalue bound on the Rayleigh quotient to obtain
\begin{align*}
\left( \bm{\eta}_t' (\bm{\lambda}_{\sP }' \bm{\lambda}_\sP)^{-1} \bm{\lambda}_s \right)^2 &
\leq \left( \bm{\eta}_t' (\bm{\lambda}_{\sP }' \bm{\lambda}_\sP)^{-1} \bm{\eta}_t \right)\left( \bm{\lambda}_s' (\bm{\lambda}_{\sP }' \bm{\lambda}_\sP)^{-1} \bm{\lambda}_s \right)\\
&\leq \left( \frac{\widebar{\eta}^2 F}{\TcE \underline{\zeta}} \right) \left( \frac{\widebar{\lambda}^2 F}{\TcE \underline{\zeta}} \right).
\end{align*}
Similarly,
\begin{align}
\left( \bm{\lambda}_t' (\bm{\lambda}_{\sP }' \bm{\lambda}_\sP)^{-1} \bm{\lambda}_s \right)^2 
&\leq \left( \frac{\widebar{\lambda}^2 F}{\TcE \underline{\zeta}} \right)^2. \label{eqn:lambda:CauchySchwarz}
\end{align}
Let
\begin{align*}
\widebar\epsilon_{jt}^{\mathcal E} = \bm{\eta}_t' (\bm{\lambda}_{\sP }' \bm{\lambda}_\sP)^{-1} \bm{\lambda}_{\mathcal E}' \bm{\epsilon}_{j}^\sP =\sum_{s\in \mathcal E} \bm{\eta}_t' (\bm{\lambda}_{\sP }' \bm{\lambda}_\sP)^{-1}
\bm{\lambda}_s \epsilon_{js}.
\end{align*}
Because $\widebar\epsilon_{jt}^{\mathcal E}$ is a linear combination of independent sub-Gaussians with variance
proxy $\widebar\sigma^2$, it follows that $\widebar\epsilon_{jt}^{\mathcal E}$ is sub-Gaussian with variance proxy
$(\widebar\eta\,\widebar\lambda F/\underline\zeta)^2\widebar\sigma^2/\TcE$. Let $\mathcal{S} = \{\bm{w} \in \bR^J : \sum_{j=1}^J w_j = 1\}$. Theorem~1.16 from \citet{rigollet2019high} implies
\begin{multline*}
\left| \bE\Big[\sum_{j = 1}^J w^*_j Y^I_{jt} - \sum_{j = 1}^J f_j Y^I_{jt} \Big]\right| \\
= \left| \bE\Big[ \sum_{j=1}^J w^*_j \widebar\epsilon_{jt}^\mathcal{E} \Big]\right| 
\leq \bE\Bigg[ \max_{\bm{w} \in \mathcal{S}} \Big|\sum_{j=1}^J w_j \widebar\epsilon_{jt}^\mathcal{E} \Big| \Bigg] 
\leq \frac{\widebar{\eta}\,\widebar{\lambda} F}{\underline{\zeta}} \sqrt{2\log{(2J)}}\frac{\widebar{\sigma}}{\sqrt{\TcE}}.
\end{multline*}

An analogous argument yields
\begin{align*}
\left| \bE\Big[\sum_{j = 1}^J v^*_j Y^N_{jt} - \sum_{j = 1}^J f_j Y^N_{jt} \Big]\right|
\leq 
\frac{\widebar{\lambda}^2 F}{\underline{\zeta}} \sqrt{2\log{(2J)}}\frac{\widebar{\sigma}}{\sqrt{\TcE}}, 
\end{align*}
which completes the proof of the theorem. 

Suppose now Assumption~\ref{asp:ApproximateFit} holds (but Assumption~\ref{asp:PerfectFit} does not). To obtain a bound on the bias, we bound the first two terms in \eqref{equation:bias5terms}.
Recall that 
\begin{align*}
\left| \bm{\eta}_t' (\bm{\lambda}_{\sP }' \bm{\lambda}_\sP)^{-1} \bm{\lambda}_s \right| \leq \frac{ \widebar{\lambda} \widebar{\eta} F}{\TcE \underline{\zeta}}.
\end{align*}
Therefore, the absolute value of each element in vector $(\bm{\gamma}'_t - \bm{\eta}_t' (\bm{\lambda}_{\sP }' \bm{\lambda}_\sP)^{-1} \bm{\lambda}_{\sP }' \bm{\theta}_\sP)$ is bounded by $\widebar{\gamma} + \widebar{\theta}  \dfrac{\widebar{\lambda} \widebar{\eta} F}{\underline{\zeta}}$.
Cauchy–Schwarz inequality and Assumption~\ref{asp:ApproximateFit} imply
\begin{align*}
\Bigg|(\bm{\gamma}'_t - \bm{\eta}_t' (\bm{\lambda}_{\sP }' \bm{\lambda}_\sP)^{-1} \bm{\lambda}_{\sP }' \bm{\theta}_\sP) & \Bigg(\sum_{j = 1}^J w^*_j \bm{Z}_j - \sum_{j = 1}^J f_j \bm{Z}_j\Bigg)\Bigg| \\
\leq & \ \Big(\widebar{\gamma} + \widebar{\theta}  \dfrac{\widebar{\lambda} \widebar{\eta} F}{\underline{\zeta}}   \Big)  \sqrt{R}  \Bigg\|\sum_{j = 1}^J w^*_j \bm{Z}_j - \sum_{j = 1}^J f_j \bm{Z}_j\Bigg\|_2 \\
\leq & \ \Big(\widebar{\gamma} + \widebar{\theta}   \dfrac{\widebar{\lambda} \widebar{\eta} F}{\underline{\zeta}} \Big) R d,
\end{align*}
and
\begin{align*}
\left|\bm{\eta}_t' (\bm{\lambda}_{\sP }' \bm{\lambda}_\sP)^{-1} \bm{\lambda}_{\sP }' \Bigg(\sum_{j = 1}^J w^*_j \bm{Y}^\sP_j - \sum_{j = 1}^J f_j \bm{Y}^\sP_j\Bigg)\right| \leq & \frac{\widebar{\lambda} \widebar{\eta} F}{\underline{\zeta}} d.
\end{align*}

Combining the last two displayed equations with \eqref{equation:bias5terms}, we have
\begin{align*}
\left| \bE\Bigg[\sum_{j = 1}^J w^*_j Y^I_{jt} - \sum_{j = 1}^J f_j Y^I_{jt} \Bigg]\right| 
\leq 
\Big(\widebar{\gamma} R + \frac{\widebar{\lambda} \widebar{\eta} F}{\underline{\zeta}}  (1+\widebar{\theta} R) \Big)    d + \frac{\widebar{\lambda} \widebar{\eta} F}{\underline{\zeta}} \sqrt{2\log{(2J)}}\frac{\widebar{\sigma}}{\sqrt{\TcE}}.
\end{align*}
An analogous derivation produces
\begin{align*}
\left| E\Bigg[\sum_{j = 1}^J v^*_j Y^N_{jt} - \sum_{j = 1}^J f_j Y^N_{jt} \Bigg]\right|
\leq 
\Big(\widebar{\theta} R + \frac{\widebar{\lambda}^2 F}{\underline{\zeta}} (1+\widebar{\theta} R) \Big)    d + \frac{\widebar{\lambda}^2 F}{\underline{\zeta}} \sqrt{2\log{(2J)}}\frac{\widebar{\sigma}}{\sqrt{\TcE}},
\end{align*}
which finishes the proof of the theorem. 
\Halmos\end{proof}

\subsection{Proof of Theorem~\ref{thm:ExactPValue}}

\begin{proof}[Proof of Theorem~\ref{thm:ExactPValue}.]
Recall that
\begin{align*}
\widehat u_t = \sum_{j=1}^J w_j^* Y_{jt}-\sum_{j=1}^J v_j^* Y_{jt},
\end{align*}
for $t \in \mathcal{B} \cup \{T_0+1, \ldots, T\}$.
For $t \in\{T_0+1, \ldots, T\}$, $\widehat u_t$ are the post-intervention estimates of the treatment effects; and for $t \in \mathcal{B}$, $\widehat u_t$ are the placebo treatment effects estimated for the blank periods. Let 
\begin{align*}
u_t = \sum_{j=1}^J w^*_j \epsilon_{jt} - \sum_{j=1}^J v^*_j \epsilon_{jt}
\end{align*}
for $t \in \mathcal{B}$, and 
\begin{align*}
u_t = \sum_{j=1}^J w^*_j \xi_{jt} - \sum_{j=1}^J v^*_j \epsilon_{jt}
\end{align*}
for $t\in\{T_0+1, \ldots, T\}$. 
The null hypothesis \eqref{equation:null} and the assumptions of Theorem~\ref{thm:ExactPValue} imply that $\{u_t\}_{t\in \mathcal B\cup \{T_0+1, \ldots, T\}}$ is a sequence of exchangeable random variables. 
Additionally, Assumption~\ref{asp:FactorModel} and the null hypothesis \eqref{equation:null} imply
\begin{align}
\widehat u_t = \sum_{j=1}^J w^*_j Y_{jt} - \sum_{j=1}^J v^*_j Y_{jt} 
= \bm{\theta}_t' \sum_{j=1}^J (w^*_j - v^*_j) \bm{Z}_j + \bm{\lambda}_t' \sum_{j=1}^J (w^*_j - v^*_j) \bm{\mu}_j + u_t, \label{eqn:ExactPValue:intermediate}
\end{align}
for $t\in \mathcal B\cup \{T_0+1, \ldots, T\}$.
The result of the theorem then follows from Theorem~D.1 in \citet{chernozhukov2021exact}.
\Halmos\end{proof}

\subsection{Proof of Theorem \ref{thm:CI:Coverage}}
\label{sec:proofs:CI}
\subsubsection{A Technical Lemma}

We first define the following quantity and present a technical lemma.
Let $\bm{\epsilon}_* = (\epsilon_{1*}, \epsilon_{2*}, ..., \epsilon_{J*})$ be an i.i.d. copy of $(\epsilon_{1t}, \epsilon_{2t}, ..., \epsilon_{Jt})$ the idiosyncratic noises.
Using the definition of $\bm{\epsilon}_*$ and conditioning on the weights $(\bm{w}^*, \bm{v}^*)$, we define, for any $q \in \bR$,
\begin{align}
P_{\cE, q} = \Pr \Bigg( \Big\vert \sum_{j=1}^J w^*_j \epsilon_{j*} - \sum_{j=1}^J v^*_j \epsilon_{j*} \Big\vert \leq q \Bigg). \label{eqn:EstimationPeriods:TrueProb}
\end{align}

\begin{lemma}
\label{lem:CI:main}
Assume there exist parameters $\epsilon_{\cB}$ and $\epsilon_{\cT}$, as well as events $\mathcal{C}_{\cB}$ and $\mathcal{C}_{\cT}$, such that the following two conditions hold:
\begin{enumerate}
\item There exists a high probability event $\mathcal{C}_\cB$ such that conditional on this event, for any weights $(\bm{w}^*, \bm{v}^*)$ and any $q \in \bR$,
\begin{align}
\bigg\vert \frac{1}{T_0 - \TcE} \sum_{t \in \cB} \bI\Big\{ \Big\vert \sum_{j=1}^J w^*_j Y_{jt} - \sum_{j=1}^J v^*_j Y_{jt} \Big\vert \leq q \Big\} - P_{\cE, q} \bigg\vert \leq \epsilon_{\cB}. \label{eqn:Condition1}
\end{align}
\item Recall that $\tau_t = \sum_{j=1}^J f_j(Y_{jt}^I - Y_{jt}^N)$. There exists a high probability event $\mathcal{C}_\cT$ such that conditional on this event, for any weights $(\bm{w}^*, \bm{v}^*)$, any $q \in \bR$, and any $t \in \{T_0+1, ..., T\}$,
\begin{align}
\bigg\vert \Pr\Big( \Big\vert \sum_{j=1}^J w^*_j Y_{jt} - \sum_{j=1}^J v^*_j Y_{jt} - \tau_t \Big\vert \leq q \Big) - P_{\cE, q} \bigg\vert \leq \epsilon_{\cT}. \label{eqn:Condition2}
\end{align}
\end{enumerate}
Assume that the joint event $\mathcal{C}_\cB \cap \mathcal{C}_\cT$ happens with probability at least $1 - \delta_\cB(\epsilon_\cB) - \delta_\cT(\epsilon_\cT)$, where we use $\delta_\cB(\epsilon_\cB)$ and $\delta_\cT(\epsilon_\cT)$ to stand for two quantities that each depends on $\epsilon_\cB$ and $\epsilon_\cT$, respectively.
In addition, assume that $\big\vert \sum_{j=1}^J w^*_j \epsilon_{j*} - \sum_{j=1}^J v^*_j \epsilon_{j*} \big\vert$ has a continuous distribution.
Then, for any $\alpha \in (0,1)$ and any $t \in \{T_0+1, ..., T\}$,
\begin{align}
\bigg\vert \Pr\Big( \Big\vert \sum_{j=1}^J w^*_j Y_{jt} - \sum_{j=1}^J v^*_j Y_{jt} - \tau_t \Big\vert \leq \widehat{q}_{1-\alpha} \Big) - (1-\alpha)\bigg\vert \leq \epsilon_{\cB} + \epsilon_{\cT} + \delta_{\cB}(\epsilon_\cB) + \delta_{\cT}(\epsilon_\cT). \label{eqn:CI:approximate}
\end{align}
\end{lemma}

Note that Lemma~\ref{lem:CI:main} does not require Assumptions~\ref{asp:FactorModel}--\ref{asp:PerfectFit}.
But for Conditions~\eqref{eqn:Condition1} and \eqref{eqn:Condition2} to hold, we will apply Assumptions~\ref{asp:FactorModel}--\ref{asp:PerfectFit}.
To prove Lemma~\ref{lem:CI:main}, we borrow the proof techniques from \citet{oliveira2022split}.
We first define the following quantile on the probability distribution (instead of the empirical distribution),
\begin{align}
q_{1-\alpha} = \inf_{z \in \bR} \Bigg\{ \Pr\bigg( \Big\vert \sum_{j=1}^J w^*_j \epsilon_{j*} - \sum_{j=1}^J v^*_j \epsilon_{j*} \Big\vert \leq z \bigg) \geq 1-\alpha \Bigg\} \label{eqn:TrueQuantile}
\end{align}
Intuitively, $\widehat{q}_{1-\alpha}$ as defined in \eqref{eqn:EmpiricalQuantile} approximates $q_{1-\alpha}$ as defined in \eqref{eqn:TrueQuantile}.

\begin{proof}[Proof of Lemma~\ref{lem:CI:main}.]
This proof proceeds in two parts.

\noindent \textbf{Part 1}:
Consider the event 
\begin{align*}
\cE_1 = \bigg\{ \widehat{q}_{1-\alpha} \geq q_{1-\alpha-\epsilon_{\cB}}\bigg\}.
\end{align*}
We aim to show that event $\cE_1$ occurs given event $\mathcal{C}_\cB$.
For any positive integer $k \in \bN$, we can use Condition~\eqref{eqn:Condition1} to show that conditional on event $\mathcal{C}_\cB$,
\begin{align*}
\frac{1}{T_0 - \TcE} \sum_{t \in \cB} \bI\Big\{ \Big\vert \sum_{j=1}^J w^*_j Y_{jt} - & \sum_{j=1}^J v^*_j Y_{jt} \Big\vert \leq q_{1-\alpha-\epsilon_{\cB}} - \frac{1}{k} \Big\} \\
\leq & \Pr \Bigg( \Big\vert \sum_{j=1}^J w^*_j \epsilon_{j*} - \sum_{j=1}^J v^*_j \epsilon_{j*} \Big\vert \leq q_{1-\alpha-\epsilon_{\cB}} - \frac{1}{k} \Bigg) + \epsilon_{\cB} \\
< & 1 - \alpha - \epsilon_{\cB} + \epsilon_{\cB} \\
= & 1 - \alpha \\
\leq & \frac{1}{T_0-\TcE} \sum_{t \in \cB} \bI\Big\{ \Big\vert \sum_{j=1}^J w^*_j Y_{jt} - \sum_{j=1}^J v^*_j Y_{jt} \Big\vert \leq \widehat{q}_{1-\alpha} \Big\},
\end{align*}
where the first inequality is due to Condition~\eqref{eqn:Condition1};
the second inequality is due to the infimum part of \eqref{eqn:TrueQuantile} (because $q_{1-\alpha-\epsilon_{\cB}} - \frac{1}{k} < q_{1-\alpha-\epsilon_{\cB}}$ which is the infimum value such that the probability in \eqref{eqn:TrueQuantile} is greater or equal to $1-\alpha$);
the last inequality is due to the definition of $\widehat{q}_{1-\alpha}$ in \eqref{eqn:EmpiricalQuantile}.

The above inequality suggests that for any $k \in \bN$, the event
\begin{multline*}
\cE_k^{(\leq)} = \Bigg\{ \frac{1}{T_0 - \TcE} \sum_{t \in \cB} \bI\Big\{ \Big\vert \sum_{j=1}^J w^*_j Y_{jt} - \sum_{j=1}^J v^*_j Y_{jt} \Big\vert \leq q_{1-\alpha-\epsilon_{\cB}} - \frac{1}{k} \Big\} \\
\leq \frac{1}{T_0-\TcE} \sum_{t \in \cB} \bI\Big\{ \Big\vert \sum_{j=1}^J w^*_j Y_{jt} - \sum_{j=1}^J v^*_j Y_{jt} \Big\vert \leq \widehat{q}_{1-\alpha} \Big\}\Bigg\}
\end{multline*}
happens conditional on event $\mathcal{C}_\cB$.
Since the left hand side of the inequality inside event $\cE_k^{(\leq)}$, which is $\frac{1}{T_0 - \TcE} \sum_{t \in \cB} \bI\Big\{ \Big\vert \sum_{j=1}^J w^*_j Y_{jt} - \sum_{j=1}^J v^*_j Y_{jt} \Big\vert \leq q_{1-\alpha-\epsilon_{\cB}} - \frac{1}{k} \Big\}$, is increasing in $k$, so the probability $\cE_k^{(\leq)}$ decreases in $k$.
Given that the lower bound of $\Pr(\cE_k^{(\leq)})$ exists, the limit of $\lim_{k \to +\infty} \Pr(\cE_k^{(\leq)})$ exists, i.e., 
\begin{align*}
1 - \delta_{\cB} \leq \lim_{k \to +\infty} \Pr(\cE_k^{(\leq)}) = \Pr(\cE_{\infty}^{(\leq)}),
\end{align*}
where we use $\Pr(\cE_{\infty}^{(\leq)})$ to stand for the limiting event
\begin{multline*}
\cE_{\infty}^{(\leq)} = \Bigg\{ \frac{1}{T_0 - \TcE} \sum_{t \in \cB} \bI\Big\{ \Big\vert \sum_{j=1}^J w^*_j Y_{jt} - \sum_{j=1}^J v^*_j Y_{jt} \Big\vert \leq q_{1-\alpha-\epsilon_{\cB}} \Big\} \\
\leq \frac{1}{T_0-\TcE} \sum_{t \in \cB} \bI\Big\{ \Big\vert \sum_{j=1}^J w^*_j Y_{jt} - \sum_{j=1}^J v^*_j Y_{jt} \Big\vert \leq \widehat{q}_{1-\alpha} \Big\}\Bigg\}.
\end{multline*}
This means that, event $\cE_1 = \{ \widehat{q}_{1-\alpha} \geq q_{1-\alpha-\epsilon_{\cB}}\}$ happens conditional on event $\mathcal{C}_\cB$.
Due to the assumption of Lemma~\ref{lem:CI:main}, event $\mathcal{C}_\cT \cap \cE_1$ happens with probability at least $1-\delta_{\cB}(\epsilon_\cB)-\delta_{\cT}(\epsilon_\cT)$.

Next we have, for any $t \in \{T_0+1, ..., T\}$ in the experimental periods,
\begin{align*}
& \Pr\Bigg( \Big\vert \sum_{j=1}^J w^*_j Y_{jt} - \sum_{j=1}^J v^*_j Y_{jt} - \tau_t \Big\vert \leq \widehat{q}_{1-\alpha} \Bigg) \\
\geq & \Pr\Bigg( \Big\vert \sum_{j=1}^J w^*_j Y_{jt} - \sum_{j=1}^J v^*_j Y_{jt} - \tau_t \Big\vert \leq \widehat{q}_{1-\alpha} \bigcap (\mathcal{C}_\cT \cap \cE_1) \Bigg) - \delta_\cB(\epsilon_\cB) - \delta_\cT(\epsilon_\cT) \\
\geq & \Pr\Bigg( \Big\vert \sum_{j=1}^J w^*_j Y_{jt} - \sum_{j=1}^J v^*_j Y_{jt} - \tau_t \Big\vert \leq q_{1-\alpha-\epsilon_{\cB}} \Bigg) - \delta_\cB(\epsilon_\cB) - \delta_\cT(\epsilon_\cT) \\
\geq & \Pr\Bigg( \Big\vert \sum_{j=1}^J w^*_j \epsilon_{j*} - \sum_{j=1}^J v^*_j \epsilon_{j*} \Big\vert \leq q_{1-\alpha-\epsilon_{\cB}} \Bigg) - \epsilon_\cT - \delta_\cB(\epsilon_\cB) - \delta_\cT(\epsilon_\cT) \\
\geq & 1 - \alpha - \epsilon_\cB - \epsilon_\cT - \delta_\cB(\epsilon_\cB) - \delta_\cT(\epsilon_\cT).
\end{align*}
where the second inequality is because the probability decreases if we decrease from $\widehat{q}_{1-\alpha}$ to $q_{1-\alpha-\epsilon_{\cB}}$;
the third inequality is due to Condition~\eqref{eqn:Condition2};
the last inequality is due to the definition of $q_{1-\alpha-\epsilon_{\cB}}$ in \eqref{eqn:TrueQuantile}.

\noindent \textbf{Part 2}:
Consider the event 
\begin{align*}
\cE_2 = \bigg\{ \widehat{q}_{1-\alpha} \leq q_{1-\alpha+\epsilon_{\cB}}\bigg\}.
\end{align*}
We wish to show that event $\cE_1$ happens conditional on event $\mathcal{C}_\cB$.
We use Condition~\eqref{eqn:Condition2} to show that conditional on event $\mathcal{C}_\cB$,
\begin{align*}
\frac{1}{T_0 - \TcE} \sum_{t \in \cB} \bI\Big\{ \Big\vert \sum_{j=1}^J w^*_j Y_{jt} - & \sum_{j=1}^J v^*_j Y_{jt} \Big\vert \leq q_{1-\alpha+\epsilon_{\cB}} \Big\} \\
\geq & \Pr \Bigg( \Big\vert \sum_{j=1}^J w^*_j \epsilon_{j*} - \sum_{j=1}^J v^*_j \epsilon_{j*} \Big\vert \leq q_{1-\alpha+\epsilon_{\cB}} \Bigg) - \epsilon_\cB\\
\geq & 1 - \alpha + \epsilon_{\cB} - \epsilon_{\cB} \\
= & 1 - \alpha,
\end{align*}
where the first inequality is due to Condition~\eqref{eqn:Condition1};
the second inequality is due to the definition of $q_{1-\alpha+\epsilon_{\cB}}$ in \eqref{eqn:TrueQuantile};

Due to \eqref{eqn:EmpiricalQuantile}, since $\widehat{q}_{1-\alpha}$ is the smallest value satisfying this condition, we have that event $\cE_2 = \{\widehat{q}_{1-\alpha} \leq q_{1-\alpha+\epsilon_{\cB}}\}$ happens conditional on event $\mathcal{C}_\cB$.
Due to the assumption of Lemma~\ref{lem:CI:main}, event $\mathcal{C}_\cT \cap \cE_2$ happens with probability at least $1-\delta_{\cB}(\epsilon_\cB)-\delta_{\cT}(\epsilon_\cT)$.

Then, for any $t \in \{T_0+1, ..., T\}$ in the experimental periods,
\begin{align*}
& \Pr\Bigg( \Big\vert \sum_{j=1}^J w^*_j Y_{jt} - \sum_{j=1}^J v^*_j Y_{jt} - \tau_t \Big\vert \leq \widehat{q}_{1-\alpha} \Bigg) \\
\leq & \Pr\Bigg( \Big\vert \sum_{j=1}^J w^*_j Y_{jt} - \sum_{j=1}^J v^*_j Y_{jt} - \tau_t \Big\vert \leq \widehat{q}_{1-\alpha} \bigcap (\mathcal{C}_\cT \cap \cE_2) \Bigg) + \delta_\cB(\epsilon_\cB) + \delta_\cT(\epsilon_\cT) \\
\leq & \Pr\Bigg( \Big\vert \sum_{j=1}^J w^*_j Y_{jt} - \sum_{j=1}^J v^*_j Y_{jt} - \tau_t \Big\vert \leq q_{1-\alpha+\epsilon_{\cB}} \Bigg) + \delta_\cB(\epsilon_\cB) + \delta_\cT(\epsilon_\cT) \\
\leq & \Pr\Bigg( \Big\vert \sum_{j=1}^J w^*_j \epsilon_{j*} - \sum_{j=1}^J v^*_j \epsilon_{j*} \Big\vert \leq q_{1-\alpha+\epsilon_{\cB}} \Bigg) + \epsilon_\cT + \delta_\cB(\epsilon_\cB) + \delta_\cT(\epsilon_\cT) \\
\leq & 1 - \alpha + \epsilon_\cB + \epsilon_\cT + \delta_\cB(\epsilon_\cB) + \delta_\cT(\epsilon_\cT).
\end{align*}
where the second inequality is because the probability increases if we increase from $\widehat{q}_{1-\alpha}$ to $q_{1-\alpha+\epsilon_{\cB}}$;
the third inequality is due to Condition~\eqref{eqn:Condition2};
the last inequality is due to the definition of $q_{1-\alpha+\epsilon_{\cB}}$ in \eqref{eqn:TrueQuantile}.
\Halmos\end{proof}

\subsubsection{Proof of Theorem~\ref{thm:CI:Coverage}}

In this section, we use Lemma~\ref{lem:CI:main} to prove Theorem~\ref{thm:CI:Coverage}.
Instead of proving exactly Theorem~\ref{thm:CI:Coverage}, we prove Theorem~\ref{thm:CI:Coverage:withConstants} below with all the constants provided.
Then, setting $z_\cB = (T_0 - \TcE)^{-\frac{1}{2}}$ and $z_\cE = \TcE^{-\frac{1}{2}}$ we prove Theorem~\ref{thm:CI:Coverage}.

\begin{theorem}
\label{thm:CI:Coverage:withConstants}
Assume that Assumptions~\ref{asp:FactorModel}--~\ref{asp:PerfectFit} hold. 
Assume there exists a constant $\kappa<\infty$, such that for all $j=1, \ldots, J$, $t=1, \ldots, T$, $\epsilon_{jt}$ are continuously distributed with the probability density function upper bounded by $\kappa$.
Assume that for $t=T_0+1, \ldots, T$, and $j=1, \ldots, J$, $\xi_{jt}$ has the same distribution as $\epsilon_{jt}$.
Then the confidence interval defined in \eqref{eqn:CI} approximately achieves point-wise coverage, i.e., for any $\alpha \in (0,1)$ and any $t \in \{T_0+1,...,T\}$,
\begin{multline*}
\bigg\vert \Pr\Big( \tau_t \in \widehat{C}_{1-\alpha}(Y_{1t}, Y_{2t},...,Y_{Jt}) \Big) - (1-\alpha)\bigg\vert \\
\leq \sqrt{\frac{1}{2(T_0-\TcE)} \log{\Big(\frac{2}{z_\cB}\Big)}} + \kappa \sqrt{\frac{8 e J \widebar{\sigma}^2 \widebar{\lambda}^2 \widebar{\eta}^2 F^2}{\underline{\zeta}^2 \TcE} \log{\Big(\frac{2J}{z_\cE}\Big)}} + 2 \kappa \sqrt{\frac{8 e J \widebar{\sigma}^2 \widebar{\lambda}^4 F^2}{\underline{\zeta}^2 \TcE} \log{\Big(\frac{2J}{z_\cE}\Big)}} + z_\cB + 3 z_\cE. 
\end{multline*}
where $z_{\cB}$ and $z_{\cE}$ are arbitrary positive constants.
\end{theorem}

\begin{proof}[Proof of Theorem~\ref{thm:CI:Coverage:withConstants}.]
We outline the proof of Theorem~\ref{thm:CI:Coverage:withConstants} as follows.
We first define four events.
We then check Conditions \eqref{eqn:Condition1} under the first two events and \eqref{eqn:Condition2} under the last two events.
Finally, we apply Lemma~\ref{lem:CI:main} and conclude the proof.

\noindent \textbf{Step 1:}
We define the following four events. 
First, in the blank periods and conditioning on the weights $(\bm{w}^*, \bm{v}^*)$ that we obtain from the fitting periods, 
\begin{multline*}
\mathcal{E}_1 = \Bigg\{ \Bigg\vert\frac{1}{T_0 - \TcE} \sum_{t \in \cB} \bI\Big\{ \Big\vert \sum_{j=1}^J w^*_j Y_{jt} - \sum_{j=1}^J v^*_j Y_{jt} \Big\vert \leq q \Big\} \\
- \frac{1}{T_0 - \TcE} \sum_{t \in \cB} \Pr\Big( \Big\vert \sum_{j=1}^J w^*_j Y_{jt} - \sum_{j=1}^J v^*_j Y_{jt} \Big\vert \leq q \Big) \Bigg\vert \leq \sqrt{\frac{1}{2(T_0-\TcE)} \log{\Big(\frac{2}{z_\cB}\Big)}} \Bigg\}.
\end{multline*}
Due to Hoeffding inequality for bounded random variables (conditioning on the weights $(\bm{w}^*, \bm{v}^*)$, these indicators are independent), we have that event $\cE_1$ happens with probability $1-z_\cB$.

Second, in the blank periods,
\begin{align*}
\mathcal{E}_2 & = \left\{ \forall t\in \mathcal B, \left| \bm{\lambda}_t' (\bm{\lambda}_{\sP }' \bm{\lambda}_\sP)^{-1} \bm{\lambda}_{\sP }' \sum_{j=1}^J (w^*_j - v^*_j) \bm{\epsilon}_j^\sP \right| \leq \sqrt{\frac{8 \widebar{\sigma}^2 \widebar{\lambda}^4 F^2}{\underline{\zeta}^2 \TcE} \log{\Big(\frac{2J}{z_\cE}\Big)}} \right\}\\
& = \left\{ \max_{t\in \mathcal B} \left| \bm{\lambda}_t' (\bm{\lambda}_{\sP }' \bm{\lambda}_\sP)^{-1} \bm{\lambda}_{\sP }' \sum_{j=1}^J (w^*_j - v^*_j) \bm{\epsilon}_j^\sP \right| \leq \sqrt{\frac{8 \widebar{\sigma}^2 \widebar{\lambda}^4 F^2}{\underline{\zeta}^2 \TcE} \log{\Big(\frac{2J}{z_\cE}\Big)}} \right\}.
\end{align*}
Note that,
\begin{align*}
\max_{t\in \mathcal B} \Big| \bm{\lambda}_t' &(\bm{\lambda}_{\sP }' \bm{\lambda}_\sP)^{-1} \bm{\lambda}_{\sP }' \sum_{j=1}^J (w^*_j - v^*_j) \bm{\epsilon}_j^\sP \Big|\\
& \leq \max_{t\in \mathcal B} \sum_{j=1}^J |w^*_j - v^*_j| \sum_{s\in \mathcal E} |\bm{\lambda}_t' (\bm{\lambda}_{\sP }' \bm{\lambda}_\sP)^{-1} \bm{\lambda}_s||\epsilon_{js}|\\
& \leq \sum_{j=1}^J |w^*_j - v^*_j| \sum_{s\in \mathcal E} \frac{\widebar\lambda^2F}{\TcE\underline{\zeta}}|\epsilon_{js}|,
\end{align*}
where the second inequality is due to \eqref{eqn:lambda:CauchySchwarz}, and because $|w^*_j - v^*_j| \geq 0$ and $|\epsilon_{js}| \geq 0$.
Therefore,
\begin{align*}
\Pr(\cE_2) & \geq 1-\Pr\Bigg(\sum_{j=1}^J \frac{|w^*_j - v^*_j|}{2} \sum_{s\in \mathcal E} \frac{\widebar\lambda^2F}{\TcE\underline{\zeta}}|\epsilon_{js}|> \sqrt{\frac{2 \widebar{\sigma}^2 \widebar{\lambda}^4 F^2}{\underline{\zeta}^2 \TcE} \log{\Big(\frac{2J}{z_\cE}\Big)}} \Bigg) \\
& \geq 1-\sum_{j=1}^J \Pr\Bigg(\sum_{s\in \mathcal E} \frac{\widebar\lambda^2F}{\TcE\underline{\zeta}}|\epsilon_{js}|> \sqrt{\frac{2 \widebar{\sigma}^2 \widebar{\lambda}^4 F^2}{\underline{\zeta}^2 \TcE} \log{\Big(\frac{2J}{z_\cE}\Big)}}\Bigg) \\
& \geq 1 - z_\cE,
\end{align*}
where the second inequality follows from union bound,
and the third inequality is the Chernoff bound for sub-Gaussian random variables.

Third, in the experimental periods,
\begin{align*}
\mathcal{E}_3 & = \left\{ \forall t \in \{T_0+1, \ldots, T\}, \left| \bm{\eta}_t' (\bm{\lambda}_{\sP }' \bm{\lambda}_\sP)^{-1} \bm{\lambda}_{\sP }' \sum_{j=1}^J (w^*_j - f_j) \bm{\epsilon}_j^\sP \right| \leq \sqrt{\frac{8 \widebar{\sigma}^2 \widebar{\lambda}^2 \widebar{\eta}^2 F^2}{\underline{\zeta}^2 \TcE} \log{\Big(\frac{2J}{z_\cE}\Big)}} \right\} \\
& = \left\{ \max_{t \in \{T_0+1, \ldots, T\}} \left| \bm{\eta}_t' (\bm{\lambda}_{\sP }' \bm{\lambda}_\sP)^{-1} \bm{\lambda}_{\sP }' \sum_{j=1}^J (w^*_j - f_j) \bm{\epsilon}_j^\sP \right| \leq \sqrt{\frac{8 \widebar{\sigma}^2 \widebar{\lambda}^2 \widebar{\eta}^2 F^2}{\underline{\zeta}^2 \TcE} \log{\Big(\frac{2J}{z_\cE}\Big)}} \right\}.
\end{align*}
Note that,
\begin{align*}
\max_{t\in \{T_0+1, \ldots, T\}} \Big| \bm{\eta}_t' &(\bm{\lambda}_{\sP }' \bm{\lambda}_\sP)^{-1} \bm{\lambda}_{\sP }' \sum_{j=1}^J (w^*_j - f_j) \bm{\epsilon}_j^\sP \Big|\\
& \leq \max_{t\in \{T_0+1, \ldots, T\}} \sum_{j=1}^J |w^*_j - f_j| \sum_{s\in \mathcal E} |\bm{\eta}_t' (\bm{\lambda}_{\sP }' \bm{\lambda}_\sP)^{-1} \bm{\lambda}_s||\epsilon_{js}|\\
& \leq \sum_{j=1}^J |w^*_j - f_j| \sum_{s\in \mathcal E} \frac{\widebar{\lambda}\widebar{\eta}F}{\TcE\underline{\zeta}}|\epsilon_{js}|,
\end{align*}
where the second inequality is due to \eqref{eqn:lambda:CauchySchwarz}, and because $|w^*_j - f_j| \geq 0$ and $|\epsilon_{js}| \geq 0$.
Therefore,
\begin{align*}
\Pr(\cE_3) & \geq 1-\Pr\Bigg(\sum_{j=1}^J \frac{|w^*_j - f_j|}{2} \sum_{s\in \mathcal E} \frac{\widebar{\lambda}\widebar{\eta}F}{\TcE\underline{\zeta}}|\epsilon_{js}|> \sqrt{\frac{2 \widebar{\sigma}^2 \widebar{\lambda}^2\widebar{\eta}^2 F^2}{\underline{\zeta}^2 \TcE} \log{\Big(\frac{2J}{z_\cE}\Big)}} \Bigg) \\
& \geq 1-\sum_{j=1}^J \Pr\Bigg(\sum_{s\in \mathcal E} \frac{\widebar{\lambda}\widebar{\eta}F}{\TcE\underline{\zeta}}|\epsilon_{js}|> \sqrt{\frac{2 \widebar{\sigma}^2 \widebar{\lambda}^2\widebar{\eta}^2 F^2}{\underline{\zeta}^2 \TcE} \log{\Big(\frac{2J}{z_\cE}\Big)}}\Bigg) \\
& \geq 1 - z_\cE,
\end{align*}
where the second inequality follows from union bound,
and the third inequality is the Chernoff bound for sub-Gaussian random variables.

Fourth, in the experimental periods,
\begin{align*}
\mathcal{E}_4 & = \left\{ \forall t \in \{T_0+1, \ldots, T\}, \left| \bm{\lambda}_t' (\bm{\lambda}_{\sP }' \bm{\lambda}_\sP)^{-1} \bm{\lambda}_{\sP }' \sum_{j=1}^J (v^*_j - f_j) \bm{\epsilon}_j^\sP \right| \leq \sqrt{\frac{8 \widebar{\sigma}^2 \widebar{\lambda}^4 F^2}{\underline{\zeta}^2 \TcE} \log{\Big(\frac{2J}{z_\cE}\Big)}} \right\}.
\end{align*}
Similar to the event $\mathcal{E}_3$, we can show that $\Pr(\cE_4) \geq 1 - z_\cE$.

\noindent \textbf{Step 2:}
Now we check Conditions~\eqref{eqn:Condition1} and~\eqref{eqn:Condition2}.
We first check Condition~\eqref{eqn:Condition1}.
In the statement of Condition~\eqref{eqn:Condition1}, let $\mathcal{C}_{\cB} = \cE_1 \cap \cE_2$.
Note that
\begin{multline*}
\sum_{j=1}^J w^*_j \bm{Y}_j^\sP - \sum_{j=1}^J v^*_j \bm{Y}_j^\sP = \bm{\theta}_\sP \left(\sum_{j=1}^J w^*_j \bm{Z}_j - \sum_{j=1}^J v^*_j \bm{Z}_j \right) \\
+ \bm{\lambda}_\sP \left(\sum_{j=1}^J w^*_j \bm{\mu}_j - \sum_{j=1}^J v^*_j \bm{\mu}_j \right) + \left(\sum_{j=1}^J w^*_j \bm{\epsilon}_j^\sP - \sum_{j=1}^J v^*_j \bm{\epsilon}_j^\sP\right).
\end{multline*}
Assumption \ref{asp:PerfectFit} implies
\begin{align*}
\sum_{j=1}^J w^*_j \bm{\mu}_j - \sum_{j=1}^J v^*_j \bm{\mu}_j 
= 
- (\bm{\lambda}_{\sP }' \bm{\lambda}_\sP)^{-1} \bm{\lambda}_{\sP }'
\left(\sum_{j=1}^J w^*_j \bm{\epsilon}_j^\sP - \sum_{j=1}^J v^*_j \bm{\epsilon}_j^\sP\right).
\end{align*}
For $t\in \mathcal B$, we have
\begin{multline*}
\sum_{j=1}^J w^*_j Y_{jt} - \sum_{j=1}^J v^*_j Y_{jt} = \sum_{j=1}^J w^*_j Y_{jt}^N - \sum_{j=1}^J v^*_j Y_{jt}^N \\
= \sum_{j=1}^J w^*_j \epsilon_{jt} - \sum_{j=1}^J v^*_j \epsilon_{jt} - \bm{\lambda}_t' (\bm{\lambda}_{\sP }' \bm{\lambda}_\sP)^{-1} \bm{\lambda}_{\sP }' \sum_{j=1}^J (w^*_j - v^*_j) \bm{\epsilon}_j^\sP.
\end{multline*} 

Conditional on event $\cE_2$, we have for any $t \in \cB$ in the blank periods,
\begin{align*}
\Big\vert \sum_{j=1}^J w^*_j Y_{jt} - \sum_{j=1}^J v^*_j Y_{jt} \Big\vert = & \Big\vert \sum_{j=1}^J w^*_j \epsilon_{jt} - \sum_{j=1}^J v^*_j \epsilon_{jt} - \bm{\lambda}_t' (\bm{\lambda}_{\sP }' \bm{\lambda}_\sP)^{-1} \bm{\lambda}_{\sP }' \sum_{j=1}^J (w^*_j - v^*_j) \bm{\epsilon}_j^\sP \Big\vert \\
\leq & \Big\vert \sum_{j=1}^J w^*_j \epsilon_{jt} - \sum_{j=1}^J v^*_j \epsilon_{jt} \Big\vert + \Big\vert \bm{\lambda}_t' (\bm{\lambda}_{\sP }' \bm{\lambda}_\sP)^{-1} \bm{\lambda}_{\sP }' \sum_{j=1}^J (w^*_j - v^*_j) \bm{\epsilon}_j^\sP \Big\vert \\
\leq & \Big\vert \sum_{j=1}^J w^*_j \epsilon_{jt} - \sum_{j=1}^J v^*_j \epsilon_{jt} \Big\vert + \sqrt{\frac{8 \widebar{\sigma}^2 \widebar{\lambda}^4 F^2}{\underline{\zeta}^2 \TcE} \log{\Big(\frac{2J}{z_\cE}\Big)}}
\end{align*}
From the above inequality, for any $t \in \cB$, due to Lemma~\ref{lem:BoundedDensity}-1 and Lemma~\ref{lem:BoundedDensity:ut}, the probability density of $\big\vert \sum_{j=1}^J w^*_j \epsilon_{jt} - \sum_{j=1}^J v^*_j \epsilon_{jt} \big\vert$ is upper bounded by $\kappa\sqrt{eJ}$, where $e \approx 2.718$ is the base of the natural logarithm.
This implies that, conditional on event $\cE_2$, for any $t \in \cB$ and any $q \in \bR$,
\begin{align*}
& \Bigg\vert \Pr\bigg( \Big\vert \sum_{j=1}^J w^*_j Y_{jt} - \sum_{j=1}^J v^*_j Y_{jt} \Big\vert \leq q \bigg) - \Pr\bigg( \Big\vert \sum_{j=1}^J w^*_j \epsilon_{j*} - \sum_{j=1}^J v^*_j \epsilon_{j*} \Big\vert \leq q \bigg) \Bigg\vert \\
= & \Bigg\vert \Pr\bigg( \Big\vert \sum_{j=1}^J w^*_j Y_{jt} - \sum_{j=1}^J v^*_j Y_{jt} \Big\vert \leq q \bigg) - \Pr\bigg( \Big\vert \sum_{j=1}^J w^*_j \epsilon_{jt} - \sum_{j=1}^J v^*_j \epsilon_{jt} \Big\vert \leq q \bigg) \Bigg\vert \\
\leq & \kappa \sqrt{\frac{8 e J \widebar{\sigma}^2 \widebar{\lambda}^4 F^2}{\underline{\zeta}^2 \TcE} \log{\Big(\frac{2J}{z_\cE}\Big)}}.
\end{align*}
This means that, conditional on event $\cE_2$, for any $t \in \cB$ and any $q \in \bR$,
\begin{multline*}
\Bigg\vert \frac{1}{T_0 - \TcE} \sum_{t \in \cB} \Pr\Big( \Big\vert \sum_{j=1}^J w^*_j Y_{jt} - \sum_{j=1}^J v^*_j Y_{jt} \Big\vert \leq q \Big) - \Pr\bigg( \Big\vert \sum_{j=1}^J w^*_j \epsilon_{j*} - \sum_{j=1}^J v^*_j \epsilon_{j*} \Big\vert \leq q \bigg) \Bigg\vert \\
\leq \kappa \sqrt{\frac{8 e J \widebar{\sigma}^2 \widebar{\lambda}^4 F^2}{\underline{\zeta}^2 \TcE} \log{\Big(\frac{2J}{z_\cE}\Big)}}.
\end{multline*}
To conclude checking Condition~\eqref{eqn:Condition1}, we see that conditional on event $\mathcal{C}_{\cB} = \cE_1 \cap \cE_2$, for any weights $(\bm{w}^*, \bm{v}^*)$ and $q \in \bR$,
\begin{multline*}
\Bigg\vert \frac{1}{T_0 - \TcE} \sum_{t \in \cB} \bI\Big\{ \Big\vert \sum_{j=1}^J w^*_j Y_{jt} - \sum_{j=1}^J v^*_j Y_{jt} \Big\vert \leq q \Big\} - \Pr\bigg( \Big\vert \sum_{j=1}^J w^*_j \epsilon_{j*} - \sum_{j=1}^J v^*_j \epsilon_{j*} \Big\vert \leq q \bigg) \Bigg\vert \\
\leq \sqrt{\frac{1}{2(T_0-\TcE)} \log{\Big(\frac{2}{z_\cB}\Big)}} + \kappa \sqrt{\frac{8 e J \widebar{\sigma}^2 \widebar{\lambda}^4 F^2}{\underline{\zeta}^2 \TcE} \log{\Big(\frac{2J}{z_\cE}\Big)}}.
\end{multline*}

We then move on to check Condition~\eqref{eqn:Condition2}.
In the statement of Condition~\eqref{eqn:Condition2}, let $\mathcal{C}_\cT = \cE_3 \cap \cE_4$.
For any $t \in \{T_0+1,...,T\}$ in the experimental periods, we have
\begin{align*}
& \sum_{j=1}^J w^*_j Y_{jt} - \sum_{j=1}^J v^*_j Y_{jt} - \tau_t \\
= & \sum_{j=1}^J (w^*_j - f_j) Y_{jt}^I - \sum_{j=1}^J (v^*_j - f_j) Y_{jt}^N\\
= & \sum_{j=1}^J w^*_j \epsilon_{jt} - \sum_{j=1}^J v^*_j \epsilon_{jt} - \bm{\eta}_t' (\bm{\lambda}_{\sP }' \bm{\lambda}_\sP)^{-1} \bm{\lambda}_{\sP }' \sum_{j=1}^J (w^*_j - f_j) \bm{\epsilon}_j^\sP + \bm{\lambda}_t' (\bm{\lambda}_{\sP }' \bm{\lambda}_\sP)^{-1} \bm{\lambda}_{\sP }' \sum_{j=1}^J (v^*_j - f_j) \bm{\epsilon}_j^\sP.
\end{align*} 
where the third equality is using Assumption~\ref{asp:PerfectFit} and using the assumption that $\xi_{jt}$ has the same distribution as $\epsilon_{jt}$ for $t=T_0+1, \ldots, T$, and $j=1, \ldots, J$.

Conditional on event $\cE_3 \cap \cE_4$, we have for any $t \in \{T_0+1,...,T\}$ in the experimental periods,
\begin{multline*}
\Big\vert \sum_{j=1}^J w^*_j Y_{jt} - \sum_{j=1}^J v^*_j Y_{jt} - \tau_t \Big\vert \\
\leq \Big\vert \sum_{j=1}^J w^*_j \epsilon_{jt} - \sum_{j=1}^J v^*_j \epsilon_{jt} \Big\vert + \sqrt{\frac{8 \widebar{\sigma}^2 \widebar{\lambda}^2 \widebar{\eta}^2 F^2}{\underline{\zeta}^2 \TcE} \log{\Big(\frac{2J}{z_\cE}\Big)}} + \sqrt{\frac{8 \widebar{\sigma}^2 \widebar{\lambda}^4 F^2}{\underline{\zeta}^2 \TcE} \log{\Big(\frac{2J}{z_\cE}\Big)}}.
\end{multline*}

Following the same argument, we see that conditional on event $\mathcal{C}_\cT = \cE_3 \cap \cE_4$, for any $t \in \{T_0+1,..., T\}$ and any $q \in \bR$,
\begin{multline*}
\Bigg\vert \Pr\Big( \Big\vert \sum_{j=1}^J w^*_j Y_{jt} - \sum_{j=1}^J v^*_j Y_{jt} - \tau_t \Big\vert \leq q \Big) - \Pr\bigg( \Big\vert \sum_{j=1}^J w^*_j \epsilon_{j*} - \sum_{j=1}^J v^*_j \epsilon_{j*} \Big\vert \leq q \bigg) \Bigg\vert \\
\leq \kappa \sqrt{\frac{8 e J \widebar{\sigma}^2 \widebar{\lambda}^2 \widebar{\eta}^2 F^2}{\underline{\zeta}^2 \TcE} \log{\Big(\frac{2J}{z_\cE}\Big)}} + \kappa \sqrt{\frac{8 e J \widebar{\sigma}^2 \widebar{\lambda}^4 F^2}{\underline{\zeta}^2 \TcE} \log{\Big(\frac{2J}{z_\cE}\Big)}}.
\end{multline*}

\noindent \textbf{Step 3:}
Now we apply Lemma~\ref{lem:CI:main}.
Note that, the joint event $\mathcal{C}_\cB \cap \mathcal{C}_\cT = \cE_1 \cap \cE_2 \cap \cE_3 \cap \cE_4$ happens with probability at least $1 - z_\cB - 3 z_\cE$.
Due to Lemma~\ref{lem:CI:main}, 
\begin{multline*}
\bigg\vert \Pr\Big( \Big\vert \sum_{j=1}^J w^*_j Y_{jt} - \sum_{j=1}^J v^*_j Y_{jt} - \tau_t \Big\vert \leq \widehat{q}_{1-\alpha} \Big) - (1-\alpha)\bigg\vert \\
\leq \sqrt{\frac{1}{2(T_0-\TcE)} \log{\Big(\frac{2}{z_\cB}\Big)}} + \kappa \sqrt{\frac{8 e J \widebar{\sigma}^2 \widebar{\lambda}^2 \widebar{\eta}^2 F^2}{\underline{\zeta}^2 \TcE} \log{\Big(\frac{2J}{z_\cE}\Big)}} + 2 \kappa \sqrt{\frac{8 e J \widebar{\sigma}^2 \widebar{\lambda}^4 F^2}{\underline{\zeta}^2 \TcE} \log{\Big(\frac{2J}{z_\cE}\Big)}} + z_\cB + 3 z_\cE.
\end{multline*}
Because $\big\vert \sum_{j=1}^J w^*_j Y_{jt} - \sum_{j=1}^J v^*_j Y_{jt} - \tau_t \big\vert \leq \widehat{q}_{1-\alpha}$ is equivalent to $\tau_t \in \widehat{C}_{1-\alpha}(Y_{1t}, Y_{2t},...,Y_{Jt})$, this implies
\begin{multline*}
\bigg\vert \Pr\Big( \tau_t \in \widehat{C}_{1-\alpha}(Y_{1t}, Y_{2t},...,Y_{Jt}) \Big) - (1-\alpha)\bigg\vert \\
\leq \sqrt{\frac{1}{2(T_0-\TcE)} \log{\Big(\frac{2}{z_\cB}\Big)}} + \kappa \sqrt{\frac{8 e J \widebar{\sigma}^2 \widebar{\lambda}^2 \widebar{\eta}^2 F^2}{\underline{\zeta}^2 \TcE} \log{\Big(\frac{2J}{z_\cE}\Big)}} + 2 \kappa \sqrt{\frac{8 e J \widebar{\sigma}^2 \widebar{\lambda}^4 F^2}{\underline{\zeta}^2 \TcE} \log{\Big(\frac{2J}{z_\cE}\Big)}} + z_\cB + 3 z_\cE.
\end{multline*}
\Halmos \end{proof}

\end{appendices}

\clearpage

\begin{appendices}

\begin{center}
{\textbf{Online Appendix}}\\[-0.9ex]
{\textbf{Synthetic Controls for Experimental Design}}\\
Alberto Abadie and Jinglong Zhao\\[-0.9ex]
\today
\end{center}
\vspace{2mm}
\setcounter{page}{1}                         
\pagenumbering{arabic}
\renewcommand*{\thepage}{OA.\arabic{page}}
\setcounter{section}{0}                      
\renewcommand{\thesection}{OA.\arabic{section}}   
\setcounter{equation}{0}
\renewcommand{\theequation}{OA.\arabic{equation}}
\setcounter{theorem}{0}
\renewcommand{\thetheorem}{OA.\arabic{theorem}}
\setcounter{figure}{0}
\renewcommand{\thefigure}{OA.\arabic{figure}}
\setcounter{table}{0}
\renewcommand{\thetable}{OA.\arabic{table}}
\medskip

\section{Designs Based on Penalized and Bias-corrected Synthetic Control Methods}
\label{sec:adaptdesigns}

Consider the design problem in \eqref{eqn:OptTreated},
\begin{align}
\underbrace{\left\|\,\widebar{\bm X}-\sum_{j=1}^J w_j\bm{X}_j \right\|^2}_{\text{\normalsize (a)}} + \xi \sum_{j=1}^J w_j \underbrace{\left\|\bm{X}_j- \sum_{i=1}^J v_{ij} \bm{X}_i \right\|^2}_{\text{\normalsize (b)}}.
\label{eqn:adaptobj}
\end{align}
To apply the penalized synthetic control method of \cite{abadie2021penalized} to this design, we replace the term (a) in \eqref{eqn:adaptobj} with
\begin{align}
\label{eqn:pen1}
\left\|\,\widebar{\bm X}-\sum_{j=1}^J w_j\bm{X}_j   \right\|^2 + \lambda_1 \sum_{j=1}^J w_j \|\widebar{\bm X}-\bm X_j\|^2,
\end{align}
and the terms (b) with
\begin{align}
\label{eqn:pen2}
\left\|\bm{X}_j- \sum_{i=1}^J v_{ij} \bm{X}_i \right\|^2 + \lambda_2 \sum_{i=1}^J v_{ij}\|\bm X_j-\bm X_i\|^2. 
\end{align}
Here, $\lambda_1$ and $\lambda_2$ are positive constants that penalize discrepancies between the target values of the predictors ($\widebar{\bm{X}}$ in \eqref{eqn:pen1} and $\bm{X}_j$ in \eqref{eqn:pen2}) and the values of the predictors for the units that contribute to their synthetic counterparts. 

All designs of Section \ref{sec:SCDesign} depend on terms akin to (a) and (b) in \eqref{eqn:adaptobj}. 
These terms can be adapted as in \eqref{eqn:pen1} and \eqref{eqn:pen2} to implement the penalized synthetic control design of \cite{abadie2021penalized}. 

For all the designs in Section~\ref{sec:SCDesign}, the bias-corrected estimator of \cite{abadie2021penalized} is
\[
\widehat\tau_t^{BC} = \sum_{j=1}^J w^*_j(Y_{jt}-\widehat\mu_{0t}(\bm X_j)) - \sum_{j=1}^J v^*_j(Y_{jt}-\widehat\mu_{0t}(\bm X_j)),
\]
where $t\geq T_0+1$ and the terms $\widehat\mu_{0t}(\bm X_j)$ are the fitted values of a regression of untreated outcomes, $Y_{jt}^N$, on unit's characteristics, $\bm X_j$. To avoid over-fitting biases, $\widehat\mu_{0t}(\bm X_j)$ can be cross-fitted for the untreated.

\section{Approximate Validity when \texorpdfstring{$\bm{\lambda}_t$}{Factor Loadings} are not Exchangeable}
\label{sec:AsymptoticValidity}

Recall that in Theorem~\ref{thm:ExactPValue} we have shown that when ${\bm \lambda}_t$ are exchangeable for $t\in\mathcal B\cup \{T_0+1, \ldots, T\}$ the $p$-value in \eqref{eqn:pValue} is exact.
In this section, we discuss the case when ${\bm \lambda}_t$ are not necessarily exchangeable.
We show below in Theorem~\ref{thm:AsymptoticValidity} that the $p$-value in \eqref{eqn:pValue} is approximately valid when $\TcE$ is large.

\begin{theorem}
\label{thm:AsymptoticValidity}
Assume that Assumptions~\ref{asp:FactorModel} --~\ref{asp:PerfectFit} hold. 
Assume there exists a constant $\kappa<\infty$, such that for $j=1, \ldots, J$, $t=1, \ldots, T$, $\epsilon_{jt}$ are continuously distributed with (a version of) the probability density function upper bounded by $\kappa$.
Then, under the null hypothesis \eqref{equation:null}, the $p$-values of equation \eqref{eqn:pValue} are approximately valid. In particular, there is an event $\mathcal C$, such that conditional on $\mathcal{C}$, for any $\alpha \in (0,1]$, we have
\begin{align*}
\alpha - 2 z_2 - \frac{1}{|\Pi|} \leq \Pr(\widehat{p} \leq \alpha) \leq \alpha + 2 z_2,
\end{align*}
and the event $\mathcal{C}$ happens with probability at least
\begin{align*}
\Pr(\mathcal C) \geq 1 - 2 J \exp\left( - \frac{z_1^2 \underline{\zeta}^2}{8 \widebar{\sigma}^2 \widebar{\lambda}^4 F^2} \TcE \right) - \frac{z_1}{z_2}\, 4 e \sqrt{2J (\min\{T-T_0, T_0-\TcE\})^3}\, \kappa,
\end{align*}
where $z_1, z_2$ are arbitrary positive constants.
\end{theorem}

A limitation of the result in Theorem~\ref{thm:AsymptoticValidity} is that there are values of the parameters of the data generating for which the result of the theorem provides a tight bound on test size only for large values of $\TcE$. 
We prove Theorem~\ref{thm:AsymptoticValidity} next.

\subsection{Definitions}

First, define $\minT = \min\{T-T_0, T_0-\TcE\}$.
Next, recall that
\begin{align*}
\widehat u_t = \sum_{j=1}^J w_j^* Y_{jt}-\sum_{j=1}^J v_j^* Y_{jt},
\end{align*}
for $t \in \mathcal{B} \cup \{T_0+1, \ldots, T\}$.
For $t \in\{T_0+1, \ldots, T\}$, $\widehat u_t$ are the post-intervention estimates of the treatment effects; and for $t \in \mathcal{B}$, $\widehat u_t$ are the placebo treatment effects estimated for the blank periods.

Let 
\begin{align}
u_t = \sum_{j=1}^J w^*_j \epsilon_{jt} - \sum_{j=1}^J v^*_j \epsilon_{jt} \label{eqn:defn:ut:blank}
\end{align}
for $t \in \mathcal{B}$, and 
\begin{align}
u_t = \sum_{j=1}^J w^*_j \xi_{jt} - \sum_{j=1}^J v^*_j \epsilon_{jt} \label{eqn:defn:ut:experimental}
\end{align}
for $t\in\{T_0+1, \ldots, T\}$. For each $\pi \in \Pi$, similar to our definition of $\widehat{\bm{e}}_{\pi}$, define the $(T-T_0)$-dimensional vector
\begin{align*}
\bm{e}_{\pi} = (u_{\pi(1)}, u_{\pi(2)}, ..., u_{\pi(T-T_0)}).
\end{align*}
In addition, let $\bm{e} = (u_1,\ldots, u_{T-T_0})= (\tau_{T_0+1}, \ldots, \tau_{T})$. 
It is useful to observe that, under the null hypothesis in \eqref{equation:null}, the random variables $u_t$ for $t\in\mathcal B\cup \{T_0+1, \ldots, T\}$ are independent and identically distributed.

Next, define the following two functions.
Let
\begin{align*}
\widehat{F}(x) = \frac{1}{|\Pi|}\sum_{\pi \in \Pi} \bI\left\{S(\widehat{\bm{e}}_{\pi}) < x\right\},
\end{align*}
and
\begin{align*}
\tilde{F}(x) = \frac{1}{|\Pi|}\sum_{\pi \in \Pi} \bI\left\{S(\bm{e}_{\pi}) < x\right\}.
\end{align*}

The proof of Theorem~\ref{thm:AsymptoticValidity} proceeds in four steps.
In step one, we define a high probability event, $\mathcal{C}_1$, such that  $u_t$ and $\widehat{u}_t$ are close to each other under $\mathcal{C}_1$.
In step two, we define a high probability event, $\mathcal{C}_2$, such that
many components of $\{S(\bm{e}_{\pi})\}_{\pi\in\Pi}$ are well-separated from $S(\bm{e})$
under $\mathcal{C}_2$.
In step three, we show that, conditional on $\mathcal{C}_1$ and $\mathcal{C}_2$, the ordering of $S(\bm{e}_{\pi})$ and $S(\bm{e})$ will be the same as the ordering of $S(\widehat{\bm{e}}_{\pi})$ and $S(\widehat{\bm{e}})$ for most $\pi\in\Pi$, which implies that $\widehat{F}(S(\widehat{\bm{e}}))$ and $\tilde{F}(S(\bm{e}))$ are also close to each other.
In step four, we conclude the proof by linking $\widehat{F}(S(\widehat{\bm{e}}))$ to the estimated $p$-value, and  $\tilde{F}(S(\bm{e}))$ to the nominal level $\alpha$.

\subsection{Lemmas for the Proof of Theorem \ref{thm:AsymptoticValidity}}

For each continuously distributed random variable $X$ with a density $f_X$, define $\Lambda_X$ to be the smallest upper bound on the probability density $f_X$.

\begin{lemma}[Corollary~2, \citet{bobkov2014bounds}]
\label{lem:BoundedSumDensity}
Let $X_1, X_2, \ldots, X_n$ be independent and continuously distributed random variables with densities $f_{X_1}, f_{X_2}, \ldots, f_{X_n}$.
For any $k \in \{1,2,\ldots,n\}$, let $\Lambda_{X_k}$ be the smallest upper bound on the probability density $f_{X_k}$.
For any $a_1, a_2, \ldots, a_n$, let $X = a_1 X_1 + a_2 X_2 + \ldots + a_n X_n$.
Suppose for any $k \in \{1,2,\ldots,n\}$, $\Lambda_{X_k} \leq \kappa$; and if $\sum_{k=1}^n a_k^2 = 1$,
\begin{align*}
\Lambda_{X} \leq \sqrt{e} \kappa.
\end{align*}
\end{lemma}

\begin{lemma}
\label{lem:BoundedDensity}
Let $X$ be a continuously distributed random variable with a density $f_X$.
Let $\Lambda_X$ be the smallest upper bound on the probability density $f_X$.
\begin{enumerate}
\item The random variable $|X|$ has a density $f_{|X|}$ bounded by $\Lambda_{|X|} \leq 2\Lambda_X$;
\item For any constant $a \ne 0$, the random variable $aX$ has a density $f_{aX}$ bounded by $\Lambda_{aX} \leq \Lambda_X/|a|$.
\end{enumerate}
\end{lemma}

\begin{proof}[Proof of Lemma~\ref{lem:BoundedDensity}.]
To prove 1, note that for any $v\geq 0$,
\begin{align*}
f_{|X|}(v) = f_X(v) + f_X(-v) \leq 2 \Lambda_X.
\end{align*}

To prove 2, note that for any $v\geq 0$,
\begin{align*}
f_{aX} (v)= \frac{1}{|a|} f_X(v/a) \leq \frac{1}{|a|} \Lambda_X.
\end{align*}
\Halmos\end{proof}

\begin{lemma}
\label{lem:BoundedDensity:ut}
Recall that $u_t$ is defined as \eqref{eqn:defn:ut:blank} and \eqref{eqn:defn:ut:experimental}, for the blank periods and the experimental periods, respectively. Under the null hypothesis \eqref{equation:null}, the probability density of $u_t$ can be bounded by
\begin{align*}
\Lambda_{u_t} \leq \frac{1}{2} \sqrt{e J} \kappa.
\end{align*}
\end{lemma}

\begin{proof}[Proof of Lemma~\ref{lem:BoundedDensity:ut}.]
This proof consists of two steps.
In Step 1, we prove a version of the lemma after conditioning on  $(\bm{w}^*, \bm{v}^*)$.
In Step 2, we apply the law of total probability to obtain a bound on the unconditional density of $u_t$.

\noindent Step 1. 
We condition on $(\bm{w}^*, \bm{v}^*)$ and write $u_t \vert (\bm{w}^*, \bm{v}^*)$ to indicate that we are conditional on $(\bm{w}^*, \bm{v}^*)$.

Fix any $t \in \mathcal{B}\cup\{T_0+1,\ldots,T\}$.
Using Lemma~\ref{lem:BoundedSumDensity} \citep[Corollary~2]{bobkov2014bounds}, let there be $J$ variables $\epsilon_{jt}$ for any $j \in \{1,2,\ldots,J\}$.
For any $j \in \{1,2,\ldots,J\}$, define $a_j = \frac{w_j^* - v_j^*}{\sqrt{\sum_{j=1}^J (w_j^* - v_j^*)^2}}$ such that $\sum_{j=1}^J a_j^2 = 1$.
Using $a_j$, we can write $u_t$ as $u_t = \sqrt{\sum_{j=1}^J (w_j^* - v_j^*)^2} \cdot \sum_{j=1}^J a_j \epsilon_{jt}$.
When $J$ is even,
\begin{align*}
\Lambda_{u_t\vert (\bm{w}^*, \bm{v}^*)} \leq & \frac{1}{\sqrt{\sum_{j=1}^J (w_j^* - v_j^*)^2}} \cdot \Lambda_{\sum_{j=1}^J a_j \epsilon_{jt}} \\
\leq & \frac{1}{\sqrt{\sum_{j=1}^J (w_j^* - v_j^*)^2}} \cdot \sqrt{e} \kappa \\
\leq & \frac{1}{\sqrt{\sum_{j=1}^J (\frac{2}{J})^2}} \cdot \sqrt{e} \kappa \\
= & \frac{\sqrt{J}}{2} \sqrt{e} \kappa,
\end{align*}
where the first inequality is due to Lemma~\ref{lem:BoundedDensity} Part~2; the second inequality is due to Lemma~\ref{lem:BoundedSumDensity}; the third inequality is due to convexity and Jensen's inequality, and the worst case is taken when $w^*_j = 2/J$ for one half of total units and $v^*_j = 2/J$ for the other half. 
When $J$ is odd,
\begin{align*}
\Lambda_{u_t\vert (\bm{w}^*, \bm{v}^*)} \leq & \frac{1}{\sqrt{\sum_{j=1}^J (w_j^* - v_j^*)^2}} \cdot \Lambda_{\sum_{j=1}^J a_j \epsilon_{jt}} \\
\leq & \frac{1}{\sqrt{\sum_{j=1}^J (w_j^* - v_j^*)^2}} \cdot \sqrt{e} \kappa \\
\leq & \frac{1}{\sqrt{\frac{J+1}{2} (\frac{2}{J+1})^2 + \frac{J-1}{2} (\frac{2}{J-1})^2}} \cdot \sqrt{e} \kappa \\
= & \sqrt{\frac{J^2-1}{J}} \cdot \frac{\sqrt{e} \kappa}{2}, \\
\leq & \frac{\sqrt{J}}{2} \sqrt{e} \kappa,
\end{align*}
where the first inequality is due to Lemma~\ref{lem:BoundedDensity} Part~2; the second inequality is due to Lemma~\ref{lem:BoundedSumDensity}; the third inequality is due to convexity and Jensen's inequality, and the worst case is taken when $w^*_j = 2/(J+1)$ for $(J+1)/2$ of total units and $v^*_j = 2/(J-1)$ for the other $(J-1)/2$ of total units. 

\noindent Step 2.
Using the law of total probability, we show 
\begin{align*}
f_{u_t}(u) = & \ \int_{(\bm{w}^*, \bm{v}^*)} f\big(u \vert (\bm{w}^*, \bm{v}^*)\big) \ \text{d} P(\bm{w}^*, \bm{v}^*) \\
\leq & \ \int_{(\bm{w}^*, \bm{v}^*)} \frac{\sqrt{J}}{2} \sqrt{e} \kappa \ \text{d} P(\bm{w}^*, \bm{v}^*) \\
= & \frac{\sqrt{J}}{2} \sqrt{e} \kappa,
\end{align*}
where we use $P(\bm{w}^*, \bm{v}^*)$ to stand for the joint distribution of $(\bm{w}^*, \bm{v}^*)$.
\Halmos\end{proof}

\subsection{Proof of Theorem~\ref{thm:AsymptoticValidity}}

\begin{proof}[Proof of Theorem~\ref{thm:AsymptoticValidity}.]
(Step one.)
Note that
\begin{multline*}
\sum_{j=1}^J w^*_j \bm{Y}_j^\sP - \sum_{j=1}^J v^*_j \bm{Y}_j^\sP = \bm{\theta}_\sP \left(\sum_{j=1}^J w^*_j \bm{Z}_j - \sum_{j=1}^J v^*_j \bm{Z}_j \right) \\
+ \bm{\lambda}_\sP \left(\sum_{j=1}^J w^*_j \bm{\mu}_j - \sum_{j=1}^J v^*_j \bm{\mu}_j \right) + \left(\sum_{j=1}^J w^*_j \bm{\epsilon}_j^\sP - \sum_{j=1}^J v^*_j \bm{\epsilon}_j^\sP\right).
\end{multline*}
Assumption \ref{asp:PerfectFit} implies
\begin{align*}
\sum_{j=1}^J w^*_j \bm{\mu}_j - \sum_{j=1}^J v^*_j \bm{\mu}_j 
= 
- (\bm{\lambda}_{\sP }' \bm{\lambda}_\sP)^{-1} \bm{\lambda}_{\sP }'
\left(\sum_{j=1}^J w^*_j \bm{\epsilon}_j^\sP - \sum_{j=1}^J v^*_j \bm{\epsilon}_j^\sP\right).
\end{align*}
Under the null hypothesis \eqref{equation:null}, it follows that 
\begin{align*}
\widehat{u}_t = & \sum_{j=1}^J w^*_j Y_{jt} - \sum_{j=1}^J v^*_j Y_{jt} \\
= & - \bm{\lambda}_t' (\bm{\lambda}_{\sP }' \bm{\lambda}_\sP)^{-1} \bm{\lambda}_{\sP }' \sum_{j=1}^J (w^*_j - v^*_j) \bm{\epsilon}_j^\sP + u_t, \end{align*}
for $t\in \mathcal B\cup \{T_0+1, \ldots, T\}$. 
We next define an event
\begin{align*}
\mathcal{C}_1 & = \left\{ \forall t\in \mathcal B\cup \{T_0+1, \ldots, T\}, \left| \bm{\lambda}_t' (\bm{\lambda}_{\sP }' \bm{\lambda}_\sP)^{-1} \bm{\lambda}_{\sP }' \sum_{j=1}^J (w^*_j - v^*_j) \bm{\epsilon}_j^\sP \right| \leq z_1 \right\}\\
& = \left\{ \max_{t\in \mathcal B\cup \{T_0+1, \ldots, T\}} \left| \bm{\lambda}_t' (\bm{\lambda}_{\sP }' \bm{\lambda}_\sP)^{-1} \bm{\lambda}_{\sP }' \sum_{j=1}^J (w^*_j - v^*_j) \bm{\epsilon}_j^\sP \right| \leq z_1 \right\}.
\end{align*}
Note that,
\begin{align*}
\max_{t\in \mathcal B\cup \{T_0+1, \ldots, T\}} \Big| \bm{\lambda}_t' &(\bm{\lambda}_{\sP }' \bm{\lambda}_\sP)^{-1} \bm{\lambda}_{\sP }' \sum_{j=1}^J (w^*_j - v^*_j) \bm{\epsilon}_j^\sP \Big|\\
& \leq \max_{t\in \mathcal B\cup \{T_0+1, \ldots, T\}} \sum_{j=1}^J |w^*_j - v^*_j| \sum_{s\in \mathcal E} |\bm{\lambda}_t' (\bm{\lambda}_{\sP }' \bm{\lambda}_\sP)^{-1} \bm{\lambda}_s||\epsilon_{js}|\\
& \leq \sum_{j=1}^J |w^*_j - v^*_j| \sum_{s\in \mathcal E} \frac{\widebar\lambda^2F}{\TcE\underline{\zeta}}|\epsilon_{js}|,
\end{align*}
where the second inequality is due to \eqref{eqn:lambda:CauchySchwarz}, and because $|w^*_j - v^*_j| \geq 0$ and $|\epsilon_{js}| \geq 0$.
Therefore,
\begin{align*}
\Pr(\mathcal{C}_1)&\geq 1-\Pr\Bigg(\sum_{j=1}^J \frac{|w^*_j - v^*_j|}{2} \sum_{s\in \mathcal E} \frac{\widebar\lambda^2F}{\TcE\underline{\zeta}}|\epsilon_{js}|> \frac{z_1}{2}\Bigg)\\
& \geq 1-\sum_{j=1}^J \Pr\Bigg(\sum_{s\in \mathcal E} \frac{\widebar\lambda^2F}{\TcE\underline{\zeta}}|\epsilon_{js}|> \frac{z_1}{2}\Bigg)\\
& \geq 1 - 2J \exp\left( - \frac{z_1^2 \underline{\zeta}^2}{8 \widebar{\sigma}^2 \widebar{\lambda}^4 F^2} \TcE \right),
\end{align*}
where the second inequality follows from union bound,
and the third inequality is the Chernoff bound for sub-Gaussian random variables.

(Step two.) Define $\tilde{z}_1 = 2 z_1 >0$, and $\minT = \min\{T-T_0, T_0-\TcE\}$.
For each $k \in\{0,1,2,...,\minT\}$, we define the following sets of permutations.
First, define $\Pi_0 = \{\pi_0\}$, where $\pi_0$ is defined as the set of post-intervention indices 
$\pi_0 = \{T_0+1,\ldots,T\}$.
Then, for any $k \in \{1,2,\ldots,\minT\}$, define
\begin{align*}
\Pi_k = \left\{ \pi \in \Pi \bigg| |\pi \setminus \pi_0| = k \right\}
\end{align*}
to be the set of $(T-T_0)$-combinations with exactly $k$ many indices from the blank periods.
Using the above definitions, we can decompose $\Pi$ into
\begin{align}
\label{eqn:Pi:decompose}
\Pi = \bigcup_{k=0}^{\minT} \Pi_k.
\end{align}

Then, for any $k \in \{1,2,...,\minT\}$ and $\pi \in \Pi_k$, we focus on the following indicator
\begin{align*}
\bI\bigg\{ \bigg| \sum_{t \in \pi \setminus \pi_0} |u_t| - \sum_{t \in \pi_0 \setminus \pi} |u_t| \bigg| \leq 2 k z_1 \bigg\}.
\end{align*}

The above indicator involves $2k$ instances of $|u_t|$'s. Intuitively, it is obtained by canceling out common terms in $S(\bm{e}_\pi)$ and $S(\bm{e})$.

Below we focus on the properties of the sum of such indicators.
First, focus on the probability density of $\Big| \sum_{t \in \pi \setminus \pi_0} |u_t| - \sum_{t \in \pi_0 \setminus \pi} |u_t| \Big|$.
We have 
\begin{align*}
\Lambda_{\left| \sum_{t \in \pi \setminus \pi_0} |u_t| - \sum_{t \in \pi_0 \setminus \pi} |u_t| \right|} & \leq 2 \Lambda_{\sum_{t \in \pi \setminus \pi_0} |u_t| - \sum_{t \in \pi_0 \setminus \pi} |u_t|} \\
& \leq 2 \sqrt{2k} \Lambda_{\sum_{t \in \pi \setminus \pi_0} \frac{1}{\sqrt{2k}}|u_t| - \sum_{t \in \pi_0 \setminus \pi} \frac{1}{\sqrt{2k}}|u_t|} \\
& \leq 2 \sqrt{2k} \sqrt{e} \Lambda_{|u_t|} \\
& \leq 2 \sqrt{2k} \sqrt{e} \sqrt{eJ} \kappa \\
& = 2 \sqrt{2Jk} e \kappa,
\end{align*}
where the first inequality is due to Lemma~\ref{lem:BoundedDensity}-1; the second inequality is due to Lemma~\ref{lem:BoundedDensity}-2; the third inequality is due to Lemma~\ref{lem:BoundedSumDensity}; the last inequality is due to Lemma~\ref{lem:BoundedDensity:ut} and~\ref{lem:BoundedDensity}-1.
We obtain
\begin{align*}
\Pr\bigg( \bigg| \sum_{t \in \pi \setminus \pi_0} |u_t| - \sum_{t \in \pi_0 \setminus \pi} |u_t| \bigg| \leq 2 k z_1  \bigg) \leq 4 e \sqrt{2Jk^3} z_1 \kappa.
\end{align*}

Next, due to Markov inequality, for any constant $z_2 > 0$, we have
\begin{align*}
\Pr\Bigg( \sum_{k=1}^{\minT}& \sum_{\pi \in \Pi_k} \bI\bigg\{ \bigg| \sum_{t \in \pi \setminus \pi_0} |u_t| - \sum_{t \in \pi_0 \setminus \pi} |u_t| \bigg| \leq 2 k z_1 \bigg\} \geq |\Pi| z_2 \Bigg) \\
& \leq  \frac{1}{|\Pi| z_2} \sum_{k=1}^{\minT} \sum_{\pi \in \Pi_k} \bE\Bigg[ \bI\bigg\{ \bigg| \sum_{t \in \pi \setminus \pi_0} |u_t| - \sum_{t \in \pi_0 \setminus \pi} |u_t| \bigg| \leq 2 k z_1 \bigg\} \Bigg] \\
& \leq \frac{\sum_{k=1}^{\minT} |\Pi_k| 4 e \sqrt{2Jk^3} z_1 \kappa}{|\Pi| z_2}.
\end{align*}

To conclude step two, define the event
\begin{align}
\mathcal{C}_2 = \Bigg\{ \sum_{k=1}^{\minT} \sum_{\pi \in \Pi_k} \bI\bigg\{ \bigg| \sum_{t \in \pi \setminus \pi_0} |u_t| - \sum_{t \in \pi_0 \setminus \pi} |u_t| \bigg| \leq 2 k z_1 \bigg\} < |\Pi| z_2 \Bigg\}.
\label{equation:C2}
\end{align}
The probability that event $\mathcal{C}_2$ happens is at least
\begin{align*}
\Pr(\mathcal{C}_2) \geq 1 - \frac{\sum_{k=1}^{\minT} |\Pi_k| \sqrt{k^3}}{|\Pi|} \frac{4 e \sqrt{2J} z_1 \kappa}{z_2}.
\end{align*}

(Step three.)
Conditional on event $\mathcal{C}_2$, fewer than $|\Pi| z_2$ of the absolute value terms in \eqref{equation:C2} are such that $\Big| \sum_{t \in \pi \setminus \pi_0} |u_t| - \sum_{t \in \pi_0 \setminus \pi} |u_t| \Big|\leq 2 k z_1$. For all the others, $\Big| \sum_{t \in \pi \setminus \pi_0} |u_t| - \sum_{t \in \pi_0 \setminus \pi} |u_t| \Big| > 2 k z_1$.

Conditional on event $\mathcal{C}_1$, we know that $|\widehat{u}_t - u_t| \leq z_1$ for any $t \in \mathcal{B}\cup\{T_0+1,\ldots,T\}$.
So we have that $\sum_{t \in \pi \setminus \pi_0} |u_t| - \sum_{t \in \pi_0 \setminus \pi} |u_t| > 2 k z_1$ implies
\begin{align*}
S(\widehat{\bm{e}}_\pi) - S(\widehat{\bm{e}}) & = \frac{1}{T-T_0} \sum_{t \in \pi} \left| \widehat{u}_t \right| - \frac{1}{T-T_0} \sum_{t \in \pi_0} \left| \widehat{u}_t \right| \\
& = \frac{1}{T-T_0} \left(\sum_{t \in \pi \setminus \pi_0} |\widehat{u}_t| - \sum_{t \in \pi_0 \setminus \pi} |\widehat{u}_t|\right) \\
& \geq \frac{1}{T-T_0} \left(\sum_{t \in \pi \setminus \pi_0} (|u_t| - z_1) - \sum_{t \in \pi_0 \setminus \pi} (|u_t| + z_1)\right) \\
& > \frac{1}{T-T_0} \left( 2k z_1 - 2k z_1\right) \\
& = 0,
\end{align*}
where the first equality is due to definition $S(\bm{e}_\pi) = \frac{1}{T-T_0} \sum_{t\in\pi} \left| u_t \right|$.
Similarly, $\mathcal{C}_1$ and
$\sum_{t \in \pi \setminus \pi_0} |u_t| - \sum_{t \in \pi_0 \setminus \pi} |u_t| < - 2 k z_1$ imply
\begin{align*}
S(\widehat{\bm{e}}_\pi) - S(\widehat{\bm{e}}) & = \frac{1}{T-T_0} \left(\sum_{t \in \pi \setminus \pi_0} |\widehat{u}_t| - \sum_{t \in \pi_0 \setminus \pi} |\widehat{u}_t|\right) \\
& \leq \frac{1}{T-T_0} \left(\sum_{t \in \pi \setminus \pi_0} (|u_t| + z_1) - \sum_{t \in \pi_0 \setminus \pi} (|u_t| - z_1)\right) \\
& < \frac{1}{T-T_0} \left( -2k z_1 + 2k z_1\right) \\
& = 0.
\end{align*}
Combining both cases, we know that conditional on $\mathcal{C}_1$ and when $\Big| \sum_{t \in \pi \setminus \pi_0} |u_t| - \sum_{t \in \pi_0 \setminus \pi} |u_t| \Big| > 2 k z_1$, the ordering of $S(\bm{e}_{\pi})$ and $S(\bm{e})$ is the same as the ordering of $S(\widehat{\bm{e}}_{\pi})$ and $S(\widehat{\bm{e}})$.
As a result, for those $\pi$ such that $\Big| \sum_{t \in \pi \setminus \pi_0} |u_t| - \sum_{t \in \pi_0 \setminus \pi} |u_t| \Big| > 2 k z_1$, we have $\bI\left\{ S(\widehat{\bm{e}}_\pi) \geq S(\widehat{\bm{e}}) \right\} = \bI\left\{ S(\bm{e}_\pi) \geq S(\bm{e}) \right\}$.
There are at most $|\Pi|z_2$ many $\pi$'s that contribute to the following summation,
\begin{align*}
\bigg| \sum_{k=1}^{\minT} \sum_{\pi \in \Pi_k} \bigg( \bI\left\{ S(\widehat{\bm{e}}_\pi) \geq S(\widehat{\bm{e}}) \right\} - \bI\left\{ S(\bm{e}_\pi) \geq S(\bm{e}) \right\} \bigg) \bigg| < |\Pi|z_2.
\end{align*}
Note that $S(\widehat{\bm{e}}_{\pi_0}) = S(\widehat{\bm{e}})$ and $S(\bm{e}_{\pi_0}) = S(\bm{e})$, so $\bI\left\{ S(\widehat{\bm{e}}_{\pi}) > S(\widehat{\bm{e}}) \right\} = \bI\left\{ S(\bm{e}_{\pi_0}) > S(\bm{e}) \right\}$ is always true.
Combining $\pi_0$ we have
\begin{align}
& \bigg| \sum_{\pi \in \Pi} \bigg( \bI\left\{ S(\widehat{\bm{e}}_\pi) \geq S(\widehat{\bm{e}}) \right\} - \bI\left\{ S(\bm{e}_\pi) \geq S(\bm{e}) \right\} \bigg) \bigg| \nonumber \\
= & \bigg| \sum_{k=0}^{\minT} \sum_{\pi \in \Pi_k} \bigg( \bI\left\{ S(\widehat{\bm{e}}_\pi) \geq S(\widehat{\bm{e}}) \right\} - \bI\left\{ S(\bm{e}_\pi) \geq S(\bm{e}) \right\} \bigg) \bigg| \nonumber \\
< & |\Pi|z_2. \label{eqn:ComparingTwoSimulatedCDFs}
\end{align}

We conclude step three using the following block of inequalities.
For any $\alpha \in (0,1]$,
\begin{align}
& \left| \Pr\left( 1 - \widehat{F}(S(\widehat{\bm{e}})) \leq \alpha \right) - \Pr\left( 1 - \tilde{F}(S(\bm{e})) \leq \alpha \right) \right| \nonumber \\
= & \left| \Pr\bigg( 1 - \frac{1}{|\Pi|} \sum_{\pi \in \Pi} \bI\{S(\widehat{\bm{e}}_\pi) < S(\widehat{\bm{e}})\} \leq \alpha \bigg) - \Pr\bigg( 1 - \frac{1}{|\Pi|} \sum_{\pi \in \Pi} \bI\{S(\bm{e}_\pi) < S(\bm{e})\} \leq \alpha \bigg) \right| \nonumber \\
= & \left| \Pr\bigg( \sum_{\pi \in \Pi} \bI\{S(\widehat{\bm{e}}_\pi) \geq S(\widehat{\bm{e}})\} \leq \alpha |\Pi| \bigg) - \Pr\bigg( \sum_{\pi \in \Pi} \bI\{S(\bm{e}_\pi) \geq S(\bm{e})\} \leq \alpha |\Pi| \bigg) \right| \nonumber \\
= & \Bigg| \bE\bigg[ \bI\bigg\{ \sum_{\pi \in \Pi} \bI\{S(\widehat{\bm{e}}_\pi) \geq S(\widehat{\bm{e}})\} \leq \alpha |\Pi| \bigg\}\bigg] - \bE\bigg[ \bI\bigg\{ \sum_{\pi \in \Pi} \bI\{S(\bm{e}_\pi) \geq S(\bm{e})\} \leq \alpha |\Pi| \bigg\}\bigg] \Bigg| \nonumber \\
\leq & \bE \left| \bI\bigg\{ \sum_{\pi \in \Pi} \bI\{S(\widehat{\bm{e}}_\pi) \geq S(\widehat{\bm{e}})\} \leq \alpha |\Pi| \bigg\} - \bI\bigg\{ \sum_{\pi \in \Pi} \bI\{S(\bm{e}_\pi) \geq S(\bm{e})\} \leq \alpha |\Pi| \bigg\} \right| \nonumber \\
\leq & \Pr\bigg( \bigg| \alpha |\Pi| - \sum_{\pi \in \Pi} \bI\{S(\bm{e}_\pi) \geq S(\bm{e})\} \bigg| \leq \bigg| \sum_{\pi \in \Pi} \bigg( \bI\left\{ S(\widehat{\bm{e}}_\pi) \geq S(\widehat{\bm{e}}) \right\} - \bI\left\{ S(\bm{e}_\pi) \geq S(\bm{e}) \right\} \bigg) \bigg| \bigg),\label{eqn:ComparingPValue}
\end{align}
where the second inequality is due to the following: $|\bI\{a \leq c\} - \bI\{b \leq c\}| \leq \bI\{|c - b| \leq |a-b|\}$.

Conditional on events $\mathcal{C}_1$ and $\mathcal{C}_2$, we obtain
\begin{align}
\bigg| \Pr\bigg( 1 - \widehat{F}(S(\widehat{\bm{e}})) &\leq \alpha \bigg) - \Pr\bigg( 1 - \tilde{F}(S(\bm{e})) \leq \alpha \bigg) \bigg| \nonumber \\
&\leq  \Pr\bigg( \bigg| \alpha |\Pi| - \sum_{\pi \in \Pi} \bI\{S(\bm{e}_\pi) \geq S(\bm{e})\} \bigg| < |\Pi|z_2 \bigg) \nonumber \\
&\leq  \frac{\ 2 |\Pi|z_2}{|\Pi|} \nonumber \\
& =  2 z_2, \label{eqn:ComparingPValue2}
\end{align}
where the last inequality is because $\sum_{\pi \in \Pi} \bI\{S(\bm{e}_\pi) \geq S(\bm{e})\}$ is a discrete uniform distribution over $\{1,2,...,|\Pi|\}$, and that there are at most $2 |\Pi|z_2$ many integers centered around $\alpha |\Pi|$.

(Step four.)
Note that, for any $\alpha \in (0,1]$,
\begin{align*}
\alpha - \frac{1}{|\Pi|} \leq \Pr\left( 1 - \tilde{F}(S(\bm{e})) \leq \alpha \right) \leq \alpha.
\end{align*}
So conditional on events $\mathcal{C}_1$ and $\mathcal{C}_2$, \eqref{eqn:ComparingPValue2} implies
\begin{align*}
\Pr\left( 1 - \widehat{F}(S(\widehat{\bm{e}})) \leq \alpha \right) \leq \Pr\left( 1 - \tilde{F}(S(\bm{e})) \leq \alpha \right) + 2 z_2 \leq \alpha + 2 z_2
\end{align*}
and
\begin{align*}
\Pr\left( 1 - \widehat{F}(S(\widehat{\bm{e}})) \leq \alpha \right) \geq \Pr\left( 1 - \tilde{F}(S(\bm{e})) \leq \alpha \right) - 2 z_2 \geq \alpha - 2 z_2 - \frac{1}{|\Pi|}.
\end{align*}
Combining both parts, conditional on $\mathcal{C} = \mathcal{C}_1 \cap \mathcal{C}_2$, we have
\begin{align*}
\alpha - 2 z_2 - \frac{1}{|\Pi|} \leq \Pr(\widehat{p} \leq \alpha) = \Pr\left( 1 - \widehat{F}(S(\widehat{\bm{e}})) \leq \alpha \right) \leq \alpha + 2 z_2,
\end{align*}
and $\mathcal{C}$ happens with probability at least
\begin{align}
\Pr(\mathcal{C}_1 \cap \mathcal{C}_2) 
&\geq  (1-\Pr(\mathcal{C}_1)) + (1-\Pr(\mathcal{C}_2)) - 1 \nonumber \\
& \geq  1 - 2 J \exp\left( - \frac{z_1^2 \underline{\zeta}^2}{8 \widebar{\sigma}^2 \widebar{\lambda}^4 F^2} \TcE \right) - \frac{\sum_{k=1}^{\minT} |\Pi_k| \sqrt{k^3}}{|\Pi|} \cdot \frac{z_1}{z_2} \cdot 4 e \sqrt{2J} \kappa, \label{eqn:thm:final}
\end{align} 
which finishes the proof.
\Halmos\end{proof}

\section{Estimating the Average Effect of Treatment on the Treated Units}
\label{sec:ATET:UnbiasedEstimator}

In Section~\ref{sec:formal}, we have shown formal results of the bias bounds in estimating the average treatment effect.
In this section, we present similar results for estimating the average effect of treatment on the treated units.
Similar to Assumption~\ref{asp:PerfectFit}, we begin with the assumption of perfect fit.

\begin{assumption}
\label{asp:ATET:PerfectFit}
With probability one, {\it (i)}
\begin{align*}
\sum_{j = 1}^J w^*_j \bm{Z}_{j} = \sum_{j = 1}^J v^*_j \bm{Z}_{j},
\end{align*}
and {\it (ii)}
\begin{align*}
\sum_{j = 1}^{J} w^*_j \bm{Y}^{\mathcal E}_j=\sum_{j = 1}^{J} v^*_j \bm{Y}^{\mathcal E}_j.
\end{align*}
\end{assumption}

In practice, Assumption~\ref{asp:PerfectFit} may only hold approximately.
The next assumption accommodates settings with imperfect fit. 

\begin{assumption}
\label{asp:ATET:ApproximateFit}
There exists a positive constant $d > 0$, such that with probability one, 
\begin{align}
\Big\| \sum_{j = 1}^J w^*_j \bm{Z}_{j} - \sum_{j = 1}^J v^*_j \bm{Z}_{j} \Big\|_2^2 \leq R d^2, \qquad \Big\|\sum_{j = 1}^{J} w^*_j \bm{Y}^{\mathcal E}_j - \sum_{j = 1}^{J} v^*_j \bm{Y}^{\mathcal E}_j \Big\|_2^2 \leq T_\mathcal{E} d^2. \label{eqn:ATET:ApproximateFit}
\end{align}
\end{assumption}

Using the above assumptions, we are able to provide the following bias bounds.

\begin{theorem}
\label{thm:ATET:UnbiasedEstimator}
If Assumptions~\ref{asp:FactorModel},~\ref{asp:ModelPrimitives}, and~\ref{asp:ATET:PerfectFit} hold, then for any $t \geq T_0+1$,
\begin{align*}
|\bE \left[ \widehat{\tau}^T_t - \tau^T_t \right]| \leq 
\frac{\widebar{\lambda}^2 F}{\underline{\zeta}} 2 \sqrt{2\log{(2J)}} \frac{\widebar{\sigma}}{\sqrt{T_\mathcal{E}}}. 
\end{align*}
If Assumptions~\ref{asp:FactorModel},~\ref{asp:ModelPrimitives}, and~\ref{asp:ATET:ApproximateFit} hold, then for any $t \geq T_0+1$,
\begin{align*}
|\bE \left[ \widehat\tau_t - \tau_t \right]| \leq 
\Big(\widebar{\theta}R + \frac{\widebar{\lambda}^2 F}{\underline{\zeta}} (1+\widebar{\theta} R)\Big) d + 
\frac{\widebar{\lambda}^2 F}{\underline{\zeta}} 2 \sqrt{2\log{(2J)}} \frac{\widebar{\sigma}}{\sqrt{T_\mathcal{E}}}.
\end{align*}
\end{theorem}

We provide the following result on inference.

\begin{theorem}
\label{thm:ATET:ExactPValue}
Suppose that Assumptions~\ref{asp:FactorModel} and \ref{asp:ATET:PerfectFit}{\it (i)} hold. 
Assume that $\{\bm{\lambda}_t\}_{t \in \mathcal{B} \cup \{T_0+1,...,T\}}$ is a sequence of exchangeable random variables independent of $\{\epsilon_{jt}\}_{t \in \mathcal{B} \cup \{T_0+1,...,T\}}$ and $\{\xi_{jt}\}_{t \in \{T_0+1,...,T\}}$.
Assume also that for each $j=1, \ldots, J$, $\{\epsilon_{jt}\}_{t \in \mathcal{B} \cup \{T_0+1,...,T\}}$ and $\{\xi_{jt}\}_{t \in \{T_0+1,...,T\}}$ are two sequences of exchangeable random variables, respectively.
Then under the null hypothesis \eqref{equation:null}, we have
\begin{align*}
\alpha - \frac{1}{|\Pi|} \leq \Pr(\widehat{p} \leq \alpha) \leq \alpha,
\end{align*}
for any $\alpha\in [0,1]$, where $\Pr(\widehat{p} \leq \alpha)$ is taken over the distributions of $\{\xi_{jt},\epsilon_{jt},\bm{\lambda}_t\}$.

If Assumption~\ref{asp:FactorModel} holds but \ref{asp:ATET:PerfectFit}{\it (i)} is violated, the same result holds if $\{({\bm \theta}_t, {\bm\lambda}_t)\}_{t \in \mathcal{B} \cup \{T_0+1,...,T\}}$ is a sequence of exchangeable random variables independent of $\{\epsilon_{jt}\}_{t \in \mathcal{B} \cup \{T_0+1,...,T\}}$ and $\{\xi_{jt}\}_{t \in \{T_0+1,...,T\}}$,
and if for each $j=1, \ldots, J$, $\{\epsilon_{jt}\}_{t \in \mathcal{B} \cup \{T_0+1,...,T\}}$ and $\{\xi_{jt}\}_{t \in \{T_0+1,...,T\}}$ are two sequences of exchangeable random variables, respectively.
Here, $\Pr(\widehat{p} \leq \alpha)$ is taken over the distributions of 
$\{\xi_{jt},\epsilon_{jt},\bm{\lambda}_t, {\bm\theta}_t\}$.
\end{theorem}

\subsection{Simulation Results for Average Treatment Effects on the Treated}
\label{sec:ATET1000Runs}

In this section, we estimate the average treatment effects on the treated units by conducting simulations following the simulation setup as in Section~\ref{sec:designs}.
Recall that we compare the performance of different synthetic control designs over 1000 simulations that independently generate the model primitives (i.e., the factor loadings, covariates, and error terms) of Assumption~\ref{asp:FactorModel}. The data generating process for each one of the 1000 simulations is the same as in Section~\ref{sec:simu:main}.
The five varieties of the synthetic control design are described in Section~\ref{sec:designs}.

We report the average treatment effects on the treated units in Table~\ref{tbl:MultiRuns:DifferentFormulations:ATET} as well as the average treatment effects on the treated units under the nonlinear model in Table~\ref{tbl:MultiRuns:Nonlinear:ATET}.

\subsubsection{Average Treatment Effects on the Treated}

The first five columns in Table~\ref{tbl:MultiRuns:DifferentFormulations:ATET} in the Online Appendix report averages of $\tau_t^T$, the average effect of treatment on the treated units. These quantities depend on the weights for the treated units, which are different across different formulations of the synthetic control design. 
The next five columns report averages of $\widehat{\tau}_t$. They are the same as in Table~\ref{tbl:MultiRuns:DifferentFormulations:ATE}, yet we use them as estimators for $\tau_t^T$ in Table~\ref{tbl:MultiRuns:DifferentFormulations:ATET}. 
The last two columns of Table~\ref{tbl:MultiRuns:DifferentFormulations:ATET} report averages across simulations of the mean absolute error and the root mean square error, defined as in \eqref{equation:MAERMSE} but with $\tau_t^T$ replacing $\tau_t$.  

The results in Table~\ref{tbl:MultiRuns:DifferentFormulations:ATET} are qualitatively similar to those for $\tau_t$ in Table~\ref{tbl:MultiRuns:DifferentFormulations:ATE}, with one notable exception. As expected, for intermediate and large values of $\beta$, the $\NewFormulation$ design outperforms the other designs when the goal is to estimate $\tau_t^T$. This is because in the $\NewFormulation$ design the synthetic control weights are targeted to $\tau_t^T$ (and more so as $\beta$ becomes large).

\subsubsection{Performance with Nonlinearities}

We now study the behavior of the estimators based on synthetic control designs under deviations from the linear model in \eqref{eqn:FactorModelN} and \eqref{eqn:FactorModelI}.  
We consider a nonlinear data generating process as defined in \eqref{eqn:NonlinearN} and \eqref{eqn:NonlinearI}.

Table~\ref{tbl:MultiRuns:Nonlinear:ATET} reports the results for $\tau_t^T$.
In comparison to the results in Tables~\ref{tbl:MultiRuns:DifferentFormulations:ATET}, we now see that the \textit{Unit-level} and \textit{Penalized} designs can easily match and in some cases improve the performance of the \textit{Unconstrained} design, especially for the estimation of $\tau_t^T$. 
By fitting each treated unit with a unit-specific synthetic control, the \textit{Unit-level} design can ameliorate interpolation biases induced by the aggregation of ${\bm X}_j$. 
Like in Table \ref{tbl:MultiRuns:DifferentFormulations:ATET}, the \textit{Weakly targeted} design easily outperforms the unconstrained estimator for large values of $\beta$.

\begin{sidewaystable}
\caption{Average Treatment Effects on the Treated (Averages over 1000 Simulations)}
\label{tbl:MultiRuns:DifferentFormulations:ATET}
\footnotesize
\begin{tabular}{llrrrrrcrrrrrccc}
\hline
                         &                  & \multicolumn{5}{c}{$\tau^T_t$}             &  & \multicolumn{5}{c}{$\widehat\tau_t$}       &  & $\mathit{MAE}^T$ & $\mathit{RMSE}^T$ \\ \cline{3-7} \cline{9-13}
                         &                  & $t=26$ & $t=27$ & $t=28$ & $t=29$ & $t=30$ &  & $t=26$ & $t=27$ & $t=28$ & $t=29$ & $t=30$ &  &                  &                   \\ [1ex]
$\textit{Unconstrained}$ &                  & -13.58 & -10.98 & -8.34  & -5.00  & -2.46  &  & -13.57 & -10.97 & -8.37  & -5.06  & -2.52  &  & 1.01             & 1.18              \\ [1ex]
$\textit{Constrained}$   & $\widebar{m}=1$  & -13.80 & -11.09 & -8.37  & -4.82  & -2.50  &  & -13.61 & -10.97 & -8.39  & -4.86  & -2.41  &  & 3.02             & 3.57              \\
                         & $\widebar{m}=2$  & -13.53 & -10.89 & -8.42  & -4.93  & -2.40  &  & -13.58 & -10.90 & -8.43  & -5.01  & -2.40  &  & 1.80             & 2.13              \\
                         & $\widebar{m}=3$  & -13.39 & -10.92 & -8.35  & -4.97  & -2.54  &  & -13.56 & -11.00 & -8.38  & -5.05  & -2.52  &  & 1.39             & 1.64              \\
                         & $\widebar{m}=4$  & -13.54 & -11.05 & -8.42  & -4.95  & -2.52  &  & -13.59 & -11.06 & -8.40  & -4.99  & -2.50  &  & 1.22             & 1.44              \\
                         & $\widebar{m}=5$  & -13.57 & -11.03 & -8.42  & -5.02  & -2.45  &  & -13.57 & -11.01 & -8.37  & -5.02  & -2.48  &  & 1.11             & 1.30              \\
                         & $\widebar{m}=6$  & -13.61 & -11.06 & -8.36  & -4.99  & -2.46  &  & -13.51 & -10.95 & -8.29  & -5.01  & -2.47  &  & 1.03             & 1.22              \\
                         & $\widebar{m}=7$  & -13.58 & -10.97 & -8.34  & -5.00  & -2.46  &  & -13.57 & -10.96 & -8.37  & -5.06  & -2.52  &  & 1.01             & 1.18              \\ [1ex]
$\NewFormulation$        & $\beta = 0.01$   & -13.59 & -10.99 & -8.31  & -5.00  & -2.51  &  & -13.58 & -10.95 & -8.38  & -4.99  & -2.53  &  & 1.31             & 1.55              \\
                         & $\beta = 0.1$    & -13.59 & -10.98 & -8.37  & -5.01  & -2.51  &  & -13.57 & -11.00 & -8.34  & -4.98  & -2.52  &  & 1.07             & 1.26              \\
                         & $\beta = 1$      & -13.55 & -10.99 & -8.35  & -5.01  & -2.48  &  & -13.56 & -10.98 & -8.32  & -4.93  & -2.44  &  & 0.99             & 1.16              \\
                         & $\beta = 10$     & -13.45 & -10.96 & -8.35  & -5.06  & -2.50  &  & -13.57 & -10.98 & -8.38  & -5.01  & -2.51  &  & 0.99             & 1.15              \\
                         & $\beta = 100$    & -13.49 & -10.90 & -8.33  & -5.01  & -2.48  &  & -13.60 & -10.98 & -8.39  & -5.07  & -2.52  &  & 1.00             & 1.16              \\ [1ex]
$\textit{Unit-level}$    & $\xi = 0.01$     & -13.60 & -10.98 & -8.35  & -4.98  & -2.51  &  & -13.60 & -10.95 & -8.39  & -5.04  & -2.53  &  & 1.09             & 1.29              \\
                         & $\xi = 0.1$      & -13.56 & -10.99 & -8.38  & -4.95  & -2.50  &  & -13.58 & -10.97 & -8.35  & -4.97  & -2.47  &  & 1.08             & 1.28              \\
                         & $\xi = 1$        & -13.57 & -11.02 & -8.31  & -4.91  & -2.48  &  & -13.57 & -10.99 & -8.39  & -4.99  & -2.49  &  & 1.40             & 1.66              \\
                         & $\xi = 10$       & -13.78 & -10.98 & -8.37  & -4.87  & -2.52  &  & -13.60 & -10.93 & -8.45  & -5.05  & -2.52  &  & 1.96             & 2.33              \\
                         & $\xi = 100$      & -13.81 & -10.90 & -8.39  & -4.80  & -2.55  &  & -13.61 & -10.86 & -8.48  & -5.02  & -2.54  &  & 2.35             & 2.78              \\ [1ex]
$\textit{Penalized}$     & $\lambda = 0.01$ & -13.59 & -10.99 & -8.35  & -5.00  & -2.47  &  & -13.59 & -10.98 & -8.35  & -5.05  & -2.48  &  & 1.05             & 1.23              \\
                         & $\lambda = 0.1$  & -13.57 & -10.98 & -8.36  & -4.91  & -2.42  &  & -13.64 & -11.03 & -8.43  & -5.03  & -2.50  &  & 1.36             & 1.61              \\
                         & $\lambda = 1$    & -13.69 & -11.01 & -8.33  & -4.80  & -2.48  &  & -13.67 & -10.96 & -8.41  & -4.87  & -2.45  &  & 2.19             & 2.59              \\
                         & $\lambda = 10$   & -13.88 & -11.06 & -8.23  & -4.73  & -2.40  &  & -13.68 & -11.04 & -8.37  & -4.79  & -2.45  &  & 3.81             & 4.50              \\
                         & $\lambda = 100$  & -13.90 & -11.10 & -8.27  & -4.75  & -2.46  &  & -13.64 & -10.94 & -8.42  & -4.86  & -2.50  &  & 4.24             & 5.00              \\ [1ex] \hline
\end{tabular}
\end{sidewaystable}

\begin{sidewaystable}
\caption{Average Treatment Effects on the Treated, Nonlinear Model (Averages over 1000 Simulations)}
\label{tbl:MultiRuns:Nonlinear:ATET}
\footnotesize
\begin{tabular}{llrrrrrcrrrrrccc}
\hline
                         &                  & \multicolumn{5}{c}{$\tau^T_t$}             &  & \multicolumn{5}{c}{$\widehat\tau_t$}       &  & $\mathit{MAE}^T$ & $\mathit{RMSE}^T$ \\ \cline{3-7} \cline{9-13}
                         &                  & $t=26$ & $t=27$ & $t=28$ & $t=29$ & $t=30$ &  & $t=26$ & $t=27$ & $t=28$ & $t=29$ & $t=30$ &  &                  &                   \\ [1ex]
$\textit{Unconstrained}$ &                  & -13.30 & -10.65 & -7.98  & -5.47  & -2.41  &  & -13.44 & -10.92 & -8.18  & -5.85  & -2.78  &  & 2.31             & 3.01              \\ [1ex]
$\textit{Constrained}$   & $\widebar{m}=1$  & -13.41 & -10.81 & -8.01  & -5.25  & -2.61  &  & -15.70 & -13.18 & -10.50 & -7.76  & -4.78  &  & 3.53             & 4.28              \\
                         & $\widebar{m}=2$  & -13.19 & -10.88 & -7.93  & -5.42  & -2.47  &  & -14.27 & -11.86 & -8.90  & -6.44  & -3.34  &  & 2.86             & 3.61              \\
                         & $\widebar{m}=3$  & -13.17 & -10.82 & -8.07  & -5.54  & -2.44  &  & -13.69 & -11.38 & -8.38  & -5.95  & -2.97  &  & 2.52             & 3.25              \\
                         & $\widebar{m}=4$  & -13.29 & -10.82 & -8.03  & -5.59  & -2.31  &  & -13.58 & -11.09 & -8.23  & -5.89  & -2.75  &  & 2.44             & 3.16              \\
                         & $\widebar{m}=5$  & -13.23 & -10.77 & -7.88  & -5.58  & -2.30  &  & -13.37 & -10.97 & -8.14  & -5.79  & -2.88  &  & 2.39             & 3.10              \\
                         & $\widebar{m}=6$  & -13.36 & -10.71 & -7.99  & -5.50  & -2.36  &  & -13.54 & -11.03 & -8.31  & -5.86  & -2.86  &  & 2.32             & 3.02              \\
                         & $\widebar{m}=7$  & -13.31 & -10.68 & -7.97  & -5.46  & -2.39  &  & -13.49 & -10.94 & -8.17  & -5.86  & -2.78  &  & 2.30             & 3.01              \\ [1ex]
$\NewFormulation$        & $\beta = 0.01$   & -13.14 & -10.67 & -7.97  & -5.47  & -2.45  &  & -11.66 & -9.02  & -6.37  & -3.87  & -1.00  &  & 2.68             & 3.36              \\
                         & $\beta = 0.1$    & -13.20 & -10.66 & -7.92  & -5.47  & -2.44  &  & -12.08 & -9.60  & -6.87  & -4.31  & -1.47  &  & 2.31             & 2.94              \\
                         & $\beta = 1$      & -13.20 & -10.71 & -7.91  & -5.47  & -2.46  &  & -12.51 & -10.13 & -7.35  & -4.81  & -1.91  &  & 1.91             & 2.43              \\
                         & $\beta = 10$     & -13.27 & -10.67 & -7.96  & -5.45  & -2.49  &  & -13.03 & -10.51 & -7.81  & -5.25  & -2.32  &  & 1.33             & 1.66              \\
                         & $\beta = 100$    & -13.29 & -10.77 & -8.04  & -5.47  & -2.53  &  & -13.28 & -10.72 & -8.00  & -5.43  & -2.59  &  & 1.14             & 1.40              \\ [1ex]
$\textit{Unit-level}$    & $\xi = 0.01$     & -13.20 & -10.76 & -7.98  & -5.44  & -2.48  &  & -11.76 & -9.15  & -6.51  & -3.91  & -1.15  &  & 2.68             & 3.35              \\
                         & $\xi = 0.1$      & -13.35 & -10.81 & -8.07  & -5.53  & -2.55  &  & -13.11 & -10.59 & -7.82  & -5.15  & -2.29  &  & 1.98             & 2.51              \\
                         & $\xi = 1$        & -13.40 & -10.77 & -8.05  & -5.47  & -2.56  &  & -13.74 & -11.12 & -8.42  & -5.75  & -2.84  &  & 1.47             & 1.81              \\
                         & $\xi = 10$       & -13.35 & -10.66 & -7.99  & -5.38  & -2.71  &  & -13.74 & -11.20 & -8.55  & -5.89  & -3.09  &  & 1.49             & 1.80              \\
                         & $\xi = 100$      & -13.39 & -10.63 & -7.95  & -5.37  & -2.69  &  & -13.79 & -11.16 & -8.54  & -5.90  & -3.08  &  & 1.56             & 1.89              \\ [1ex]
$\textit{Penalized}$     & $\lambda = 0.01$ & -13.32 & -10.69 & -8.07  & -5.50  & -2.41  &  & -13.40 & -10.93 & -8.32  & -5.82  & -2.82  &  & 2.25             & 2.93              \\
                         & $\lambda = 0.1$  & -13.17 & -10.72 & -8.09  & -5.59  & -2.55  &  & -13.33 & -10.79 & -8.13  & -5.56  & -2.65  &  & 1.87             & 2.39              \\
                         & $\lambda = 1$    & -13.40 & -10.80 & -7.98  & -5.35  & -2.66  &  & -13.32 & -10.84 & -8.15  & -5.39  & -2.60  &  & 1.97             & 2.44              \\
                         & $\lambda = 10$   & -13.41 & -10.89 & -8.00  & -5.32  & -2.64  &  & -13.39 & -10.82 & -7.95  & -5.34  & -2.58  &  & 2.79             & 3.46              \\
                         & $\lambda = 100$  & -13.42 & -10.87 & -8.02  & -5.26  & -2.63  &  & -13.35 & -10.82 & -8.00  & -5.29  & -2.57  &  & 3.06             & 3.80              \\ [1ex] \hline
\end{tabular}
\end{sidewaystable}

\clearpage

\subsection{Proofs of Theorem~\ref{thm:ATET:UnbiasedEstimator} and Theorem~\ref{thm:ATET:ExactPValue}}

\begin{proof}[Proof of Theorem~\ref{thm:ATET:UnbiasedEstimator}.]
For any period $t = T_0+1, \ldots, T$ we decompose $(\widehat{\tau}^T_t - \tau^T_t)$ as follows,
\begin{align}
\widehat{\tau}^T_t - \tau^T_t = \sum_{j=1}^J w^*_j Y^N_{jt} - \sum_{j=1}^J v^*_j Y^N_{jt}. \label{eqn:ATET_final}
\end{align}
From \eqref{eqn:FactorModelN}, we obtain
\begin{multline}
\sum_{j = 1}^J w^*_j Y^N_{jt} - \sum_{j = 1}^J v^*_j Y^N_{jt} = \bm{\theta}_t' \Big(\sum_{j = 1}^J w^*_j\bm{Z}_j - \sum_{j = 1}^J v^*_j\bm{Z}_j\Big) \\
+ \bm{\lambda}_t' \Big(\sum_{j = 1}^J w^*_j \bm{\mu}_j - \sum_{j = 1}^J v^*_j \bm{\mu}_j\Big) + \Big(\sum_{j = 1}^J w^*_j \epsilon_{jt} - \sum_{j = 1}^J v^*_j \epsilon_{jt}\Big). \label{eqn:ATET:interStep1}
\end{multline}
Similarly, using expression  \eqref{eqn:FactorModelN}, we obtain
\begin{multline*}
\sum_{j = 1}^J w^*_j \bm{Y}_{j}^\sP - \sum_{j = 1}^J v^*_j \bm{Y}_{j}^\sP = \bm{\theta}_\sP  \Big(\sum_{j = 1}^J w^*_j \bm{Z}_j - \sum_{j = 1}^J v^*_j \bm{Z}_j\Big) \\
+ \bm{\lambda}_\sP \Big(\sum_{j = 1}^J w^*_j \bm{\mu}_j - \sum_{j = 1}^J v^*_j \bm{\mu}_j\Big) + \Big(\sum_{j = 1}^J w^*_j \bm{\epsilon}_{j}^\sP - \sum_{j = 1}^J v^*_j \bm{\epsilon}_j^\sP\Big),
\end{multline*}
where $\bm{\theta}_\sP$ is the $(\TcE\times R)$ matrix with rows equal to the $\bm\theta_t$'s indexed by $\mathcal E$, and $\bm{\epsilon}_{j}^\sP$ is defined analogously. 
Pre-multiplying by $\bm{\lambda}_t' (\bm{\lambda}_{\sP}' \bm{\lambda}_\sP)^{-1} \bm{\lambda}_{\sP}'$ yields
\begin{align}
\bm{\lambda}_t' (\bm{\lambda}_{\sP }' \bm{\lambda}_\sP)^{-1} \bm{\lambda}_{\sP }' & \Bigg(\sum_{j = 1}^J w^*_j \bm{Y}_{j}^\sP - \sum_{j = 1}^J v^*_j \bm{Y}_{j}^\sP\Bigg) \nonumber \\
= & \bm{\lambda}_t' (\bm{\lambda}_{\sP }' \bm{\lambda}_\sP)^{-1} \bm{\lambda}_{\sP }' \bm{\theta}_\sP \Bigg(\sum_{j = 1}^J w^*_j \bm{Z}_j - \sum_{j = 1}^J v^*_j \bm{Z}_j\Bigg) \nonumber \\
+ & \bm{\lambda}_t' \Bigg(\sum_{j = 1}^J w^*_j \bm{\mu}_j - \sum_{j = 1}^J v^*_j \bm{\mu}_j\Bigg) \nonumber \\
+ & \bm{\lambda}_t' (\bm{\lambda}_{\sP }' \bm{\lambda}_\sP)^{-1} \bm{\lambda}_{\sP }' \Bigg(\sum_{j = 1}^J w^*_j \bm{\epsilon}_{j}^\sP - \sum_{j = 1}^J v^*_j \bm{\epsilon}_j^\sP\Bigg). \label{eqn:ATET:interStep2}
\end{align}
Equations \eqref{eqn:ATET:interStep1} and \eqref{eqn:ATET:interStep2} imply
\begin{align}
\sum_{j = 1}^J w^*_j Y^N_{jt} - \sum_{j = 1}^J v^*_j Y^N_{jt} = \ & (\bm{\theta}'_t - \bm{\lambda}_t' (\bm{\lambda}_{\sP }' \bm{\lambda}_\sP)^{-1} \bm{\lambda}_{\sP }' \bm{\theta}_\sP) \Big(\sum_{j = 1}^J w^*_j \bm{Z}_j - \sum_{j = 1}^J v^*_j \bm{Z}_j\Big) \nonumber \\
& + \bm{\lambda}_t' (\bm{\lambda}_{\sP }' \bm{\lambda}_\sP)^{-1} \bm{\lambda}_{\sP }' \Big(\sum_{j = 1}^J w^*_j \bm{Y}^\sP_j - \sum_{j = 1}^J v^*_j \bm{Y}^\sP_j\Big) \nonumber \\
& - \bm{\lambda}_t' (\bm{\lambda}_{\sP }' \bm{\lambda}_\sP)^{-1} \bm{\lambda}_{\sP }' \Big(\sum_{j = 1}^J w^*_j \bm{\epsilon}_{j}^\sP - \sum_{j = 1}^J v^*_j \bm{\epsilon}_{j}^\sP\Big)\nonumber\\ 
& + \Big(\sum_{j = 1}^J w^*_j \epsilon_{j t} - \sum_{j = 1}^J v^*_j \epsilon_{j t}\Big). \label{equation:ATETbias4terms}
\end{align}

If Assumption~\ref{asp:PerfectFit} holds, \eqref{equation:ATETbias4terms} becomes
\begin{align}
\sum_{j = 1}^J w^*_j Y^N_{jt} - \sum_{j = 1}^J v^*_j Y^N_{jt} = \ & - \bm{\lambda}_t' (\bm{\lambda}_{\sP }' \bm{\lambda}_\sP)^{-1} \bm{\lambda}_{\sP }' \Big(\sum_{j = 1}^J w^*_j \bm{\epsilon}_{j}^\sP - \sum_{j = 1}^J v^*_j \bm{\epsilon}_{j}^\sP\Big)\nonumber\\ 
& + \Big(\sum_{j = 1}^J w^*_j \epsilon_{j t} - \sum_{j = 1}^J v^*_j \epsilon_{j t}\Big). \label{equation:ATETbias2terms}
\end{align}
Only the first term on the right-hand side of \eqref{equation:ATETbias2terms} has a non-zero mean (because the weights $w^*_j$ and $v^*_j$, depend on the error terms $\bm{\epsilon}_{j}^\sP$). Therefore,
\begin{multline}
\left| E\Bigg[\sum_{j = 1}^J w^*_j Y^N_{jt} - \sum_{j = 1}^J v^*_j Y^N_{jt} \Bigg]\right| = \left| \bE\left[\bm{\lambda}_t' (\bm{\lambda}_{\sP }' \bm{\lambda}_\sP)^{-1} \bm{\lambda}_{\sP }' \Big(\sum_{j = 1}^J w^*_j \bm{\epsilon}_{j}^\sP - \sum_{j = 1}^J v^*_j \bm{\epsilon}_{j}^\sP\Big) \right] \right| \\
\leq \left| \bE\left[\bm{\lambda}_t' (\bm{\lambda}_{\sP }' \bm{\lambda}_\sP)^{-1} \bm{\lambda}_{\sP }' \sum_{j = 1}^J w^*_j \bm{\epsilon}_{j}^\sP \right] \right| + \left| \bE\left[\bm{\lambda}_t' (\bm{\lambda}_{\sP }' \bm{\lambda}_\sP)^{-1} \bm{\lambda}_{\sP }' \sum_{j = 1}^J v^*_j \bm{\epsilon}_{j}^\sP \right] \right|.\label{eqn:ATET:intermediate}
\end{multline}

For any $t\geq T_0+1$ and $s \in \mathcal{E}$, under Assumption \ref{asp:ModelPrimitives} (i), we apply Cauchy-Schwarz inequality and the eigenvalue bound on the Rayleigh quotient to obtain
\begin{align*}
\left( \bm{\lambda}_t' (\bm{\lambda}_{\sP }' \bm{\lambda}_\sP)^{-1} \bm{\lambda}_s \right)^2 
&\leq \left( \frac{\widebar{\lambda}^2 F}{\TcE \underline{\zeta}} \right)^2.
\end{align*}

Let
\begin{align*}
\widebar\epsilon_{jt}^{\mathcal E} = \bm{\lambda}_t' (\bm{\lambda}_{\sP }' \bm{\lambda}_\sP)^{-1} \bm{\lambda}_{\mathcal E}' \bm{\epsilon}_{j}^\sP
=\sum_{s\in \mathcal E} \bm{\lambda}_t' (\bm{\lambda}_{\sP }' \bm{\lambda}_\sP)^{-1}
\bm{\lambda}_s \epsilon_{js}.
\end{align*}
Because $\widebar\epsilon_{jt}^{\mathcal E}$ is a linear combination of independent sub-Gaussians with variance
proxy $\widebar\sigma^2$, we know $\widebar\epsilon_{jt}^{\mathcal E}$ is sub-Gaussian with variance proxy
$(\widebar{\lambda}^2 F/\underline\zeta)^2\widebar\sigma^2/\TcE$. 
Let $\mathcal{S} = \{\bm{w} \in \bR^J : \sum_{j=1}^J w_j = 1\}$ be the unit simplex. 
Theorem~1.16 from \citet{rigollet2019high} implies
\begin{align*}
\left| \bE\Big[\sum_{j = 1}^J w^*_j Y^N_{jt} - \sum_{j = 1}^J v^*_j Y^N_{jt} \Big]\right| & \leq \left| \bE\Big[ \sum_{j=1}^J w^*_j \widebar\epsilon_{jt}^\mathcal{E} \Big]\right| + \left| \bE\Big[ \sum_{j=1}^J v^*_j \widebar\epsilon_{jt}^\mathcal{E} \Big]\right| \\
& \leq \bE\Bigg[ \max_{\bm{w} \in \mathcal{S}} \Big|\sum_{j=1}^J w_j \widebar\epsilon_{jt}^\mathcal{E} \Big| \Bigg] + \bE\Bigg[ \max_{\bm{v} \in \mathcal{S}} \Big|\sum_{j=1}^J v_j \widebar\epsilon_{jt}^\mathcal{E} \Big| \Bigg] \\
& \leq \frac{\widebar{\lambda}^2 F}{\underline{\zeta}} 2\sqrt{2\log{(2J)}}\frac{\widebar{\sigma}}{\sqrt{\TcE}},
\end{align*}
which finishes the proof of the theorem. 

Suppose now Assumption~\ref{asp:ATET:ApproximateFit} holds (but Assumption~\ref{asp:ATET:PerfectFit} does not). To obtain a bound on the bias we need to bound the first two terms in \eqref{equation:ATETbias4terms}.
Recall that 
\begin{align*}
\bm{\lambda}_t' (\bm{\lambda}_{\sP }' \bm{\lambda}_\sP)^{-1} \bm{\lambda}_s \leq \frac{\widebar{\lambda}^2 F}{\TcE \underline{\zeta}}.
\end{align*}
Therefore, the absolute value of each element in vector $(\bm{\theta}'_t - \bm{\lambda}_t' (\bm{\lambda}_{\sP }' \bm{\lambda}_\sP)^{-1} \bm{\lambda}_{\sP }' \bm{\theta}_\sP)$ is bounded by $\widebar{\theta} \Big(1 + \dfrac{\widebar{\lambda}^2 F}{\underline{\zeta}} \Big)$.
Cauchy–Schwarz inequality and Assumption~\ref{asp:ATET:ApproximateFit} imply
\begin{align*}
\Bigg|(\bm{\theta}'_t - \bm{\lambda}_t' (\bm{\lambda}_{\sP }' \bm{\lambda}_\sP)^{-1} \bm{\lambda}_{\sP }' \bm{\theta}_\sP) & \Bigg(\sum_{j = 1}^J w^*_j \bm{Z}_j - \sum_{j = 1}^J v^*_j \bm{Z}_j\Bigg)\Bigg| \\
\leq & \ \widebar{\theta} \Big(1 + \dfrac{\widebar{\lambda}^2 F}{\underline{\zeta}} \Big) \sqrt{R} \Bigg\|\sum_{j = 1}^J w^*_j \bm{Z}_j - \sum_{j = 1}^J v^*_j \bm{Z}_j\Bigg\|_2 \\
\leq & \ \widebar{\theta} \Big(1 + \dfrac{\widebar{\lambda}^2 F}{\underline{\zeta}} \Big) R d,
\end{align*}
and
\begin{align*}
\left|\bm{\lambda}_t' (\bm{\lambda}_{\sP }' \bm{\lambda}_\sP)^{-1} \bm{\lambda}_{\sP }' \Bigg(\sum_{j = 1}^J w^*_j \bm{Y}^\sP_j - \sum_{j = 1}^J v^*_j \bm{Y}^\sP_j\Bigg)\right| \leq \frac{\widebar{\lambda}^2 F}{\underline{\zeta}} d.
\end{align*}

Combining the last two displayed equations with \eqref{equation:ATETbias4terms}, we have
\begin{align*}
\left| \bE\Bigg[\sum_{j = 1}^J w^*_j Y^I_{jt} - \sum_{j = 1}^J f_j Y^I_{jt} \Bigg]\right| 
\leq 
\Big(\widebar{\theta} R + \frac{\widebar{\lambda}^2 F}{\underline{\zeta}}  (1+\widebar{\theta} R) \Big) d + \frac{\widebar{\lambda}^2 F}{\underline{\zeta}} 2\sqrt{2\log{(2J)}}\frac{\widebar{\sigma}}{\sqrt{\TcE}},
\end{align*}
which finishes the proof of the theorem. 
\Halmos\end{proof}

\begin{proof}[Proof of Theorem~\ref{thm:ATET:ExactPValue}.]
Recall that
\begin{align*}
\widehat u_t = \sum_{j=1}^J w_j^* Y_{jt}-\sum_{j=1}^J v_j^* Y_{jt},
\end{align*}
for $t \in \mathcal{B} \cup \{T_0+1, \ldots, T\}$.
For $t \in\{T_0+1, \ldots, T\}$, $\widehat u_t$ are the post-intervention estimates of the treatment effects; and for $t \in \mathcal{B}$, $\widehat u_t$ are the placebo treatment effects estimated for the blank periods.
Let 
\begin{align*}
u_t = \sum_{j=1}^J w^*_j \epsilon_{jt} - \sum_{j=1}^J v^*_j \epsilon_{jt}
\end{align*}
for $t \in \mathcal{B}$, and 
\begin{align*}
u_t = \sum_{j=1}^J w^*_j \xi_{jt} - \sum_{j=1}^J v^*_j \epsilon_{jt}
\end{align*}
for $t\in\{T_0+1, \ldots, T\}$. 
The null hypothesis \eqref{equation:null} and the assumptions of Theorem~\ref{thm:ATET:ExactPValue} imply that $\{u_t\}_{t\in \mathcal B\cup \{T_0+1, \ldots, T\}}$ is a sequence of exchangeable random variables. 
Additionally, Assumption~\ref{asp:FactorModel} and the null hypothesis \eqref{equation:null} imply
\begin{align*}
\widehat u_t = \sum_{j=1}^J w^*_j Y_{jt} - \sum_{j=1}^J v^*_j Y_{jt} 
= \bm{\theta}_t' \sum_{j=1}^J (w^*_j - v^*_j) \bm{Z}_j + \bm{\lambda}_t' \sum_{j=1}^J (w^*_j - v^*_j) \bm{\mu}_j + u_t,
\end{align*}
for $t\in \mathcal B\cup \{T_0+1, \ldots, T\}$.
The result of the theorem follows now from Theorem~D.1 in \citet{chernozhukov2021exact}.
\Halmos\end{proof}

\section{Swapping Treated and Control Weights}
\label{sec:SwappingRule}

Recall that when it is possible to swap synthetic treated and synthetic control weights, we choose the treated units so that the number of units with positive weights in $\bm w^*$ is smaller than the number of units with positive weights in $\bm v^*$.
When $\|\bm w^*\|_0=\|\bm v^*\|_0$, we determine whether or not to swap using the following rule.
For the \textit{Unconstrained} design, we choose the treated group to be the one with the smallest index among the units with positive weights. 
We use the same procedure based on the lowest index for \textit{Constrained} with $\widebar m=7$ (highest value) and \textit{Penalized} with $\lambda=0.01$ (lowest value). 
Then, starting from $\widebar m=7$ and for smaller values of $\widebar m$, we assign to the treated group the set of weights that is most similar to the weights obtained for $\|\bm w^*\|_0\leq \widebar m+1$ (in terms of what units obtain positive weights). In those cases where the two sets of swappable weights for $\|\bm w^*\|_0\leq \widebar m$ are equally similar to the synthetic treated weights for $\|\bm w^*\|_0\leq \widebar m+1$, we select the set of weights with the smallest index. We follow the analogous procedure for $\lambda > 0.01$, starting from smaller values of $\lambda$. 

\section{Implementations of the Optimization Formulations}
\label{sec:ComputationalImplementation}

To computationally solve \eqref{eqn:OptOriginal}, i.e., the \textit{Unconstrained} design, we propose two methods. 
The first method is by enumeration, which takes advantage of the objective function of \eqref{eqn:OptOriginal} being separated between $\bm{w}$ and $\bm{v}$.
If we knew which units were to receive treatment and which units were to receive control, then we could decompose \eqref{eqn:OptOriginal} into two classical synthetic control problems and solve both of them efficiently.
We brute-force enumerate all the possible combinations of the treatment units and control units.
Because the two groups of treated and control units can be swapped, 
we only enumerate combinations such that the cardinality of the treated group is smaller than or equal to the cardinality of the control group.
When the cardinality of the treated group is equal to the cardinality of the control group, we prioritize the treated group to be the one with the smallest index among the units with positive weights.

The second method solves a constrained optimization problem, by converting it into the canonical form of a Quadratic Constraint Quadratic Program (QCQP), which we detail below.
The decision variables are $w_j$ and $v_j, \forall \ j = 1, \ldots, J$.
For simplicity, we write it in a vector form $\tilde{\bm{W}} = (w_1, w_2, ..., w_J, v_1, v_2, ..., v_J)$.

Let $M$ be the dimension of the predictors $\bm{X}_j$.
Let $X$ be an $M \times J$ matrix, each column of which is $\bm{X}_j$, which stands for the predictors of unit $j$.

Define $P^0 = \{ P^0_{k,l} \}_{k, l = 1, \ldots, 2J} \in \bR^{2J \times 2J}$, such that $P^0$ has only two diagonal blocks, while the two off-diagonal blocks are zero. Define for any $k, l = 1, \ldots, 2J,$
\begin{align*}
P^0_{k,l} = 
\left\{
\begin{aligned}
& \sum_{i=1}^M X_{i,k} X_{i,l}, && k, l = 1, \ldots, J; \\
& \sum_{i=1}^M X_{i,(k-J)} X_{i,(l-J)}, && k, l = J+1,\ldots,2J; \\
& 0, && \text{otherwise}.
\end{aligned}
\right.
\end{align*}
Define $\bm{q}^0 \in \bR^{2J}$, such that for any $k = 1, \ldots, 2J$
\begin{align*}
q^0_{k} = 
\left\{
\begin{aligned}
& - 2 \sum_{i=1}^M X_{i,k} \cdot (\sum_{j=1}^J f_j X_{i,j}), && k = 1, \ldots, J; \\
& - 2 \sum_{i=1}^M X_{i,k-J} \cdot (\sum_{j=1}^J f_j X_{i,j}), && k =J+1, \ldots, 2J.
\end{aligned}
\right.
\end{align*}

Further define $\bm{e}_1 = (1,1,...,1,0,0,...,0)'$ whose first $J$ elements are $1$ and last $J$ elements $0$; and $\bm{e}_2 = (0,0,...,0,1,1,...,1)'$ whose first $J$ elements are $0$ and last $J$ elements $1$.

Finally, define $P^1 = \{ P^1_{k,l} \}_{k, l = 1, \ldots, 2J} \in \bR^{2J \times 2J}$ such that $P^1$ only has non-zero values in the two off-diagonal blocks, i.e., for any $k, l = 1, \ldots, 2J,$
\begin{align*}
P^1_{k,l} = 
\left\{
\begin{aligned}
& 1, && k = l + J; \\
& 1, && k = l - J; \\
& 0, && \text{otherwise}.
\end{aligned}
\right.
\end{align*}

Using the above notations we re-write the (non-convex) QCQP as follows,
\begin{align}
\min \quad & \tbW' P^0 \tbW + \bm{q}^{0'} \tbW & \label{eqn:QCQPFormulation} \\
\mbox{\ \ \ s.t.\ \ } & \bm{e}_1' \tbW = 1, \nonumber \\
& \bm{e}_2' \tbW  = 1, \nonumber \\
& \tbW' P^1 \tbW  = 0, \nonumber \\
& \tbW  \geq \bm{0}. \nonumber
\end{align}

The first computational method (enumeration) solves two synthetic control problems in each iteration.
The synthetic control problems can be efficiently solved. We implement the synthetic control problem using the ``lsei'' function from ``limSolve'' package in R 4.0.2.
For the second computational method (quadratic programming), the problem \eqref{eqn:QCQPFormulation} is implemented using Gurobi 9.0.2 in R 4.0.2.
Since the QCQP is non-convex, the computation leads to some numerical errors up to $0.001$ in finding the treated and control weights.
So we round the treated and control weights to the nearest 2-digits in the implementation of the QCQP.
Moreover, for all the weights that are less than or equal to $0.01$, we trim the weights to zero.
This is because smaller weights suffer from greater impacts of numerical errors, and that numerical errors could cause zero weights to be non-zero, thus having a non-negligible impact on the swapping rule.

To conclude, we compare the treated and control weights calculated from both methods.
Both methods yield the same treated and control weights up to some negligible rounding error, while the first method takes longer computational time.

The other designs are computationally implemented using either one of the above two methods.
The \textit{Constrained} design is implemented using the enumeration method.
In cases when the cardinality constraint $\widebar{m}$ is small, this brute force enumeration is very efficient.
The $\NewFormulation$ design is implemented using the quadratic programming method.
In the QCQP formulation, the objective function has both a different quadratic term $P^0$ and a different linear term $\bm{q}^0$.
The \textit{Unit-Level} design is implemented using the enumeration method.
The \textit{Penalized} design is implemented using the quadratic programming method.
In the QCQP formulation, the objective function has the same quadratic term $P^0$ and a different linear term $\bm{q}^{0}$.

\clearpage

\section{Additional Illustrations Using Walmart Data}
\label{sec:Additional:Walmart}

\begin{figure}[t]\centering
\caption{Synthetic Treated Unit and Synthetic Control Unit, $\widebar{m} = 1$}
\label{fig:Walmart1Treat}
\includegraphics[width=0.9\textwidth]{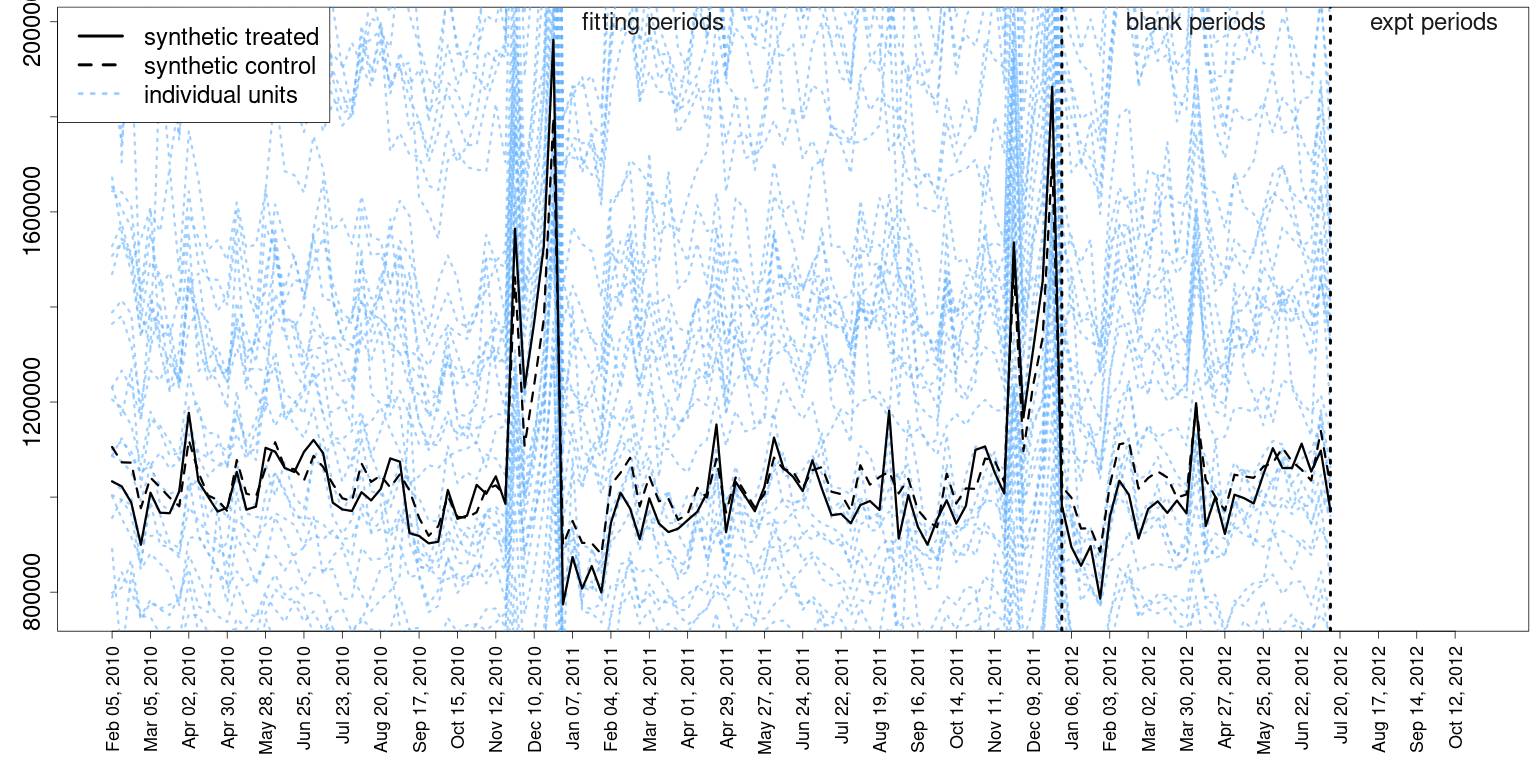}
\floatfoot{{\it Note:} The black solid line represents the synthetic treated outcome. The black dashed line represents the synthetic control outcome. The blue dashed lines are individual stores' sales.}
\end{figure}

\begin{figure}[h]\centering
\caption{Placebo Treatment Effects, $\widebar{m} = 1$}
\label{fig:WalmartResiduals1Treat}
\includegraphics[width=0.9\textwidth]{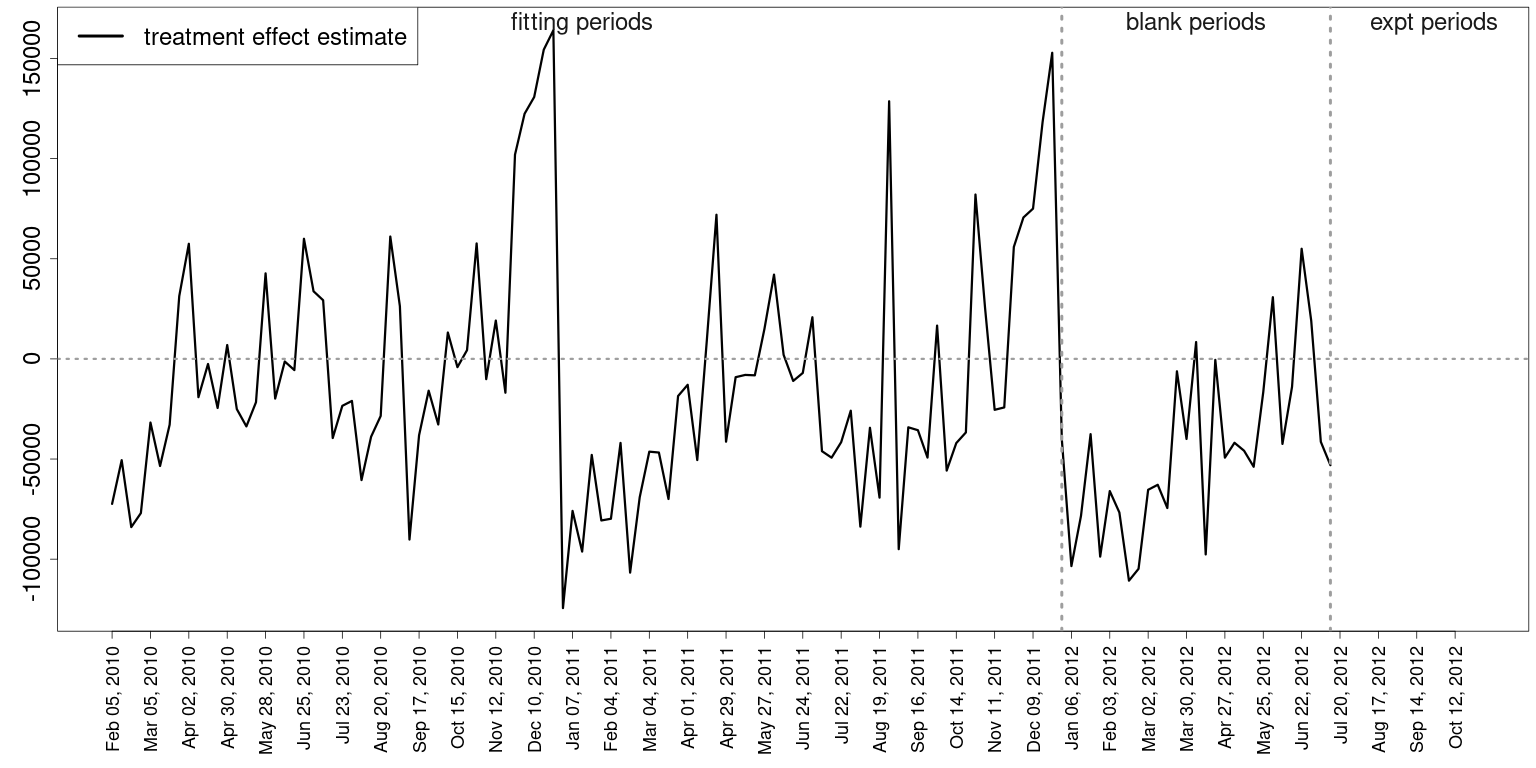}
\floatfoot{{\it Note:} This figure reports the difference between the synthetic treated and synthetic control outcomes of Figure~\ref{fig:Walmart1Treat}.}
\end{figure}

\begin{figure}[t]\centering
\caption{Synthetic Treatment Unit and Synthetic Control Unit, when $\widebar{m} = 3$.}
\label{fig:Walmart3Treat}
\includegraphics[width=0.9\textwidth]{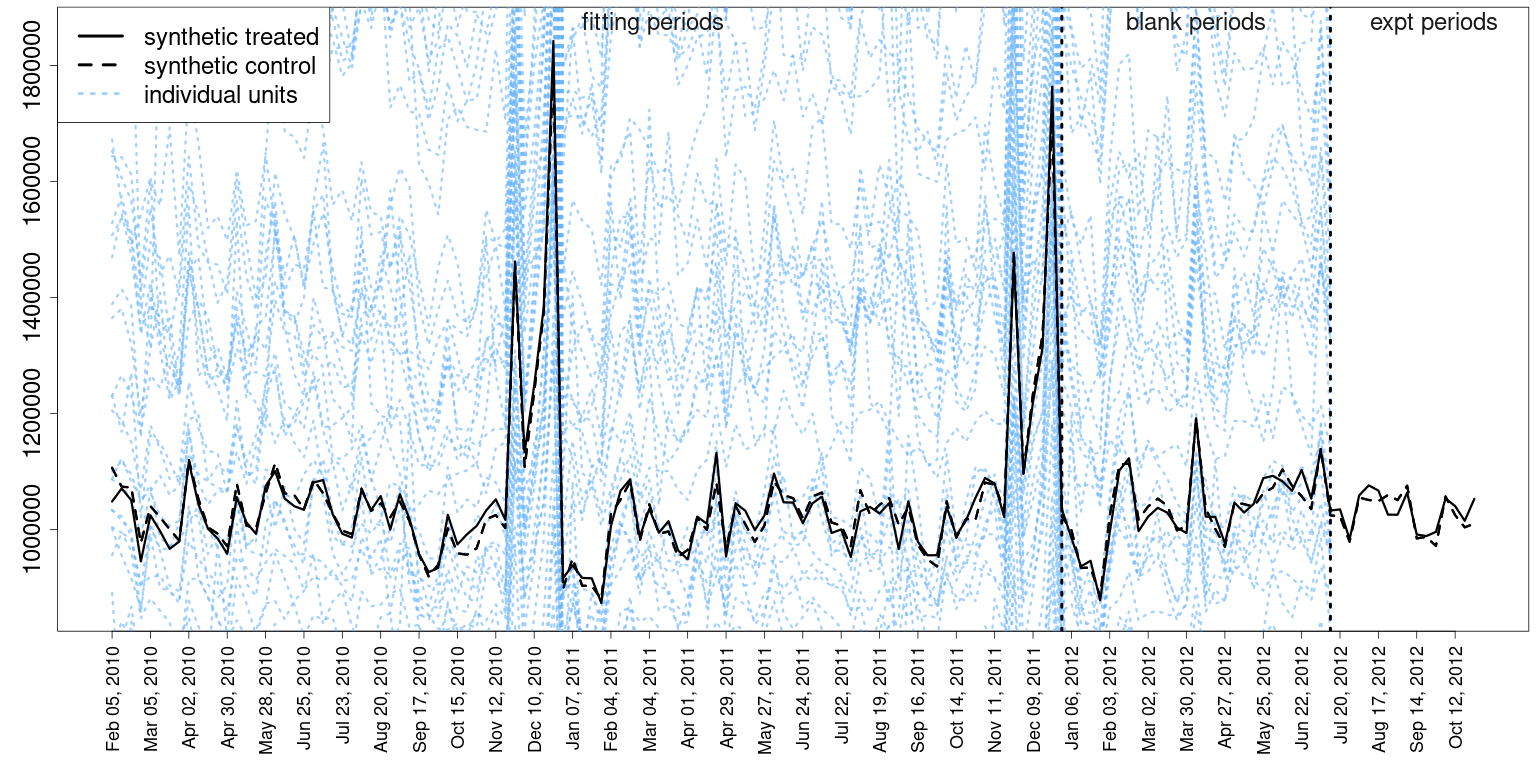}
\floatfoot{{\it Note:} The black solid line represents the synthetic treated outcome. The black dashed line represents the synthetic control outcome. The blue dashed lines are individual stores' sales.}
\end{figure}

\begin{figure}[h]\centering
\caption{Treatment Effect Estimate, when $\widebar{m} = 3$.}
\label{fig:WalmartResiduals3Treat}
\includegraphics[width=0.9\textwidth]{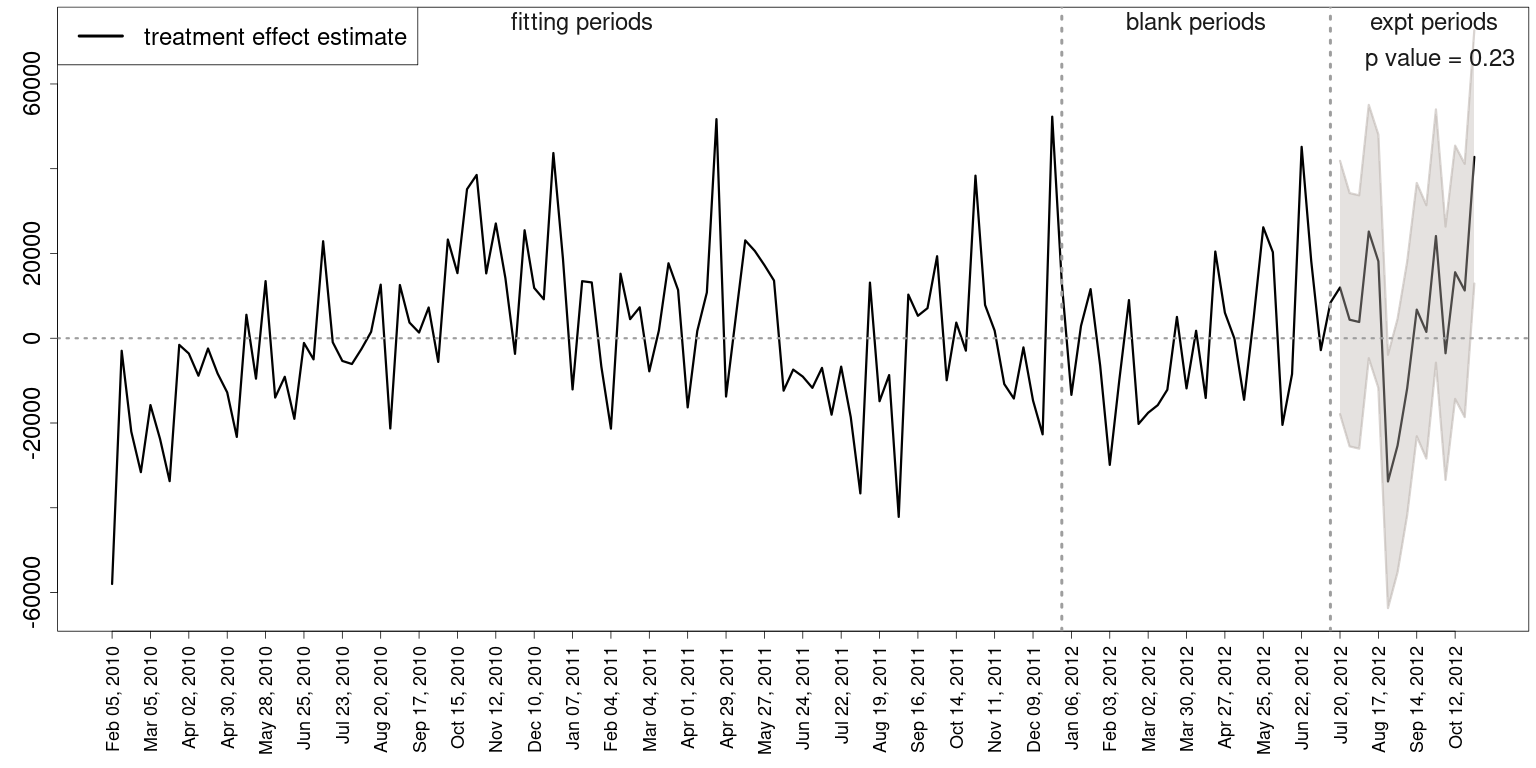}
\floatfoot{{\it Note:} This figure reports the difference between the synthetic treated and synthetic control outcomes of Figure~\ref{fig:Walmart3Treat}. For the experimental periods, this is the treatment effect estimate.}
\end{figure}

In this section, we present results for $\widebar m=1$ and $\widebar m=3$. Using only one treated unit ($\widebar m =1$) fails to produce a good fit between the treated and synthetic control unit in the fitting periods. For the case of $\widebar m =1$, Figures~\ref{fig:Walmart1Treat} and \ref{fig:WalmartResiduals1Treat} reveal a substantial gap with a clear seasonal trend between the two synthetic units. Figures~\ref{fig:Walmart3Treat} and \ref{fig:WalmartResiduals3Treat} report results for $\widebar m =3$. Increasing $\widebar m$ from $\widebar m =2$ to $\widebar m =3$ results in a minor improvement in fit, and leaves estimation results substantively unchanged. 

\clearpage

\clearpage

\section{Additional Simulation Results}
\label{sec:additional:simulation}

In Section~\ref{sec:simu:main} the idiosyncratic shocks are i.i.d. Normal with variance $\sigma^2=1$.
Figures~\ref{fig:ComputationalN(0,5)Noise} and~\ref{fig:ComputationalN(0,10)Noise} report results for $\sigma^2=5$ and $\sigma^2=10$, respectively. Figures~\ref{fig:ResidualsN(0,5)Noise} and~\ref{fig:ResidualsN(0,10)Noise} report differences between the outcomes for the synthetic treated and the synthetic control units for the same values for $\sigma^2$.
As the value of $\sigma^2$ increases, the quality of the post-treatment estimation and inference deteriorates, and the $p$-value for the null hypotheses of in \eqref{equation:null} increases. The deterioration in pre-treatment fit in Figures~\ref{fig:ComputationalN(0,5)Noise} and~\ref{fig:ComputationalN(0,10)Noise} provides a diagnosis of the accuracy of the respective estimates.

\clearpage

\begin{figure}[t]\centering
\includegraphics[width=0.9\textwidth]{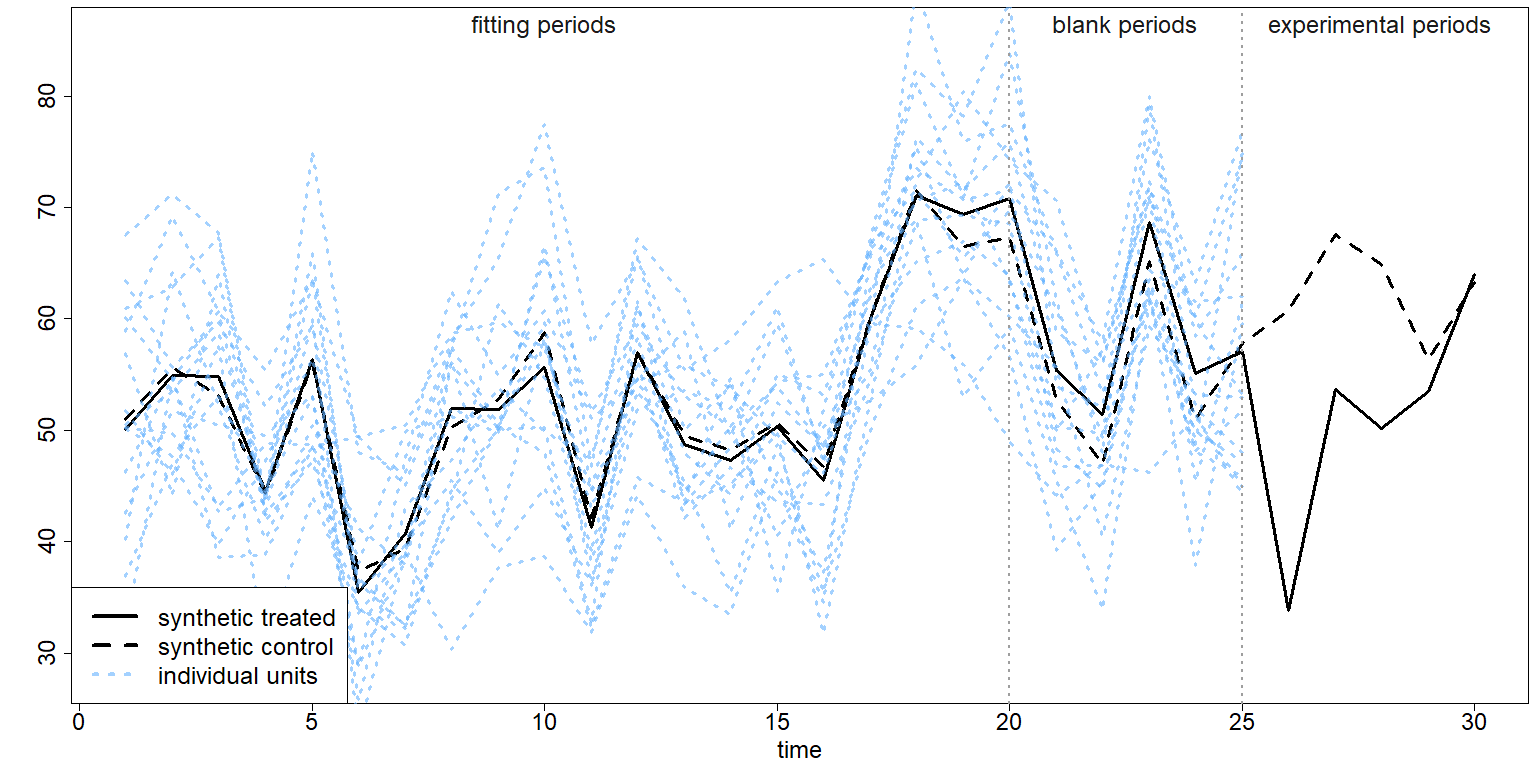}
\caption{Synthetic Treatment Unit and Synthetic Control Unit, when $\sigma^2 = 5$.}
\label{fig:ComputationalN(0,5)Noise}
\end{figure}

\begin{figure}[t]\centering
\includegraphics[width=0.9\textwidth]{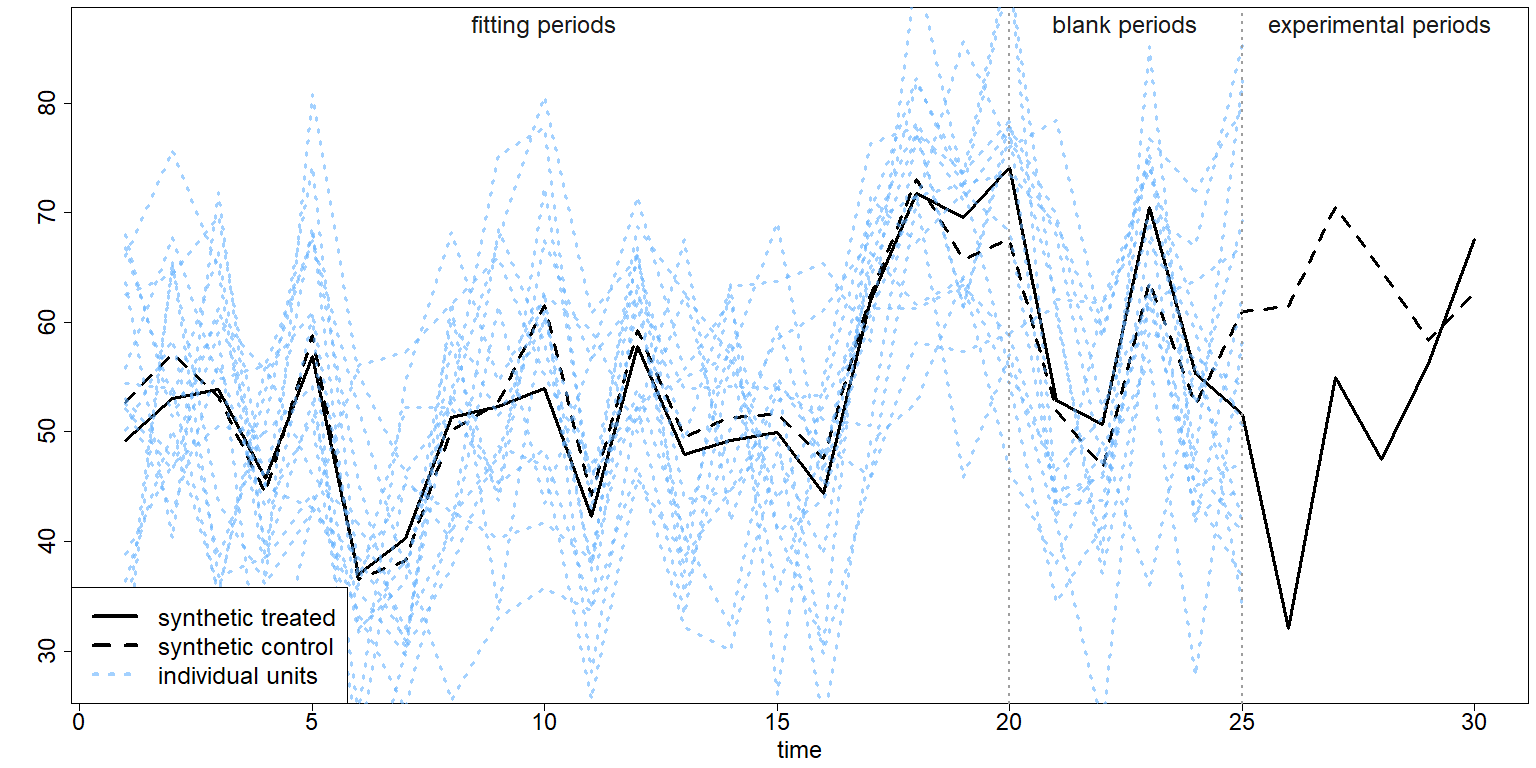}
\caption{Synthetic Treatment Unit and Synthetic Control Unit, when $\sigma^2 = 10$.}
\label{fig:ComputationalN(0,10)Noise}
\end{figure}

\clearpage
\begin{figure}[t]\centering
\includegraphics[width=0.9\textwidth]{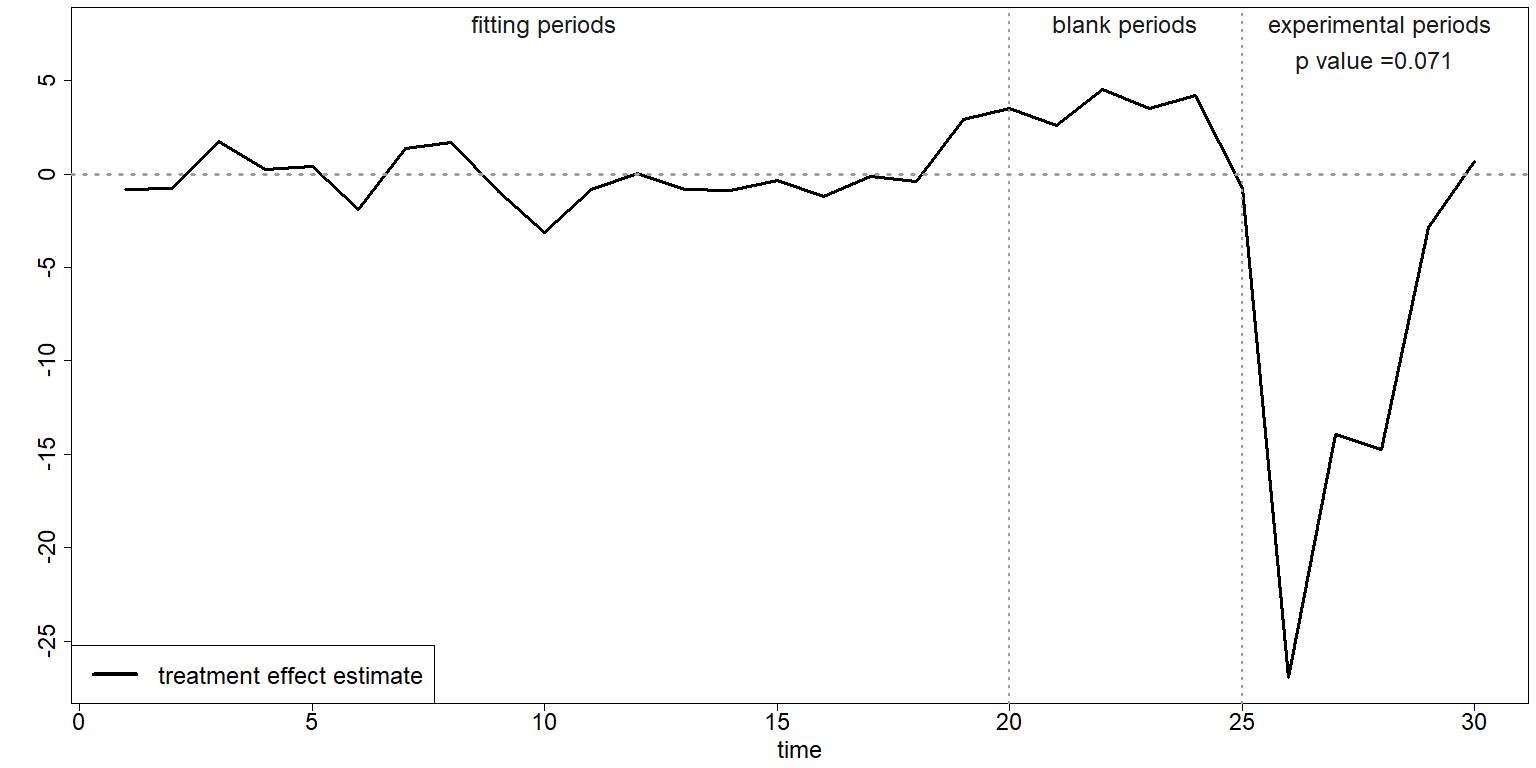}
\caption{Treatment Effect Estimate, when $\sigma^2 = 5$.}
\label{fig:ResidualsN(0,5)Noise}
\end{figure}

\begin{figure}[t]\centering
\includegraphics[width=0.9\textwidth]{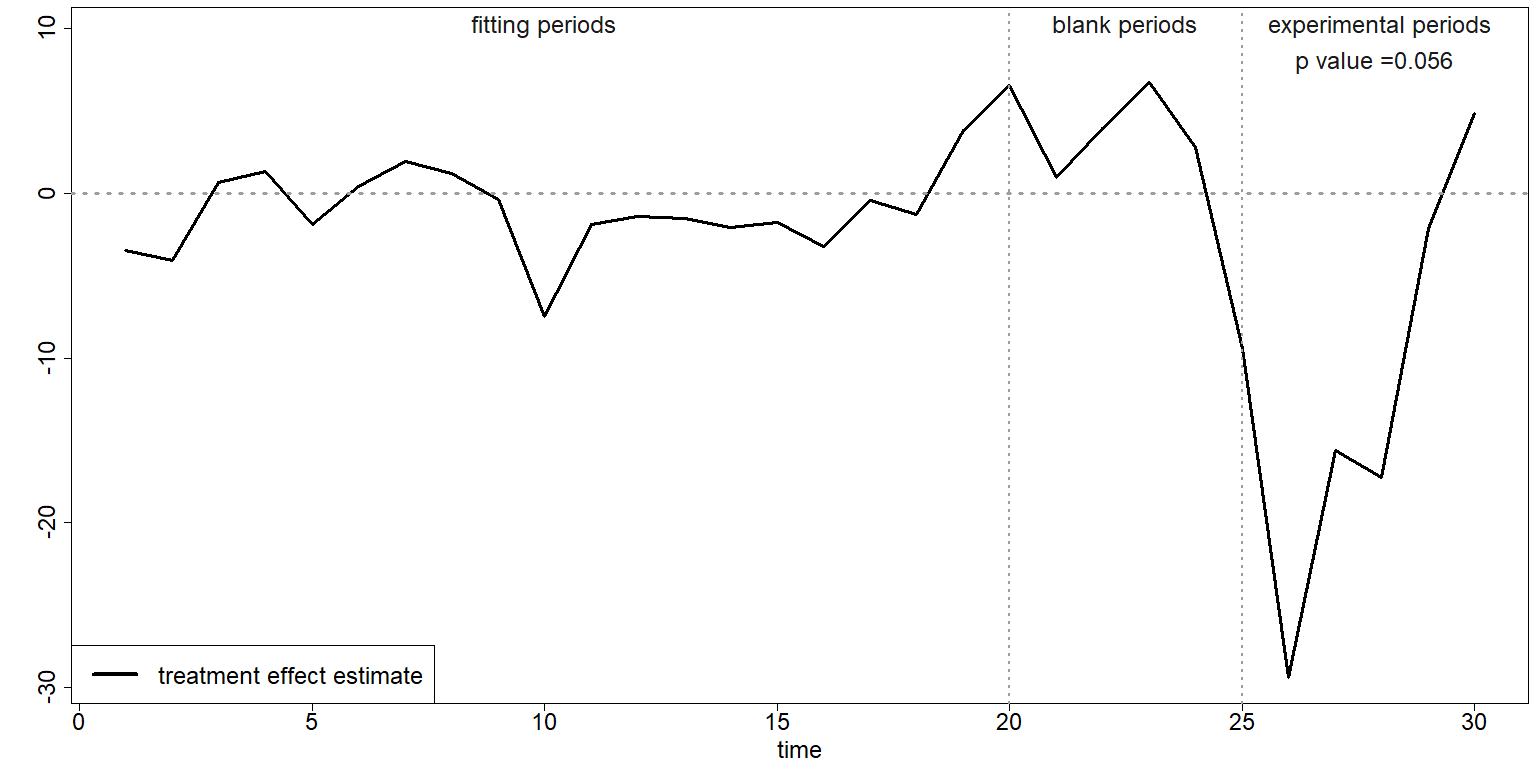}
\caption{Treatment Effect Estimate, when $\sigma^2 = 10$.}
\label{fig:ResidualsN(0,10)Noise}
\end{figure}

\clearpage

\end{appendices}

\end{document}